\documentclass[a4paper,11pt]{article}
\usepackage{jheppub}
\usepackage{hyperref}
\usepackage{booktabs,multirow}
\usepackage{amsmath,amssymb,mathtools,dsfont}
\usepackage{bbold}
\usepackage{bbm}
\usepackage{comment}
\usepackage{tcolorbox}
\usepackage{slashed}
\usepackage{makecell}
\usepackage[T1]{fontenc}

\usepackage{caption}
\usepackage{subcaption}

\usepackage[dvipsnames,table]{xcolor}

\usepackage{sectsty}
\allsectionsfont{\boldmath}
\usepackage{multirow}
\usepackage{tikz}

\newcommand{\iu}{\scalebox{1}[0.65]{$\mathbb{I}$}\kern-0.34em{\raisebox{4.5pt}{$\cdot$}}\kern0.1em} 

\newcommand{\dd}{\mathrm{d}}

\allowdisplaybreaks

\title{\boldmath Generalised $\Omega$-backgrounds in twisted $\mathcal{N}=4$ super Yang-Mills}

\author[a]{Elisa Iris Marieni,}
\author[a]{Itamar Yaakov}

\affiliation[a]{STAG Research Center \& Mathematical Sciences, University of Southampton\\Highfield, Southampton SO17 1BJ, UK}

\emailAdd{e.i.marieni@soton.ac.uk}
\emailAdd{i.yaakov@soton.ac.uk}

\abstract{We introduce rigid supergravity backgrounds for 4d $\mathcal{N}=4$ super Yang-Mills which deform its topological twists and generalise the $\Omega$-deformation of $\mathcal{N}=2$ gauge theories. We derive the twisted supersymmetry variations of the $\mathcal{N}=4$ Weyl multiplet, and the $\mathcal{N}=4$ matter multiplet coupled to a rigid bosonic background, for each of the three inequivalent topological twists. Other, more exotic backgrounds are also discussed.}

\begin{document}
\maketitle
\flushbottom

\section{Introduction}
Supersymmetric Quantum Field Theories (QFTs) admit a host of protected observables and supersymmetric deformations.  Computations of these observables have led to progress in understanding the physics of supersymmetry breaking \cite{Witten:1982df}; strong coupling dynamics of $\mathcal{N}=4$ super Yang-Mills (SYM) \cite{Vafa:1994tf}; and numerous aspects of duality, both duality between QFTs and holographic duality linking a QFT with a string theory. The computation of protected observables is often made feasible by the use of supersymmetric localization, which can be used to write down the exact non-perturbative answer.\footnote{The list of relevant references is immense. Early examples include \cite{Pestun:2007rz,Kim:2009wb,Kapustin:2009kz}. Further references can be found in the review \cite{Pestun:2016zxk}.}

Perhaps the most basic protected observables are supersymmetry preserving partition functions on compact Euclidean manifolds. A subset of these observables belong to a diverse collection of \textit{supersymmetric twists}, the first of which was introduced by Witten in \cite{Witten:1988ze}. The twists are special, in that they often yield topological quantum field theories (TQFTs), or more precisely models whose observables are insensitive to small changes in the metric structure of spacetime. The resulting connections with mathematics have been nothing short of spectacular.  For instance, Witten's original work linked a topologically twisted version of $\mathcal{N}=2$ super Yang-Mills in four dimensions with gauge group $SU(2)$, henceforth the Donaldson-Witten theory and DW twist, to Donaldson's smooth invariants and to instanton Floer homology \cite{Donaldson:1983wm, Floer:1988lge}. Computations in this model via physical means \cite{Witten:1994ev}, later combined with the Seiberg-Witten solution of the infra-red dynamics of the model \cite{Seiberg:1994rs,Seiberg:1994aj}, have led to new topological invariants and invaluable progress in our understanding of low-dimensional topology \cite{Witten:1994cg}. The Donaldson-Witten twist can be applied to any 4d theory with $\mathcal{N}=2$  supersymmetry, potentially yielding more powerful invariants \cite{Labastida:1995zj,Marino:1998bm}.  

There are also partial twists whose observables are insensitive only to some deformations of the metric structure, yielding a pseudo-topological model which is sensitive only to e.g. complex structure moduli. Partial topological twists are central to class-$\mathcal{S}$ theories \cite{Gaiotto:2009we} and related constructions. Holomorphic twists \cite{Bomans:2025klo} or the Maldacena-Nu\~nez twist used in holography \cite{Maldacena:2000yy} may require manifolds with additional geometric structures. 

Topologically twisting a supersymmetric theory amounts to redefining the structure group for the frame bundle by mixing it with the R-symmetry group $G_R$. Twisting is usually performed in Euclidean signature, with the group $\text{Spin}(d)$, and all discussions in this paper will involve \textit{Euclidean} four-manifolds and the group $\text{Spin}(4)\cong SU(2)_\text {l}\times SU(2)_\text{r}$.\footnote{In the case of special holonomy manifolds, the structure group may be reduced to a subgroup of  $\text{Spin}(d)$ and there are additional options for twisting.} Twisting means that one chooses a homomorphism $\varphi:\text{Spin}(4) \rightarrow G_R$ and uses it to define a twisted rotation group embedded in the product
\begin{align}
    \text{Spin}'(4) {\hookrightarrow} \text{Spin}(4) \times G_R \, .
\end{align}
Decomposing the supercharges under the twisted rotation group may produce one or more scalar supercharge $Q$. When such a supercharge exists, one may formulate a cohomological field theory composed only of $Q$-closed observables, i.e. with $Q$  acting as a BRST operator. If the energy momentum tensor is $Q$-exact, the model is topological in the sense that it is insensitive to small deformations of the metric. Using the terminology introduced in \cite{Blau:1996bx}, the number of such topological supercharges in the twisted theory is denoted $N_T$. 

The twisting procedure can be considered from a number of interesting viewpoints, including twisted dimensional reduction \cite{Ito:2013eva}, top-down string theory constructions \cite{Bershadsky:1995qy}, or in terms of a modified differential in the Batalin–Vilkovisky formalism \cite{Elliott:2020ecf}. The  procedure can also be systematically understood as a supersymmetry preserving coupling to off-shell background supergravity \cite{Festuccia:2011ws,Dumitrescu:2012ha,Dumitrescu:2012at,Closset:2012ru,Closset:2013vra,Closset:2014uda,Klare:2013dka}. This last approach, originally due to Festuccia and Seiberg, which is far more general in nature, is the one we follow throughout this paper.

In the Festuccia–Seiberg approach, one proceeds as follows:
\begin{enumerate}
    \item Couple the theory to a background off-shell supergravity multiplet. The correct multiplet depends on both the spacetime dimension and the symmetries of the theory, and is dictated by the existence of an appropriate type of conserved supercurrent multiplet \cite{Komargodski:2010rb}. We will couple to the Weyl multiplet of 4d superconformal $\mathcal{N}=4$ supergravity, presented in \cite{Bergshoeff:1980is}.\footnote{We use conformal supergravity because its additional gauge redundancies and auxiliary fields provide the most general off-shell background multiplet. Poincar\'e supergravity can subsequently be obtained by gauge fixing \cite{Kaku:1978ea}. The construction of Poincar\'e supergravity by gauge fixing conformal supergravity is known as the superconformal method of superconformal tensor calculus. See e.g. the introduction of \cite{vanMuiden:2017qsh} and references therein.}
    \item Take the rigid limit $M_P \rightarrow \infty$, which amounts to decoupling dynamical gravity so that the metric and all other supergravity fields are fixed to classical values;
    \item  Choose the values of these background fields so that the supersymmetry variations for the supergravity fields vanish for at least one non-zero supersymmetry parameter.\footnote{Neither the equations of motion nor the reality conditions for the supergravity fields are relevant in this procedure.}
\end{enumerate}

Conventionally, one solves for the vanishing of the supergravity variations, which we refer to simply as the "supergravity equations", at a \textit{bosonic} fixed point where all fermionic supergravity fields vanish. Consequently, the bosonic variations vanish automatically, and supersymmetry requires only the fermionic variations to vanish. From this perspective, the twisted supercharge of $\mathcal{N}=2$ super Yang-Mills found by Witten can be understood as a Killing spinor of $\mathcal{N}=2$ conformal supergravity \cite{Karlhede:1988ax}. The supersymmetry-preserving background is obtained by identifying the relevant $SU(2)$ R-symmetry connection with the corresponding component of the spin connection selected by the twisting homomorphism $\varphi$. In this example, all the fields in the $\mathcal{N}=2$  Weyl multiplet except the vielbein, the $SU(2)$ R-symmetry connection, and an auxiliary field $D$ are turned off.\footnote{There are at least two conventions for the definition of the auxiliary field $D$, one of which requires it to also vanish for the twisted background.}

The supersymmetry algebra of a twisted theory with a single BRST supercharge is given schematically by \begin{align}
    [\delta_Q, \delta_Q] = \delta_{\text{gauge}} \, ,
\end{align}
where $\delta_{\text{gauge}}$  are ordinary gauge transformations of the twisted model, whose parameter depends on both dynamical and background fields. $\delta_Q$ can be used, for instance, to show that the original Donaldson-Witten theory localizes to the moduli space of $SU(2)$ instantons. 

The same theories which admit the DW twist also admit a further supersymmetric deformation, compatible with the twist, called the $\Omega$-deformation, which is parameterised by a Killing vector field $V$ \cite{Losev:1997tp,Nekrasov:2002qd, Nekrasov:2003rj}. The supersymmetry algebra in the deformed-twisted theory is, schematically, 
\begin{align}\label{eq: algebra omega def}
    [\delta_{Q}, \delta_{Q}]= \mathcal{L}_{V} + \delta_{\text{gauge}} \, ,
\end{align}
where $\mathcal{L}_{V} $ denotes the Lie derivative with respect to $V$. The $\Omega$-deformed theory is no longer a TQFT. It can be formulated only on manifolds with metrics which support Killing vectors and its observables depend explicitly on the choice of $V$. The partition function of the  $\Omega$-deformed theory can be localized to the fixed points of the action of $V$ -- a substantial improvement in terms of computability. The $\Omega$-deformed partition function on $\mathbb{R}^4$, where the DW twist itself is trivial, yields Nekrasov's instanton partition function, which can be used to recover the Seiberg-Witten solution \cite{Nekrasov:2002qd}. The $\Omega$-deformed $\mathcal{N}=2$ theory can also be understood from string and M-theory constructions \cite{Hellerman:2012zf,Angelantonj:2022dsx}.

Motivated by the success of twisted and $\Omega$-deformed $\mathcal{N}=2$ theories, we wish to systematically study $\Omega$-deformed supersymmetric backgrounds of $\mathcal{N}=4$ super Yang–Mills theory. There are three inequivalent topological twists of $\mathcal{N}=4$ super Yang-Mills, corresponding to different choices of homomorphism $\varphi$. These are the Donaldson-Witten (DW) or "half" twist \cite{Yamron:1988qc}, the Yamron or Vafa-Witten (VW) twist \cite{Yamron:1988qc, Vafa:1994tf}, and the Marcus or Geometric Langlands (GL) twist \cite{Marcus:1995mq, Kapustin:2006pk}. The various twists also have anti-twist versions, exchanging the left and right $SU(2)$ factors of $\text{Spin}(4)$, which are related to the original twists by a parity transformation. 

Analogues of $\Omega$-deformations for $\mathcal{N}=4$  twists were shown to exist on $\mathbb{R}^4$, via a twisted dimensional reduction argument, by Ito et al. \cite{Ito:2013eva, Ito:2011cr,Ito:2012hs}. In this paper, we will find $\Omega$-backgrounds for each of the twisted theories by solving the supergravity equations on an arbitrary Euclidean manifold. In order to facilitate a systematic search, we begin by twisting the appropriate fields in the Weyl multiplet, in analogy with the treatment of $\mathcal{N}=2$ in \cite{Cushing:2023rha}.  We will briefly compare our results to those of Ito et al. and to the twisted but undeformed theories. 

The rest of this paper is organised as follows. In Section \ref{sec: sugra eq}, we review the relevant features of $\mathcal{N}=4$ conformal supergravity and collect the supersymmetry variations of the Weyl and matter multiplet fields at the bosonic fixed point in Euclidean signature. We then twist the supersymmetry variations with respect to the three twists of $\mathcal{N}=4$. In Section \ref{sec: susy backgrounds}, we analytically solve the gravitino and gaugino field variations and compute the closure of the algebra for different types of backgrounds. These include backgrounds of $\Omega$-type for each one of the twists, a special background with varying gauge coupling, and a fuzzy sphere background. We did not solve the dilatino equation analytically, but rather checked it numerically on selected geometries collected in Appendix \ref{sec: metrics}. Our conventions are collected in Appendix \ref{sec: conventions}, while the lengthy variations of the twisted matter fields can be found in Appendix \ref{sec: vec mult transf}. In Appendix \ref{sec: fuzzy matching}, we connect our fuzzy sphere background to the one discussed in \cite{Kato:2011yh}.

\section{Supergravity equations}\label{sec: sugra eq}

\begin{table}[t]
    \centering
    \begin{tabular}{lccc}
    \toprule
        Local symmetry & Generator & Field & $SU(4)_R$ \\
        \midrule
        Translations & $P_a$ & $e^a_\mu$ & $\mathbf 1$ \\
        Dilatations & $D$ & $b_\mu$& $\mathbf 1$\\
        Poincar\'e SUSY & $Q^i$ & $\psi_{\mu}{}^i$& $\mathbf 4$\\
        $SU(4)$ R-symmetry & $U^i{}_j$ & $V_\mu^i{}_j$& $\mathbf{15}$\\
        \midrule
        Lorentz & $M_{ab}$ & $\omega_{\mu}{}^{ab}$& $\mathbf 1$\\
        Conformal SUSY & $S_i$ & $\phi_{\mu i}$& $\mathbf 4$\\
        Special Conformal & $K_a$ & $f^a_\mu$& $\mathbf 1$\\
        $U(1)$ bonus symmetry & $A$ & $a_\mu$& $\mathbf 1$\\
        \midrule
        &&$\Phi_\alpha$& $\mathbf 1$\\
        && $E_{ij}$& $\mathbf{\overline{10}}$\\
        Auxiliary && $T_{ab}{}^{ij}$& $\mathbf 6$\\
        fields&& $D^{ij}{}_{kl}$& $\mathbf{20}$\\
        && $\Lambda_i$& $\mathbf{\overline{4}}$\\
        && $\chi^{ij}{}_k$& $\mathbf{\overline{20'}}$\\
        \bottomrule
    \end{tabular}
    \caption{Field content of the off-shell $\mathcal N=4$ Weyl multiplet. 
The first block lists the gauge fields associated with the independent local superconformal symmetries, while the second block contains the gauge fields that become composite after imposing conventional constraints. The final block lists the auxiliary fields.}
    \label{tab: generators}
\end{table}

 $\mathcal{N}=4$ super Yang-Mills theory can be coupled to $\mathcal{N}=4$ conformal supergravity, whose Lorentzian formulation was given in \cite{Bergshoeff:1980is}. The underlying symmetry algebra for this supergravity is $SU(2,2|4) \simeq SO(4,2|4)$. The approach of superconformal tensor calculus is based on gauging this algebra. The algebra contains the generators of the $SU(2,2)$ conformal group, the  Poincar\'e and conformal supersymmetries, and the R-symmetry. Poincar\'e supersymmetries are denoted collectively by $Q$, while conformal supersymmetries are denoted by $S$. 
 
 Aside from the local $SU(2,2|4)$ symmetry, this supergravity also has a rigid $U(1)$ invariance which is not a symmetry of $\mathcal{N}=4$ SYM. In fact, it is useful to consider an alternative off-shell formulation with an even larger redundancy. As noted in \cite{Bergshoeff:1980is}, one can formulate off-shell $\mathcal{N}=4$ conformal supergravity in terms of a local $SU(2,2|4) \times U(1)$ and an additional  \textit{rigid} $SU(1,1)\cong SL(2,\mathbb{R})$  symmetry. The local $U(1)$ is often referred to as a \textit{bonus symmetry}. After gauge-fixing, it reduces to the rigid chiral $U(1)$ of the original formulation. The bonus symmetry is closely related to the action of the $SL(2,\mathbb{Z})$ S-duality group on the supercharges \cite{Kapustin:2006pk}.\footnote{In the case of simply-laced groups, the S-duality group is $SL(2,\mathbb{Z})$, but it can be a different discrete subgroup of $SL(2,\mathbb{R})$ in the non-simply-laced case. See \cite{Kapustin:2006pk} for more details.} 

The field content of $\mathcal{N}=4$ conformal supergravity, including the auxiliary fields, is summarised in table \ref{tab: generators}. The relationship between the fields and the gauging of the superconformal algebra is explained in \cite{Bergshoeff:1980is}. We will also make use of the on-shell matter multiplet of $\mathcal{N}=4$, which comprises the field content of pure $\mathcal{N}=4$ super Yang-Mills. This is a dynamical vector multiplet, depending on a choice of gauge group, whose field content is summarised in table \ref{tab: vec mult}. In Lorentzian signature, the gauge field $A_\mu$  is real, the scalars $\phi^{ij}$ satisfy a reality condition, and the spinor $\lambda_{\alpha}{}^i$ is a left-handed Weyl spinor \cite{deRoo:1984zyh,deRoo:1985np}. All the fields in the vector multiplet are adjoint valued.

\begin{table}[t]
    \centering
    \begin{tabular}{cc}
    \toprule
        Field & $SU(4)_R$ \\
        \midrule
        $A_\mu$ & $\mathbf{1}$\\
        $\lambda_{\alpha}{}^i$ & $\mathbf{4}$\\
        $\phi^{ij}$ & $\mathbf{6}$\\
        \bottomrule
    \end{tabular}
    \caption{On-shell field content of the $\mathcal{N}=4$ vector multiplet.}
    \label{tab: vec mult}
\end{table}

In the following we will use Greek letters from the middle of the alphabet $(\mu,\nu,\ldots)$ to denote coordinate indices and Latin letters $(m,n,\ldots)$ for frame indices. Indices $i,j,k,l$ denote the $SU(4)_{R}$ symmetry. We will also employ the chiral notation of \cite{Bergshoeff:1980is}, where upper and lower $SU(4)$ indices denote the defining and conjugate representations respectively. Spinor conventions, index contractions and our Euclidean Clifford algebra conventions are summarised in Appendix~\ref{sec: clifford}. Specifically, Greek indices from the beginning of the alphabet $(\alpha, \beta,...)$ denote left-handed $SU(2)$ indices, the single exception being that the field $\Phi^\alpha$, has an index structure which indicates its transformation properties under the rigid $SU(1,1)$.

\subsection{Euclidean $\mathcal{N}=4$ Equations}\label{sec: euclidean equations}
We begin by writing down the Euclidean version of the multiplets and their supersymmetry transformations. Fields in Euclidean signature are complexifications of their Lorenzian counterparts. Specifically, fields which were real in Lorentzian signature become complex, while complex fields become complex doublets of independent fields valued in complex conjugate representations of the symmetry algebra. We will denote the second field in the doublet with a bar.  

The Euclidean Weyl multiplet fields satisfy the algebraic constraints summarised in table~\ref{tab:weyl_constraints}. In the equations, we have replaced $\overline D_{ij}{}^{kl}$ by its expression in terms of $D^{ij}{}_{kl}$.
\begin{table}[t]
\centering
\begin{tabular}{ll}
\toprule Field & Algebraic Constraints \\
\midrule $T_{ab}{}^{ij}$ & \( T_{ab}{}^{ij} = -T_{ba}{}^{ij} = -T_{ab}{}^{ji}, \quad T_{ab\,ij}= \frac12 \varepsilon_{ijkl}T_{ab}{}^{kl} \) \\[1ex] $\overline T_{abij}$ & \( \overline T_{abij} = -\overline T_{baij} = -\overline T_{abji}, \quad \overline T_{ab}{}^{ij}= \frac12 \varepsilon^{ijkl}\overline T_{ab\, kl} \) \\[1ex] $E_{ij}$ & \( E_{ij}=E_{ji} \) \\[1ex] $\overline E^{ij}$ & \( \overline E^{ij} = \overline E^{ji} \) \\[1ex] $D^{ij}{}_{kl}$ & \( D^{ij}{}_{kl} = -D^{ji}{}_{kl} = -D^{ij}{}_{lk}, \qquad D^{ij}{}_{kj}=0, \qquad D^{ij}{}_{kl} = \frac14 \varepsilon^{ijmn} \varepsilon_{klpq} D^{pq}{}_{mn} \) \\[2ex] $\overline D_{ij}{}^{kl}$ & \( \overline D_{ij}{}^{kl} = -\overline D_{ji}{}^{kl} = -\overline D_{ij}{}^{lk}, \qquad \overline D_{ij}{}^{kj}=0, \qquad \overline D_{ij}{}^{kl} = \frac14 \varepsilon_{ijmn} \varepsilon^{klpq} \overline D_{pq}{}^{mn} \) \\[2ex] & \( \overline D_{kl}{}^{ij} = \frac14 \varepsilon^{ijmn} \varepsilon_{klpq} D^{pq}{}_{mn} \) \\[2ex] $\chi^{ij}{}_{k}$ & \( \chi^{ij}{}_{k} = -\chi^{ji}{}_{k}, \qquad \chi^{ij}{}_{i}=0 \) \\[1ex] $\overline\chi_{ij}{}^{k}$ & \( \overline\chi_{ij}{}^{k} = -\overline\chi_{ji}{}^{k}, \qquad \overline\chi_{ij}{}^{i}=0 \) \\[1ex] $\Phi_\alpha,\overline\Phi^\alpha$ & \( \overline\Phi^\alpha \Phi_\alpha =1 \) \\[1ex] $V_\mu{}^i{}_j,\overline V_{\mu i}{}^j$ & \( V_\mu{}^i{}_i=0, \qquad \overline V_{\mu i}{}^i=0, \qquad \overline V_{\mu i}{}^j = -(V_\mu^T)_i{}^j \) \\ \bottomrule
\end{tabular}
\caption{Algebraic constraints obeyed by the bosonic and fermionic fields of the Euclidean $\mathcal N=4$ Weyl multiplet. Notice that in Euclidean signature the Lorentzian reality conditions on $D^{ij}{}_{kl},V_\mu{}^i{}_j$ become algebraic conditions relating them to their barred counterpart.}
\label{tab:weyl_constraints}
\end{table}
The structure of the equations forces $T_{ab}{}^{ij}$ to be self-dual in the frame indices, while $\overline{T}_{ab \, ij}$ is anti-self-dual.  

Performing the path integral over the \textit{dynamical fields} of e.g. $\mathcal{N}=4$ SYM in Euclidean signature requires choosing a contour of integration: a half-dimensional subspace. No such choice is necessary for background fields, and in fact interesting observables are often defined with values for these fields which violate the standard reality conditions \cite{Festuccia:2011ws}.\footnote{As noted, these values also need not satisfy the equations of motion for any specific supergravity action. In fact, the supergravity action plays no role in the construction of rigid backgrounds, although it can affect the result for observables via classical contributions.} For simplicity, we will write the equations for the fields in the  $\mathcal{N}=4$  vector multiplet assuming that the relevant contour makes the connection $A_\mu$ real in Euclidean signature. 

The Euclidean supersymmetry equations are extracted from those in Lorentzian signature by considering both the Lorentzian signature equations and their complex conjugates. One must then complexify the fields, replacing real with complex fields and complex conjugate fields with independent complex fields. Experience shows that this procedure can fail to produce a field representation of the Euclidean supersymmetry algebra without additional choices for phases appearing in the transformations. We will make such choices at an appropriate point below.\footnote{We verify that our choices lead to a closure of the algebra in the vector multiplet. However, the choices require changing a sign in the original transformations. More standard choices may exist, but would likely necessitate a redefinition of the supergravity fields.} 

\paragraph{The equations for the Euclidean Weyl multiplet} We would like to obtain values for the supergravity fields and the supersymmetry parameters for which  all supersymmetry variations of the Euclidean Weyl multiplet vanish --  a fixed point. We consider only bosonic fixed points: configurations where all the fermionic fields are turned off. Thus, the relevant equations for us are the variations with respect to the $Q$ and $S$ supersymmetry of the gravitino $\psi_{\mu}{}^{i}$, the gaugino $\Lambda_{i}$ and the dilatino $\chi^{ij}{}_{k}$. 

The variations at a Euclidean bosonic fixed point read (see \cite{Bergshoeff:1980is, Maxfield:2016lok} for the Lorentzian version)
\begin{subequations}\label{eq: Maxfield sugra eq}
\begin{align}
    \delta \psi_{\mu}{}^{i} =& \, \, \mathcal{D}_\mu \xi^{i} - \frac{1}{4} T_{ab}{}^{ij} \gamma^{ab} \gamma_\mu \overline \xi_{j} - \frac{1}{4} \gamma_\mu \gamma^\nu \mathcal{D}_\nu \xi^{i} \, , \\
     \delta \overline\psi_{\mu}{}_{i} =& \, \, \mathcal{D}_\mu \overline \xi_{i} - \frac{1}{4} \overline{T}_{ab}{}_{ij} \gamma^{ab} \gamma_\mu \xi^{j} - \frac{1}{4} \gamma_\mu \gamma^\nu \mathcal{D}_\nu \overline \xi_{i} \, , \\
    \delta \Lambda_{i} =& \, \, 2 \varepsilon^{\alpha \beta} \Phi_{\alpha} \mathcal{D}_\mu \Phi_{\beta} \gamma^\mu \overline \xi_i + E_{ij} \xi^{j} + \frac{1}{2} \varepsilon_{ijkl} T_{ab}{}^{kl} \gamma^{ab} \xi^{j} \, ,  \\
     \delta \overline \Lambda^{i} =& \, \, 2 \varepsilon_{\alpha \beta} \overline{\Phi}^{\alpha} \mathcal{D}_\mu \overline{\Phi}^{\beta} \gamma^\mu \xi^i + \overline{E}^{ij} \overline \xi_{j} + \frac{1}{2} \varepsilon^{ijkl} \overline{T}_{ab}{}_{kl} \gamma^{ab} \overline \xi_{j} \, ,  \\
    \delta \chi^{ij}{}_{k} =& - \frac{1}{2} \mathcal{D}_\mu T_{ab}{}^{ij} \gamma^{ab} \gamma^\mu \overline \xi_{k} - R_{ab}{}^{[i}{}_{k} \gamma^{ab} \xi^{j]} - \frac{1}{2} \varepsilon^{ijlm} \mathcal{D}_\mu E_{kl} \gamma^{\mu} \overline \xi_{m} + D^{ij}{}_{kl} \xi^{l} \nonumber \\
    & + \frac{1}{4}T_{ab}{}^{ij} \gamma^{ab} \gamma^\nu \mathcal{D}_\nu \overline\xi_{k} + \frac{1}{12} \varepsilon_{klmn} T_{ab}{}^{mn} \overline{E}^{l[i} \gamma^{ab} \xi^{j]} + \frac{1}{2} E_{kl} \overline{E}^{l[i} \xi^{j]} \nonumber \\
    &- \frac{1}{4} \varepsilon^{ijlm} E_{kl} \gamma^\nu \mathcal{D}_\nu \overline \xi_m + \frac12 \varepsilon^{ijlm}\varepsilon^{\alpha \beta} \Phi_{\alpha}\mathcal{D}_\mu \Phi_{\beta} \overline{T}_{abkl} \gamma^\mu \gamma^{ab} \overline \xi_{m} - (\text{traces})\, , \\
    \delta \overline \chi_{ij}{}^{k} =& - \frac{1}{2} \mathcal{D}_\mu \overline{T}_{ab}{}_{ij} \gamma^{ab} \gamma^\mu \xi^{k} - \overline{R}_{ab}{}_{[i}{}^{k} \gamma^{ab} \overline \xi_{j]} - \frac{1}{2} \varepsilon_{ijlm} \mathcal{D}_\mu \overline E^{kl} \gamma^{\mu} \xi^{m} + \frac14 \varepsilon^{klmn}\varepsilon_{ijpq} D^{pq}{}_{mn} \overline \xi_{l} \nonumber \\
    & + \frac{1}{4} \overline{T}_{ab}{}_{ij} \gamma^{ab} \gamma^\nu \mathcal{D}_\nu \xi^{k}+ \frac{1}{12} \varepsilon^{klmn} \overline{T}_{ab}{}_{mn} E_{l[i} \gamma^{ab} \overline \xi_{j]} + \frac{1}{2} \overline{E}^{kl} E_{l[i} \overline \xi_{j]} \nonumber \\
    &- \frac{1}{4} \varepsilon_{ijlm} \overline{E}^{kl} \gamma^\nu \mathcal{D}_\nu \xi^m + \frac12 \varepsilon_{ijlm}\varepsilon_{\alpha \beta} \overline{\Phi}^{\alpha}\mathcal{D}_\mu \overline{\Phi}^{\beta} T_{ab}{}^{kl} \gamma^\mu \gamma^{ab} \xi^{m} - (\text{traces})\, ,
\end{align}
\end{subequations}
where the gravitino equations have been rescaled by a factor of 2. $\xi^i, \overline \xi_i$ are projected Dirac spinor parameters of the $Q$-supersymmetry transformations, as defined in Appendix \ref{sec: clifford}. $\xi^i=(\epsilon_{\alpha}{}^i,0)$ only depends on the left-handed Weyl spinor and $\overline \xi_i=(0, \overline \epsilon^{\dot \alpha}{}_i)$ on the right-handed one, so the equations are chiral. At the fixed point, the $S$-supersymmetry parameters $\overline \eta^i, \eta_i$  can be written in terms of $\xi^i, \overline \xi_i$ as
\begin{align}
    \overline \eta^i = \frac12 \gamma^\nu \mathcal{D}_\nu \xi^i \, ; \quad \eta_i = \frac12 \gamma^\nu \mathcal{D}_\nu \overline \xi_i \, ,
\end{align}
by using the $\gamma$-trace of the equations, and these values have been substituted back into the equations above. Unless otherwise stated, suppressed Dirac indices are contracted in their natural positions according to the conventions reviewed in Appendix~\ref{sec: clifford}.

The covariant derivatives and curvature are defined as \cite{Maxfield:2016lok} (see \cite{Bergshoeff:1980is} for the last one)
\begin{equation}
\begin{aligned}
    \mathcal{D}_\mu \xi^i &= \partial_\mu \xi^i + \frac{1}{4} \omega_\mu {}^{ab} \gamma_{ab}\xi^i - V_\mu^{i}{}_{j} \xi^j -\frac12 a_\mu \xi^i \, , \\
    \mathcal{D}_\mu T_{ab}{}^{ij} &= \partial_\mu T_{ab}{}^{ij} + 2\omega_\mu{}^{c}{}_{[a}T_{b]c}{}^{ij} + 2V_{\mu}^{[i}{}_{k} T_{ab}{}^{j]k} -a_\mu T_{ab}{}^{ij} \, ,  \\
    \mathcal{D}_\mu E_{ij} &= \partial_\mu E_{ij} + 2V^{k}_{\mu}{}_{(i}E_{j)k} -a_\mu E_{ij} \, ,\\
    R_{\mu \nu}{}^{i}{}_{j} &= 2 \partial_{[\mu}V_{\nu]}^{i}{}_{j} - 2V^{i}_{[\mu}{}_{k} V^k_{\nu]}{}_{j} = \dd V - V \wedge V \, , \\
    \mathcal{D}_{\mu} \Phi_{\alpha} &= \partial_{\mu} \Phi_{\alpha} + \text{fermions} \, .
\end{aligned}
\end{equation}
From which we can deduce also
\begin{equation}
\begin{aligned}\label{eq: conjugate relations}
\mathcal{D}_\mu \overline\xi_i &= \partial_\mu \overline\xi_i + \frac{1}{4} \omega_\mu{}^{ab} \gamma_{ab} \overline \xi_i - \overline{V}_{\mu}{}_{i}{}^{j} \overline \xi_j +\frac12  a_\mu \overline \xi_i \, , \\
    \mathcal{D}_{\mu}\overline{T}_{ab\,ij} &= \partial_\mu \overline T_{ab\, ij} + 2\omega_\mu{}^{c}{}_{[a} \overline{T}_{b]c\,ij} + 2 \overline{V}_{\mu}{}_{[i}{}^{k}\overline{T}_{ab\, j]k} + a_\mu \overline T_{ab\,ij}\, ,  \\
    \mathcal{D}_\mu \overline{E}^{ij} &= \partial_\mu \overline{E}^{ij} + 2\overline{V}_{\mu}{}_{k}{}^{(i} \overline{E}^{j) k} + a_\mu \overline E^{ij} \, ,\\
    \overline{R}_{\mu \nu}{}_{i}{}^{j} &= 2 \partial_{[\mu} \overline{V}_{\nu]}{}_{i}{}^{j} - 2 \overline{V}_{[\mu}{}_{i}{}^{k} \overline{V}_{\nu]}{}_{k}{}^{j} = \dd \overline{V} - \overline{V} \wedge \overline{V} \, , \\
    \mathcal{D}_{\mu} \overline{\Phi}^{\alpha} &=  \partial_{\mu} \overline{\Phi}^{\alpha} + \text{fermions} \, .
\end{aligned}
\end{equation}
All differential-geometric conventions, including the normalisation of the exterior derivative and the Hodge dual, are summarised in Appendix \ref{sec: diff geo}. Since we will be working at the bosonic fixed point, the covariant derivative of $\Phi, \overline{\Phi}$ is just a partial derivative. Notice that $\overline{a}_\mu=-a_\mu$ (cfr \eqref{eq: amu def}). For our specific backgrounds, we will in fact always have $a_\mu =0$. This is made possible by an appropriate choice of $\Phi$ which we will expand on later in this section.

\paragraph{Comments on the field $\Phi$.}
In Lorentzian signature, $\Phi_{\alpha}, \,\, \alpha=1,2$ is a doublet of complex fields. $\Phi^\alpha$, on the other hand is \textit{defined} by the following equation
\begin{align}\label{eq: conjugate of Phi}
    \Phi^\alpha := \eta^{\alpha \beta} \Phi_{\beta}^\ast \, ,
\end{align}
with
\begin{align}
    \eta=\text{diag}\left(1,-1\right)\, .
\end{align}
This pair of doublets is subject to the $SU(1,1)\times U(1)$ invariant constraint
\begin{align}
    \Phi^\alpha \Phi_{\alpha} = 1 \, .
\end{align}
The purely imaginary composite $U(1)$ connection is defined in Lorentzian signature as
\begin{align}
    a_\mu = \frac12 \Phi^\alpha \partial_\mu \Phi_{\alpha}-\frac12 \Phi_\alpha \partial_\mu \Phi^{\alpha} +\text{fermions} \, .
\end{align}
The role of $\Phi_\alpha$ becomes particularly transparent when comparing the two equivalent formulations of $\mathcal N=4$ conformal supergravity: one with local $SU(2,2|4)$ and a rigid $U(1)$, while the other one with local $SU(2,2|4) \times U(1)$ and a rigid $SU(1,1)\cong SL(2,\mathbb{R})$. The former can be obtained by an appropriate gauge fixing of the latter.

$\Phi_{\alpha}$ can be identified with the complex Yang-Mills coupling $\tau$ in a gauge invariant way using (see \cite{Bergshoeff:1980is} or Appendix B of \cite{Maxfield:2016lok} or Section 3.2 of \cite{Bachas:1999um}) 
\begin{align}
    \tau:=\frac{\theta}{2\pi}+\frac{4\pi i}{{g_\text{YM}}^2}= i\frac{\Phi^1-\Phi^2}{\Phi^1+\Phi^2} \, .
\end{align}
By further fixing the gauge (see \cite{Maxfield:2016lok})
\begin{equation}
    \text{Im}\left(\Phi_1-\Phi_2\right)=0\, ,
\end{equation}
one can arrange for a multiple of the Yang-Mills coupling to appear in the standard position in the matter equations as we have given them.

In Euclidean signature, we drop \eqref{eq: conjugate of Phi} and make the conjugate field independent. To stress this, we denote it with a bar $\overline{\Phi}^{\alpha}$. We further impose the following \textit{complex} constraint:
\begin{align}\label{eq: constraint on Phi}
    \overline{\Phi}^{\alpha} \Phi_{\alpha} = 1 \, .
\end{align}
The now complex composite connection is defined to be
\begin{align}\label{eq: amu def}
    a_\mu = \frac12 \overline{\Phi}^\alpha \partial_\mu \Phi_{\alpha} - \frac12 \Phi_{\alpha} \partial_\mu \overline{\Phi}^\alpha \, + \text{fermions}.
\end{align}
The matter equations \eqref{eq: matter equations euclidean} are most conveniently written in terms of \cite{deRoo:1984zyh}
\begin{equation}
    \Phi:=\overline{\Phi}^1+\overline{\Phi}^2\,\quad \overline{\Phi}:=\Phi_1-\Phi_2\, .
\end{equation}
In Lorenzian signature, these are complex conjugates while in Euclidean signature they are independent complex fields. As in Lorentzian signature, in Euclidean signature we can gauge-fix $\Phi_{\alpha}, \overline{\Phi}^\alpha$ in such a way that only the Yang-Mills coupling $g$ appears in the matter equations. In most of the backgrounds we consider, we will simply choose constant representatives for $\Phi_\alpha, \overline \Phi^\alpha$ satisfying \eqref{eq: constraint on Phi} and yielding a constant \textit{real} $g$, as this is the choice relevant for most applications. The choice leads to a simplification of the supergravity equations and of the covariant derivatives and, specifically, $a_\mu=0$. We will also consider one configuration with the field $\Phi_\alpha$ in a different gauge and which does not satisfy the reality condition but still has $a_\mu=0$. 

\paragraph{The equations for the Euclidean vector multiplet} The matter fields in Euclidean signature are subject to the following algebraic constraints
\begin{align}
    \phi^{ij} = - \phi^{ji} \, , \quad \overline \phi_{ij}= -\overline \phi_{ji} \, , \quad \overline \phi_{ij} = -\frac12 \varepsilon_{ijkl} \phi^{kl} \, .
\end{align}
Note that, just as for $\overline D_{ij}{}^{kl}$ in the gravitational equations, we may as well replace $\overline \phi_{ij}$ with its expression in terms of $\phi^{ij}$, both of which are complex fields. We do not impose any additional reality condition for the scalars, but an appropriate condition must be specified before performing the path integral in e.g. a localization computation. 

The Euclidean matter equations at the bosonic fixed point read \cite{Maxfield:2016lok,deRoo:1984zyh, deRoo:1985np}
\begin{subequations}\label{eq: matter equations euclidean}
\begin{align}
    \delta A_\mu =& \,\, g \big( \xi^i \gamma_\mu \overline \lambda_i - \overline \xi_i \gamma_\mu \lambda^i \big) \, , \\
    \delta \lambda^i =& -\frac{1}{2g} F_{\mu \nu} \gamma^{\mu \nu} \xi^i - 2 \mathcal{D}_\mu \phi^{ij} \gamma^\mu \overline \xi_j  - \frac12 \varepsilon_{jklm} \overline{E}^{ij} \phi^{lm} \xi^k \nonumber \\
    &+\frac14 \varepsilon_{jklm} T_{ab}{}^{jk} \gamma^{ab} \phi^{lm} \xi^i - g \varepsilon_{jklm} [\phi^{ij},\phi^{lm}] \xi^k - \phi^{ij} \gamma^{\mu} \mathcal{D}_\mu \overline \xi_j \, ,\\
    \delta \overline\lambda_i =& -\frac{1}{2g} F_{\mu \nu} \gamma^{\mu \nu} \overline \xi_i + \varepsilon_{ijlm} \mathcal{D}_\mu \phi^{lm} \gamma^\mu \xi^j  + E_{ij} \phi^{jk} \overline \xi_k  \nonumber \\
    &-\frac12 \overline T_{ab jk} \gamma^{ab} \phi^{jk} \overline \xi_i -g \varepsilon_{ijlm} [\phi^{lm}, \phi^{jk}] \overline \xi_k + \frac12 \varepsilon_{ijlm} \phi^{lm} \gamma^{\mu} \mathcal{D}_\mu \xi^j \, ,\\
    \delta \phi^{ij} =& \,\, 2 \xi^{[i} \lambda^{j]} + \varepsilon^{ijkl} \overline \xi_k \overline \lambda_l \,  .
\end{align}
\end{subequations}
The imaginary part of the Yang-Mills coupling, denoted $g$, is obtained from an appropriate gauge-fixing of the field $\Phi_\alpha$ \cite{Maxfield:2016lok}. $\lambda^i, \overline \lambda_i$ are projected Dirac spinors defined in Appendix \ref{sec: clifford}, where conventions for the spinor bilinear can also be found. $\lambda^i$ is a left-handed Weyl spinor and $\overline \lambda_i$ a right-handed one.

The supersymmetry algebra acting on the vector multiplet fields closes \textit{on-shell} for backgrounds satisfying the gravitational equations \eqref{eq: Maxfield sugra eq}. Specifically, it requires the use of the equations of motion for the Weyl spinors. It is sometimes possible to introduce appropriate auxiliary fields and close some of the transformations off-shell. This can be done, for instance, for all of the twisted theories and their scalar supercharges \cite{Kapustin:2006pk,Vafa:1994tf,Bonelli:2020xps}.

Covariant derivatives and curvatures are defined as \cite{deRoo:1984zyh, deRoo:1985np}
\begin{equation}
\begin{aligned}
    F_{\mu \nu} &= \partial_{\mu} A_\nu - \partial_\nu A_\mu +[A_\mu \, , \, A_\nu] \, ,\\
    \mathcal{D}_\mu \phi^{ij} &= \partial_\mu \phi^{ij} + 2 V_\mu ^{[i}{}_{k} \phi^{j]k} + [A_\mu \, , \, \phi^{ij}] \, , \\
    \mathcal{D}_\mu \lambda^i &= \partial_\mu \lambda^i +\frac14 \omega_\mu{}^{ab} \gamma_{ab} \lambda^i - V_\mu^i{}_j \lambda^j +\frac12 a_\mu \lambda^i +[A_\mu \, , \, \lambda^i] \, ,\\
    \mathcal{D}_\mu \overline \lambda_i &= \partial_\mu \overline \lambda_i +\frac14 \omega_\mu{}^{ab} \gamma_{ab} \overline \lambda_i - \overline{V}_{\mu i}{}^j \overline\lambda_j -\frac12 a_\mu \overline \lambda_i +[A^\ast_\mu \, , \, \overline \lambda_i] \, .
\end{aligned}
\end{equation}
Fields in the vector multiplet also transform according to the following gauge transformations
\begin{equation}\label{eq: gauge conventions}
    \begin{aligned}
        \delta_{\text{gauge}} A_\mu =& - \mathcal{D}_\mu \varphi \, , \\
        \delta_{\text{gauge}} \lambda^i =& [\varphi, \lambda^i] \, , \\
        \delta_{\text{gauge}} \phi^{ij} =& [\varphi, \phi^{ij}] \, ,
    \end{aligned}
\end{equation}
for a scalar gauge-parameter $\varphi$. 

A few clarifications about how equations \eqref{eq: matter equations euclidean} have been derived from \cite{Maxfield:2016lok} are in order. The first caveat is that we are using different conventions for the charge conjugation matrix $C$ (see Appendix \ref{sec: clifford}). This results in a different relative sign in the equations for $\delta A_\mu$ and $\delta \phi^{ij}$. The second caveat is that we are using an algebraic condition on $\phi^{ij}$ which differs by a sign from that of \cite{Maxfield:2016lok}. Our condition is the same as the one in the original derivation of the multiplet by de Roo \cite{deRoo:1984zyh, deRoo:1985np}, and appears also in more recent supergravity papers \cite{Butter:2019edc, Bhattacharjee:2025qro, Bhattacharjee:2026lzl}.\footnote{We have  checked that the supersymmetry algebra does not close correctly on the matter fields with the equations given in \cite{Maxfield:2016lok}, including the algebraic condition on $\phi^{ij}$} Finally, closure of the algebra in our conventions required flipping the sign in front of the commutator terms in the positive chirality part of the equations. This last passage was needed for the algebra to close with the correct signs according to our conventions \eqref{eq: gauge conventions}. The differences in convention cancel out for the commutator terms and as a result our commutators have the same sign as in \cite{Maxfield:2016lok}. However, the terms involving background supergravity fields and the term with the derivative of $\phi^{ij}$ in the negative chirality equation are different. The net effect, compared now to the matter equations of \cite{deRoo:1984zyh, deRoo:1985np}, is that we have an opposite sign in the commutator terms. Note that the precise conventions for the field strength, the covariant derivatives, and the gauge transformations are not specified in those papers, and are likely to also differ from ours. 

\subsection{Donaldson-Witten twisted equations}\label{sec: twisted DW}

The Donaldson-Witten twist is implemented by the following $SU(4)_R$ gauge fields 
\begin{align}\label{eq: SU(4) gauge fields DW}
   V_\mu{}^{i}{}_{j} = - \frac{1}{4} \omega_{\mu}{}^{ab} \begin{pmatrix}
        (\overline{\sigma}_{ab})^{\dot \alpha}{}_{\dot \beta} & 0 \\
        0 & 0
    \end{pmatrix}^{i}{}_{j} \, ; \quad \overline{V}_{\mu}{}_{i}{}^{j} = -(V_\mu ^{T})_{i}{}^{j} \, ,
\end{align}
since $V$ is anti-hermitian, the two connections are related by $\overline{V}=V^{\ast} = - V^T$.
This corresponds to the following homomorphism $\varphi$ of $\text{Spin}(4)$ into the R-symmetry group
\begin{align}
    \begin{pmatrix}
        SU(2)_r & 0 \\
        0 & SU(2)_{A}
    \end{pmatrix} \, ,
\end{align}
where $SU(2)_{A}$ is an extra flavour symmetry of this model. The $A$ indices can be lowered and raised using matrices $\varepsilon^{AB}, \varepsilon_{AB}$ obeying the same conventions as the $SU(2)$ chiral indices. Accordingly, the $SU(4)_R$ indices decompose as $i \rightarrow \dot \alpha + A$. Making use of the $SU(2)$ invariant tensors \eqref{eq: SU(2) inv tens}, the supersymmetry parameters are
\begin{align}
    \epsilon_{\alpha}{}^i = \begin{pmatrix}
        \epsilon_m^{(1)}(\sigma^{m})_{\alpha \dot \beta}\varepsilon^{\dot \alpha \dot \beta}\,  & \, \epsilon^{(2)}_{\alpha}{}^{A}
    \end{pmatrix} \, ;\qquad \overline{\epsilon}^{\dot \alpha}{}_{i} = \begin{pmatrix}
        \delta^{\dot \alpha}{}_{\dot \beta} \overline{\epsilon}^{(1)} + \overline{\epsilon}^{(1)-}_{mn}(\overline{\sigma}^{mn})^{\dot \alpha}{}_{\dot \beta}\,  & \, \overline{\epsilon}^{\dot \alpha (2)}{}_{A}
    \end{pmatrix} \, .
\end{align}
The twisted components of the supersymmetry parameters and of the bosonic Weyl multiplet fields are summarised in table \ref{tab:dw-twisted-fields}. The table records their tensorial content, while the explicit block decompositions below fix our sigma-matrix conventions, relative signs, and normalisations. We write an $SU(4)_R$ tensor in $2\times2$ block form, with the first and second blocks carrying, respectively, $\dot\alpha$ and $A$ indices. Thus, \begin{align} X^{ij} = \begin{pmatrix} X^{\dot\alpha\dot\beta} & X^{\dot\alpha B} \\ X^{A\dot\beta} & X^{AB} \end{pmatrix}. \end{align} For an antisymmetric tensor, $X^{A\dot\beta}=-X^{\dot\beta A}$, whereas for a symmetric tensor, $X_{A\dot\beta}=X_{\dot\beta A}$.
The self-dual and anti-self-dual tensor fields decompose as \begin{align} T_{mn}^{+\,ij} = \begin{pmatrix} C_{mn}^{(1,1)+}\varepsilon^{\dot\alpha\dot\beta} & T_{mn}^{(1,2)+\dot\alpha B} \\[2pt] - T_{mn}^{(1,2)+\dot\beta A} & C_{mn}^{(2,2)+}\varepsilon^{AB} \end{pmatrix} \,, \quad \overline T_{mn\,ij}^{-} = \begin{pmatrix} \overline C_{mn}^{(1,1)-} \varepsilon_{\dot\alpha\dot\beta} & \overline T_{mn \, \dot\alpha B}^{(1,2)-} \\[2pt] -\overline T_{mn\, \dot\beta A}^{(1,2)-} & \overline C_{mn}^{(2,2)-} \varepsilon_{AB} \end{pmatrix} \, . \end{align}
The symmetric scalar fields decompose as \begin{align} E_{ij} = \begin{pmatrix} E_{mn}^{(1,1)-} (\overline\sigma^{mn})_{\dot\alpha\dot\beta} & E_{\dot\alpha B}^{(1,2)} \\[2pt] E_{\dot\beta A}^{(1,2)} & E_{(AB)}^{(2,2)} \end{pmatrix} \, , \quad \overline E^{ij} = \begin{pmatrix} \overline E_{mn}^{(1,1)-} (\overline\sigma^{mn})^{\dot\alpha\dot\beta} & \overline E^{(1,2)\dot\alpha B} \\[2pt] \overline E^{(1,2)\dot\beta A} & \overline E^{(2,2)(AB)} \end{pmatrix} \, .
\end{align}
We now apply the same decomposition to the fermionic supersymmetry and to the matter fields variations. These decompositions make use of the $SU(2)$ invariant tensors \eqref{eq: SU(2) inv tens}.
\begin{table}[t]
\centering
\begin{tabular}{@{}cl@{}}
\toprule Original field & Twisted components \\ \midrule $\epsilon_\alpha{}^i$ & $\epsilon_m^{(1)},\ \epsilon_\alpha^{(2)A}$ \\ \addlinespace
$\overline\epsilon^{\dot\alpha}{}_i$ & $\overline\epsilon^{(1)},\ \overline\epsilon_{mn}^{(1)-},\ \overline\epsilon^{(2)\dot\alpha}{}_A$ \\
\addlinespace $T_{mn}^{+\,ij}$ & \makecell[l]{ $C_{mn}^{(1,1)+},\ T_{mn}^{(1,2)+\dot\alpha A},\ C_{mn}^{(2,2)+}$} \\ \addlinespace
$\overline T_{mn\,ij}^{-}$  & \makecell[l]{ $\overline C_{mn}^{(1,1)-},\ \overline T_{mn \,\dot\alpha A}^{(1,2)-},\ \overline C_{mn}^{(2,2)-}$} \\
\addlinespace $E_{ij}$  & \makecell[l]{ $E_{mn}^{(1,1)-},\ E_{\dot\alpha A}^{(1,2)},\ E_{(AB)}^{(2,2)}$} \\ \addlinespace
$\overline E^{ij}$  & \makecell[l]{ $\overline E_{mn}^{(1,1)-},\ \overline E^{(1,2)\dot\alpha A},\ \overline E^{(2,2)(AB)}$} \\ \bottomrule 
\end{tabular} 
\caption{Twisted components of the supersymmetry parameters and bosonic Weyl-multiplet fields in the Donaldson-Witten decomposition.}
\label{tab:dw-twisted-fields}
\end{table}
\paragraph{Twisted gravitino equations.} For the gravitino, the components in the first $SU(2)$ block are extracted by the projections 
\begin{subequations}\label{eq:dw-gravitino-projections} \begin{align} \delta\psi^{(1)}_{mn} &:= -\frac12 \delta\psi^{(1)}_{m\, \alpha}{}^{\dot\alpha} (\overline\sigma_n)_{\dot\alpha}{}^\alpha,  \\ \delta\overline\psi^{(1)}_m &:= \frac12 \delta\overline\psi^{(1)}_m{}^{\dot\alpha}{}_{\dot\beta} \delta^{\dot\beta}{}_{\dot\alpha},  \\ \delta\overline\psi^{(1)-}_{m\,nr} &:= -\frac18 \delta\overline\psi^{(1)}_m{}^{\dot\alpha}{}_{\dot\beta} (\overline\sigma_{nr})^{\dot\beta}{}_{\dot\alpha} \, . 
\end{align}
\end{subequations}
The components $\delta\psi^{(2)}_{m\alpha}{}^A$ and $\delta\overline\psi^{(2)\dot\alpha}{}_{mA}$ already carry irreducible twisted indices and therefore require no further projection. The resulting gravitino variations are
\begin{subequations}
\begin{align}
\delta \psi^{(1)}_{mn} ={}&\, \nabla_{m}\epsilon^{(1)}_{n} -\frac14 \Bigl( \nabla_{m}\epsilon^{(1)}_{n} -\nabla_{n}\epsilon^{(1)}_{m} +\delta_{mn}\nabla_{a}\epsilon^{(1)a} -\varepsilon_{mnab}\nabla^{a}\epsilon^{(1)b} \Bigr) \nonumber\\ &\, +C^{(1,1)+}_{mn}\,\overline\epsilon^{(1)} +4\,C^{(1,1)+}_{ma}\, \overline\epsilon^{(1)-a}{}_{n} +\frac18\, T^{(1,2)+}_{ab}{}^{\dot\alpha A} (\overline\sigma_{n}\sigma^{ab}\sigma_{m} \overline\epsilon^{(2)})_{\dot\alpha A}\, , \\[1ex]  \delta \psi^{(2)}_{m\alpha}{}^{A} ={}&\, \nabla_{m}\epsilon^{(2)}_{\alpha}{}^{A} -\frac14 \nabla_{n} (\sigma_{m}\overline\sigma^{n}\epsilon^{(2)})_{\alpha}{}^{A} \nonumber\\ &\, -\frac14 C^{(2,2)+}_{ab} (\sigma^{ab}\sigma_{m}\overline\epsilon^{(2)})_{\alpha}{}^{A} +\frac14 T^{(1,2)+}_{ab}{}^{\dot\gamma A} (\sigma^{ab}\sigma_{m})_{\alpha\dot\gamma} \,\overline\epsilon^{(1)} \nonumber\\ &\, +\frac14 T^{(1,2)+}_{ab}{}^{\dot\gamma A} (\sigma^{ab}\sigma_{m})_{\alpha\dot\beta} (\overline\sigma_{cd})^{\dot\beta}{}_{\dot\gamma} \,\overline\epsilon^{(1)-cd}\,, \\[1ex]  \delta \overline\psi^{(1)}_{m} ={}&\, \frac34\,\nabla_{m}\overline\epsilon^{(1)} +\nabla_{a}\overline\epsilon^{(1)-}_{m}{}^{a} \nonumber\\ &\, +\overline C^{(1,1)-}_{ma}\epsilon^{(1)a} -\frac18\, \overline T^{(1,2)-}_{ab\,\dot\beta A} (\overline\sigma^{ab}\overline\sigma_{m} \epsilon^{(2)})^{\dot\beta A}\, , \\[1ex]  \delta \overline\psi^{(1)-}_{m\,nr} ={}&\, \nabla_{m}\overline\epsilon^{(1)-}_{nr} -\frac1{16} \Bigl( \delta_{mn}\nabla_{r}\overline\epsilon^{(1)} -\delta_{mr}\nabla_{n}\overline\epsilon^{(1)} -\varepsilon_{manr}\nabla^{a}\overline\epsilon^{(1)} \Bigr) \nonumber\\ &\, -\frac14 \Bigl( \delta_{mn}\nabla_{a}\overline\epsilon^{(1)-a}{}_{r} -\delta_{mr}\nabla_{a}\overline\epsilon^{(1)-a}{}_{n} +\nabla_{n}\overline\epsilon^{(1)-}_{rm} -\nabla_{r}\overline\epsilon^{(1)-}_{nm} +\nabla_{m}\overline\epsilon^{(1)-}_{nr} \Bigr) \nonumber\\ &\, -\frac14 \Bigl( \delta_{mn}\overline C^{(1,1)-}_{ra}\epsilon^{(1)a} -\delta_{mr}\overline C^{(1,1)-}_{na}\epsilon^{(1)a} -\overline C^{(1,1)-}_{rm}\epsilon^{(1)}_{n} +\overline C^{(1,1)-}_{nm}\epsilon^{(1)}_{r} +\overline C^{(1,1)-}_{nr}\epsilon^{(1)}_{m} \Bigr) \nonumber\\ &\, +\frac1{32}\, \overline T^{(1,2)-}_{ab\,\dot\beta A} (\overline\sigma^{ab}\overline\sigma_{m}\epsilon^{(2)})^{\dot\alpha A} (\overline\sigma_{nr})^{\dot\beta}{}_{\dot\alpha}\, , \\[1ex]  \delta \overline\psi^{(2)}_{m}{}^{\dot\alpha}{}_{A} ={}&\, \nabla_{m}\overline\epsilon^{(2)\dot\alpha}{}_{A} -\frac14 \nabla_{n} (\overline\sigma_{m}\sigma^{n} \overline\epsilon^{(2)})^{\dot\alpha}{}_{A} \nonumber\\ &\, -\frac14 \overline C^{(2,2)-}_{ab} (\overline\sigma^{ab}\overline\sigma_{m} \epsilon^{(2)})^{\dot\alpha}{}_{A} +\frac14 \overline T^{(1,2)-}_{ab\,\dot\beta A} (\overline\sigma^{ab}\overline\sigma_{m}\sigma_{c}) ^{\dot\alpha\dot\beta} \epsilon^{(1)c}\, .
\end{align}
\end{subequations}

\paragraph{Twisted gaugino equations.} 
The twisted gaugino components are obtained analogously. In the first $SU(2)$ block, we define
\begin{subequations}\label{eq:dw-weyl-gaugino-projections} \begin{align} \delta\Lambda^{(1)}_m &:= \frac12 \delta\Lambda^{(1)}_{\alpha\dot\alpha} (\overline \sigma_m)^{\dot \alpha \alpha} \, ,  \\ \delta\overline\Lambda^{(1)} &:= \frac12 \delta\overline\Lambda^{(1)\dot\alpha\dot\beta} \varepsilon_{\dot\beta\dot\alpha} \, ,  \\ \delta\overline\Lambda^{(1)-}_{mn} &:= \frac18\delta\overline\Lambda^{(1)\dot\alpha\dot\beta} (\overline\sigma_{mn})_{\dot\beta\dot\alpha} \, , \end{align} \end{subequations}
while the components $\delta\Lambda^{(2)}_{\alpha A}$ and $\delta\overline\Lambda^{(2)\dot\alpha A}$ are already irreducible and require no additional projection. The twisted gaugino equations are thus
\begin{subequations} \begin{align} \delta \Lambda^{(1)}_{m} ={}&\, 4\,E^{(1,1)-}_{m a}\epsilon^{(1) a} +\frac12\,E^{(1,2)}_{\dot\alpha A} (\overline\sigma_{m}\epsilon^{(2)})^{\dot\alpha A} \nonumber\\ &\, -4\,C^{(2,2)+}_{m a}\epsilon^{(1) a} -\frac12\,T^{(1,2)+}_{ab}{}_{\dot\alpha A} (\sigma^{ab}\epsilon^{(2)})_{\beta}{}^{A} (\overline\sigma_{m})^{\dot\alpha\beta}, \\[1ex] \delta \Lambda^{(2)}_{\alpha A} ={}&\, E^{(2,2)}_{AB}\,\epsilon^{(2)}_{\alpha}{}^{B} +E^{(1,2)}_{\dot\gamma A} (\sigma_{c})_{\alpha}{}^{\dot\gamma}\epsilon^{(1)c} \nonumber\\ &\, -C^{(1,1)+}_{ab} (\sigma^{ab}\epsilon^{(2)})_{\alpha A} -T^{(1,2)+}_{ab}{}^{\dot\beta}{}_{A} (\sigma^{ab}\sigma_{c})_{\alpha\dot\beta}\epsilon^{(1)c}\,, \\[1ex] \delta \overline\Lambda^{(1)} ={}&\, \frac12\,\overline E^{(1,2)\dot\alpha A} \overline\epsilon^{(2)}{}_{\dot\alpha A} +4\,\overline E^{(1,1)-}_{ab} \overline\epsilon^{(1)-ab} \nonumber\\ &\, +4\,\overline C^{(2,2)-}_{ab} \overline\epsilon^{(1)-ba} +\frac12\,\overline T^{(1,2)-}_{ab\,\dot\gamma}{}^{A} (\overline\sigma^{ab}\overline\epsilon^{(2)})^{\dot\gamma}{}_{A}\,, \\[1ex] \delta \overline\Lambda^{(1)-}_{mn} ={}&\, \frac18\,\overline E^{(1,2)\dot\alpha A} \overline\epsilon^{(2)\dot\beta}{}_{A} (\overline\sigma_{mn})_{\dot\alpha\dot\beta} +\overline E^{(1,1)-}_{mn}\overline\epsilon^{(1)} -\overline C^{(2,2)-}_{mn}\overline\epsilon^{(1)} \nonumber\\ &\, +4\Bigl( \overline E^{(1,1)-}_{ma}\overline\epsilon^{(1)-}_{n}{}^{a} -\overline E^{(1,1)-}_{na}\overline\epsilon^{(1)-}_{m}{}^{a} +\frac12\varepsilon_{mnab} \overline E^{(1,1)-a}{}_{c} \overline\epsilon^{(1)-bc} \Bigr) \nonumber\\ &\, +4\Bigl( \overline C^{(2,2)-}_{ma}\overline\epsilon^{(1)-}_{n}{}^{a} -\overline C^{(2,2)-}_{na}\overline\epsilon^{(1)-}_{m}{}^{a} +\frac12\varepsilon_{mnab} \overline C^{(2,2)-a}{}_{c} \overline\epsilon^{(1)-bc} \Bigr) \nonumber\\ &\, +\frac18\,\overline T^{(1,2)-}_{ab\,\dot\beta}{}^{A} (\overline\sigma^{ab}\overline\epsilon^{(2)})^{\dot\alpha}{}_{A} (\overline\sigma_{mn})^{\dot\beta}{}_{\dot\alpha}, \\[1ex] \delta \overline\Lambda^{(2)\dot\alpha A} ={}&\, \overline E^{(2,2)AB} \overline\epsilon^{(2)\dot\alpha}{}_{B} +\overline E^{(1,2)\dot\alpha A}\overline\epsilon^{(1)} +\overline E^{(1,2)\dot\beta A} (\overline\sigma_{ab})^{\dot\alpha}{}_{\dot\beta} \overline\epsilon^{(1)-ab} \nonumber\\ &\, -\overline C^{(1,1)-}_{ab} (\overline\sigma^{ab}\overline\epsilon^{(2)})^{\dot\alpha A} -\overline T^{(1,2)-}_{ab\,\dot\beta}{}^{A} (\overline\sigma^{ab})^{\dot\alpha\dot\beta} \overline\epsilon^{(1)}  -\overline T^{(1,2)-}_{ab\,\dot\beta}{}^{A} (\overline\sigma^{ab}\overline\sigma_{cd})^{\dot\alpha\dot\beta} \overline\epsilon^{(1)-cd}\,. \end{align}
\end{subequations}
\paragraph{Twisted vector multiplet.}
\begin{table}[t]
\centering \begin{tabular}{@{}cl@{}} \toprule Original field & Twisted components \\ \midrule $A_m$ & $A_m$ \\ \addlinespace
$\lambda_\alpha{}^i$ & $\lambda_m^{(1)},\ \lambda_\alpha^{(2)A}$ \\ \addlinespace
$\overline\lambda^{\dot\alpha}{}_i$ & $\overline\lambda^{(1)},\ \overline\lambda_{mn}^{(1)-},\ \overline\lambda^{(2)\dot\alpha}{}_A$ \\ \addlinespace
$\phi^{ij}$ & \makecell[l]{ $\phi^{(1,1)},\ \phi^{(1,2)\dot\alpha A},\ \phi^{(2,2)}$} \\ \bottomrule
\end{tabular}
\caption{Twisted components of the Donaldson-Witten vector multiplet.}
\label{tab:dw-matter-fields}
\end{table}
We finally turn to the vector multiplet. The fermions decompose in the same representation as the supersymmetry parameters, while the antisymmetric scalar field is embedded as \begin{align} \phi^{ij} &= \begin{pmatrix} \phi^{(1,1)} \varepsilon^{\dot\alpha\dot\beta} & \phi^{(1,2)\dot\alpha B} \\[2pt] -\phi^{(1,2)\dot\beta A} & \phi^{(2,2)} \varepsilon^{AB} \end{pmatrix} \, . \end{align}
The twisted vector-multiplet components are extracted through the following inverse projections 
\begin{subequations}
\begin{alignat}{2}
\delta\lambda_m^{(1)} &:= -\frac12\, \delta\lambda_\alpha^{(1)\dot\alpha} (\overline\sigma_m)_{\dot\alpha}{}^\alpha, &\qquad \delta\phi^{(1,1)} &:= \frac12\, \delta\phi^{\dot\alpha\dot\beta} \varepsilon_{\dot\beta\dot\alpha}, \\ \delta\overline\lambda^{(1)} &:= \frac12\, \delta\overline\lambda^{(1)\dot\alpha}{}_{\dot\beta} \delta^{\dot\beta}{}_{\dot\alpha}, & \delta\phi^{(2,2)} &:= \frac12\, \delta\phi^{AB} \varepsilon_{BA}, \\ \delta\overline\lambda_{mn}^{(1)-} &:= -\frac18\, \delta\overline\lambda^{(1)\dot\alpha}{}_{\dot\beta} (\overline\sigma_{mn})^{\dot\beta}{}_{\dot\alpha}. \end{alignat}
\end{subequations}
The normalisation factors in these projections are fixed by requiring them to invert the corresponding twisting maps. Unlike in the preceding supergravity projections, these factors are relevant for the normalisation of the supersymmetry algebra.
The vector multiplet transformations are lengthy and are collected in Appendix \ref{app: DW vector multiplet}. The twisted field content of the vector multiplet is summarised in table \ref{tab:dw-matter-fields}.

\subsection{Vafa-Witten twisted equations}\label{sec: twisted VW}
The VW twist consists of the following definitions of the $SU(4)$ gauge fields
\begin{align}\label{eq: SU(4) gauge fields VW}
    V_\mu{}^{i}{}_{j} = - \frac{1}{4} \omega_{\mu}{}^{ab} \begin{pmatrix}
        (\overline \sigma_{ab})^{\dot \alpha}{}_{\dot \beta} & 0 \\
        0 & (\overline \sigma_{ab})^{\dot \alpha}{}_{\dot \beta}
    \end{pmatrix}^{i}{}_{j} \, ; \quad \overline{V}_{\mu}{}_{i}{}^{j} = -(V_\mu ^{T})_{i}{}^{j} \, ,
\end{align}
corresponding to the following embedding
\begin{align}\label{eq: VW embedding}
    \begin{pmatrix}
        SU(2)_r&0\\0&SU(2)_r
    \end{pmatrix} \, .
\end{align}
Since the embedding $SU(2)$'s have the same chirality, this choice commutes with an additional $SU(2)$ rotating them into one-another \cite{Vafa:1994tf}.
This means that the $SU(4)$ indices split as $i \rightarrow \dot \alpha + \dot \beta$. we can twist the Weyl spinors
\begin{align}
   \epsilon_{\alpha}{}^i = \begin{pmatrix}
       \epsilon_m^{(1)}(\sigma^{m})_{\alpha }{}^{\dot \alpha} \, & \, \epsilon_m^{(2)}(\sigma^{m})_{\alpha}{}^{\dot \alpha}
   \end{pmatrix}  \, ;\qquad \overline{\epsilon}^{\dot \alpha}{}_{i} = \begin{pmatrix}
       \delta^{\dot \alpha}{}_{\dot \beta} \overline{\epsilon}^{(1)} + \overline{\epsilon}^{(1)-}_{mn}(\overline{\sigma}^{mn})^{\dot \alpha}{}_{\dot \beta} \, & \, \delta^{\dot \alpha}{}_{\dot \beta} \overline{\epsilon}^{(2)} + \overline{\epsilon}^{(2)-}_{mn}(\overline{\sigma}^{mn})^{\dot \alpha}{}_{\dot \beta}
   \end{pmatrix}  \, .
\end{align}
Similarly
\begin{equation}
\begin{aligned}
    &T_{mn}^{+ \,ij} = \begin{pmatrix}
        C_{mn}^{(1,1)+} \varepsilon^{\dot \alpha \dot \beta} & C_{mn}^{(1,2)+} \varepsilon^{\dot \alpha \dot \beta} + K^{(1,2)+}_{mn \, rs}(\overline \sigma^{rs})^{\dot \alpha \dot \beta}\\
        -C_{mn}^{(1,2)+} \varepsilon^{\dot \beta \dot \alpha} - K^{(1,2)+}_{mn \, rs}(\overline \sigma^{rs})^{\dot \beta \dot \alpha} & C_{mn}^{(2,2)+} \varepsilon^{\dot \alpha \dot \beta}
    \end{pmatrix} \, , \\ &\overline{T}^-_{mn \, ij}= \begin{pmatrix}
        \overline C_{mn}^{(1,1)-} \varepsilon_{\dot \alpha \dot \beta} & \overline C_{mn}^{(1,2)-} \varepsilon_{\dot \alpha \dot \beta} + \overline K^{(1,2)-}_{mn \, rs}(\overline \sigma^{rs})_{\dot \alpha \dot \beta}\\
        -\overline C_{mn}^{(1,2)-} \varepsilon_{\dot \beta \dot \alpha} - \overline K^{(1,2)-}_{mn \, rs}(\overline \sigma^{rs})_{\dot \beta \dot \alpha} & \overline C_{mn}^{(2,2)-} \varepsilon_{\dot \alpha \dot \beta}
    \end{pmatrix} \, ,
\end{aligned}
\end{equation}
and
\begin{equation}
\begin{aligned}
    E_{ij}&= \begin{pmatrix}
        E^{(1,1)-}_{mn} (\overline \sigma^{mn})_{\dot \alpha \dot \beta} & E^{(1,2)-}_{mn} (\overline \sigma^{mn})_{\dot \alpha \dot \beta}+E^{(1,2)} \varepsilon_{\dot \alpha \dot \beta}\\
        E^{(1,2)-}_{mn} (\overline \sigma^{mn})_{\dot \beta \dot \alpha}+E^{(1,2)} \varepsilon_{\dot \beta \dot \alpha} & E^{(2,2)-}_{mn}(\overline \sigma^{mn})_{\dot \alpha \dot \beta}
    \end{pmatrix} \, ,\\
    \overline E^{ij}&= \begin{pmatrix}
        \overline E^{(1,1)-}_{mn} (\overline \sigma^{mn})^{\dot \alpha \dot \beta} & \overline E^{(1,2)-}_{mn} (\overline \sigma^{mn})^{\dot \alpha \dot \beta}+\overline E^{(1,2)} \varepsilon^{\dot \alpha \dot \beta}\\
        \overline E^{(1,2)-}_{mn} (\overline \sigma^{mn})^{\dot \beta \dot \alpha}+\overline E^{(1,2)} \varepsilon^{\dot \beta \dot \alpha} & \overline E^{(2,2)-}_{mn}(\overline \sigma^{mn})^{\dot \alpha \dot \beta}
    \end{pmatrix} \, .
\end{aligned}
\end{equation}
\begin{table}[t]
\centering
\begin{tabular}{@{}cl@{}}
\toprule Original field  & Twisted components \\ \midrule
$\epsilon_\alpha{}^i$ & $\epsilon_m^{(1)},\ \epsilon^{(2)}_m$ \\ \addlinespace
$\overline\epsilon^{\dot\alpha}{}_i$ & $\overline\epsilon^{(1)},\ \overline\epsilon_{mn}^{(1)-},\ \overline\epsilon^{(2)},\ \overline\epsilon_{mn}^{(2)-}$ \\ \addlinespace
$T_{mn}^{+\,ij}$  & \makecell[l]{ $C_{mn}^{(1,1)+},\ C^{(1,2)+}_{mn},\ K^{(1,2)+}_{mn \, rs},\ C_{mn}^{(2,2)+}$} \\ \addlinespace
$\overline T_{mn\,ij}^{-}$ & \makecell[l]{ $\overline C_{mn}^{(1,1)-},\ \overline C_{mn}^{(1,2)-} ,\ \overline{K}^{(1,2)-}_{mn\,rs},\ \overline C_{mn}^{(2,2)-}$} \\
\addlinespace $E_{ij}$  & \makecell[l]{ $E_{mn}^{(1,1)-},\ E_{mn}^{(1,2)-},\ E^{(1,2)},\ E_{mn}^{(2,2)-}$} \\ \addlinespace
$\overline E^{ij}$  & \makecell[l]{ $\overline E_{mn}^{(1,1)-},\ \overline E^{(1,2)-}_{mn},\ \overline{E}^{(1,2)} ,\ \overline E^{(2,2)-}_{mn}$} \\ \bottomrule 
\end{tabular} 
\caption{Twisted components of the supersymmetry parameters and bosonic Weyl-multiplet fields in the Vafa-Witten decomposition.}
\label{tab:vw-twisted-fields}
\end{table}

\paragraph{Twisted gravitino equations.}
The twisted equations can then be obtained as the following traces of the supergravity equations
\begin{subequations}
\begin{alignat}{2}
    \delta \psi^{(1)}_{mn}&:=-\frac12\delta \psi^{(1)}_{m \alpha}{}^{\dot \alpha} (\overline \sigma_n)_{\dot \alpha}{}^{\alpha} \, ,&\qquad  \delta \psi^{(2)}_{m n} &:= -\frac12\delta \psi^{(2)}_{m \alpha}{}^{\dot \alpha} (\overline \sigma_n)_{\dot \alpha}{}^{\alpha} \, ,\\
    \delta \overline \psi^{(1)}_m &:= \frac12 \delta \overline \psi^{(1)}_{m}{}^{\dot \alpha}{}_{\dot \beta} \delta^{\dot \beta}{}_{\dot \alpha} \, , &\qquad \delta \overline \psi^{(2)}_m &:= \frac12 \delta \overline \psi^{(2)}_{m}{}^{\dot \alpha}{}_{\dot \beta} \delta^{\dot \beta}{}_{\dot \alpha} \, ,\\
    \delta \overline\psi^{(1)-}_{m\, n r} &:= -\frac18 \delta \overline \psi^{(1)}_{m}{}^{\dot \alpha}{}_{\dot \beta} (\overline \sigma_{n r})^{\dot \beta}{}_{\dot \alpha}\, , &\qquad \delta \overline\psi^{(2)-}_{m\, n r} &:= -\frac18 \delta \overline\psi^{(2)}_{m}{}^{\dot \alpha}{}_{\dot \beta} (\overline \sigma_{n r})^{\dot \beta}{}_{\dot \alpha}  \, .
\end{alignat}
\end{subequations}
The twisted gravitino equations read
\begin{subequations}
\begin{align}
\delta\psi^{(1)}_{mn} ={}& \nabla_{m}\epsilon^{(1)}_{n} +C^{(1,1)+}_{mn}\,\overline\epsilon^{(1)} +4C^{(1,1)+}_{ma}\, \overline\epsilon^{(1)-a}{}_{n} \nonumber\\ &+C^{(1,2)+}_{mn}\,\overline\epsilon^{(2)} +4C^{(1,2)+}_{ma}\, \overline\epsilon^{(2)-a}{}_{n} \nonumber\\ &-4K^{(1,2)+}_{man}{}^{a}\, \overline\epsilon^{(2)} -16K^{(1,2)+}_{abn}{}^{b}\, \overline\epsilon^{(2)-}{}_{m}{}^{a} \nonumber\\ &-\frac14\Big( \nabla_{m}\epsilon^{(1)}_{n} -\nabla_{n}\epsilon^{(1)}_{m} +\delta_{mn}\nabla_{a}\epsilon^{(1)a} -\varepsilon_{mnab}\nabla^{a}\epsilon^{(1)b} \Big), \\[0.5em]  \delta\psi^{(2)}_{mn} ={}& \nabla_{m}\epsilon^{(2)}_{n} +C^{(2,2)+}_{mn}\,\overline\epsilon^{(2)} +4C^{(2,2)+}_{ma}\, \overline\epsilon^{(2)-a}{}_{n} \nonumber\\ &+C^{(1,2)+}_{mn}\,\overline\epsilon^{(1)} +4C^{(1,2)+}_{ma}\, \overline\epsilon^{(1)-a}{}_{n} \nonumber\\ &+4K^{(1,2)+}_{man}{}^{a}\, \overline\epsilon^{(1)} +16K^{(1,2)+}_{abn}{}^{b}\, \overline\epsilon^{(1)-}{}_{m}{}^{a} \nonumber\\ &-\frac14\Big( \nabla_{m}\epsilon^{(2)}_{n} -\nabla_{n}\epsilon^{(2)}_{m} +\delta_{mn}\nabla_{a}\epsilon^{(2)a} -\varepsilon_{mnab}\nabla^{a}\epsilon^{(2)b} \Big)\,, \\[0.5em]  \delta\overline\psi^{(1)}_{m} ={}& \nabla_{m}\overline\epsilon^{(1)} +\overline C^{(1,1)-}_{ma}\epsilon^{(1)a} +\overline C^{(1,2)-}_{ma}\epsilon^{(2)a} \nonumber\\ &-4\overline K^{(1,2)-}_{mab}{}^{a}\epsilon^{(2)b} -\frac14\nabla_{m}\overline\epsilon^{(1)} +\nabla_{a}\overline\epsilon^{(1)-}_{m}{}^{a}, \\[0.5em]  \delta\overline\psi^{(1)-}_{m\,nr} ={}& \nabla_{m}\overline\epsilon^{(1)-}_{nr} +2\overline K^{(1,2)-}_{mnra}\epsilon^{(2)a} -2\overline K^{(1,2)-}_{mrna}\epsilon^{(2)a} \nonumber\\ &+\delta_{mn}\, \overline K^{(1,2)-}_{rab}{}^{a}\epsilon^{(2)b} -\delta_{mr}\, \overline K^{(1,2)-}_{nab}{}^{a}\epsilon^{(2)b} \nonumber\\ &+\varepsilon_{manr}\, \overline K^{(1,2)-a}{}_{bc}{}^{b}\epsilon^{(2)c} \nonumber\\ &-\frac14\Big( \delta_{mn}\overline C^{(1,1)-}_{ra}\epsilon^{(1)a} -\delta_{mr}\overline C^{(1,1)-}_{na}\epsilon^{(1)a} -\overline C^{(1,1)-}_{rm}\epsilon^{(1)}_{n} +\overline C^{(1,1)-}_{nm}\epsilon^{(1)}_{r} +\overline C^{(1,1)-}_{nr}\epsilon^{(1)}_{m} \Big) \nonumber\\ &-\frac14\Big( \delta_{mn}\overline C^{(1,2)-}_{ra}\epsilon^{(2)a} -\delta_{mr}\overline C^{(1,2)-}_{na}\epsilon^{(2)a} -\overline C^{(1,2)-}_{rm}\epsilon^{(2)}_{n} +\overline C^{(1,2)-}_{nm}\epsilon^{(2)}_{r} +\overline C^{(1,2)-}_{nr}\epsilon^{(2)}_{m} \Big) \nonumber\\ &-\frac1{16}\Big( \delta_{mn}\nabla_{r}\overline\epsilon^{(1)} -\delta_{mr}\nabla_{n}\overline\epsilon^{(1)} -\varepsilon_{manr}\nabla^{a}\overline\epsilon^{(1)} \Big) \nonumber\\ &-\frac14\Big( \delta_{mn}\nabla_{a}\overline\epsilon^{(1)-a}{}_{r} -\delta_{mr}\nabla_{a}\overline\epsilon^{(1)-a}{}_{n} +\nabla_{n}\overline\epsilon^{(1)-}_{rm} -\nabla_{r}\overline\epsilon^{(1)-}_{nm} +\nabla_{m}\overline\epsilon^{(1)-}_{nr} \Big)\,, \\[0.5em]  \delta\overline\psi^{(2)}_{m} ={}& \nabla_{m}\overline\epsilon^{(2)} +\overline C^{(2,2)-}_{ma}\epsilon^{(2)a} +\overline C^{(1,2)-}_{ma}\epsilon^{(1)a} \nonumber\\ &+4\overline K^{(1,2)-}_{mab}{}^{a}\epsilon^{(1)b} -\frac14\nabla_{m}\overline\epsilon^{(2)} +\nabla^{a}\overline\epsilon^{(2)-}_{ma}\,, \\[1ex]  \delta\overline\psi^{(2)-}_{m\,nr} ={}& \nabla_{m}\overline\epsilon^{(2)-}_{nr} -2\overline K^{(1,2)-}_{mnra}\epsilon^{(1)a} +2\overline K^{(1,2)-}_{mrna}\epsilon^{(1)a} \nonumber\\ &-\delta_{mn}\, \overline K^{(1,2)-}_{rab}{}^{a}\epsilon^{(1)b} +\delta_{mr}\, \overline K^{(1,2)-}_{nab}{}^{a}\epsilon^{(1)b} \nonumber\\ &-\varepsilon_{mnra}\, \overline K^{(1,2)-ab}{}_{cb}\epsilon^{(1)c} \nonumber\\ &-\frac14\Big( \delta_{mn}\overline C^{(2,2)-}_{ra}\epsilon^{(2)a} -\delta_{mr}\overline C^{(2,2)-}_{na}\epsilon^{(2)a} -\overline C^{(2,2)-}_{rm}\epsilon^{(2)}_{n} +\overline C^{(2,2)-}_{nm}\epsilon^{(2)}_{r} +\overline C^{(2,2)-}_{nr}\epsilon^{(2)}_{m} \Big) \nonumber\\ &-\frac14\Big( \delta_{mn}\overline C^{(1,2)-}_{ra}\epsilon^{(1)a} -\delta_{mr}\overline C^{(1,2)-}_{na}\epsilon^{(1)a} -\overline C^{(1,2)-}_{rm}\epsilon^{(1)}_{n} +\overline C^{(1,2)-}_{nm}\epsilon^{(1)}_{r} +\overline C^{(1,2)-}_{nr}\epsilon^{(1)}_{m} \Big) \nonumber\\ &-\frac1{16}\Big( \delta_{mn}\nabla_{r}\overline\epsilon^{(2)} -\delta_{mr}\nabla_{n}\overline\epsilon^{(2)} -\varepsilon_{manr}\nabla^{a}\overline\epsilon^{(2)} \Big) \nonumber\\ &-\frac14\Big( \delta_{mn}\nabla^{a}\overline\epsilon^{(2)-}_{ar} -\delta_{mr}\nabla^{a}\overline\epsilon^{(2)-}_{an} +\nabla_{n}\overline\epsilon^{(2)-}_{rm} -\nabla_{r}\overline\epsilon^{(2)-}_{nm} +\nabla_{m}\overline\epsilon^{(2)-}_{nr} \Big)\,. 
\end{align} 
\end{subequations}

\paragraph{Twisted gaugino equations.}
The twisted equations can then be obtained as the following traces of the supergravity equations
\begin{subequations}
\begin{alignat}{2}
    \delta \Lambda^{(1)}_{m}&:=\frac12 \delta \Lambda^{(1)}_{\alpha \dot \alpha} (\overline \sigma_m)^{\dot \alpha \alpha} \, , &\qquad  \delta \Lambda^{(2)}_{m}&:=\frac12\delta \Lambda^{(2)}_{\alpha \dot \alpha} (\overline \sigma_m)^{\dot \alpha \alpha} \,,\\
    \delta \overline \Lambda^{(1)} &:= \frac12\delta \overline \Lambda^{(1) \dot \alpha \dot \beta} \varepsilon_{\dot \beta \dot \alpha} \, , &\qquad \delta \overline \Lambda^{(2)} &:= \frac12\delta \overline \Lambda^{(2) \dot \alpha \dot \beta} \varepsilon_{\dot \beta \dot \alpha} \, ,\\ \delta \overline \Lambda^{(1)-}_{m n}&:=\frac18\delta \overline \Lambda^{(1) \dot \alpha \dot \beta}  (\overline \sigma_{m n})_{\dot \beta \dot \alpha} \, , &\qquad \delta \overline \Lambda^{(2)-}_{m n}&:=\frac18\delta \overline \Lambda^{(2) \dot \alpha \dot \beta}  (\overline \sigma_{m n})_{\dot \beta \dot \alpha} \, .
\end{alignat}
\end{subequations}
Thus, one finds
\begin{subequations} \begin{align} \delta \Lambda^{(1)}_{m} ={}&\; 4\,E^{(1,1)-}_{ma}\,\epsilon^{(1)a} +4\,E^{(1,2)-}_{ma}\,\epsilon^{(2)a} +E^{(1,2)}\,\epsilon^{(2)}_{m} \nonumber\\ &\; -4\,C^{(2,2)+}_{ma}\,\epsilon^{(1)a} +4\,C^{(1,2)+}_{ma}\,\epsilon^{(2)a} +16\,K^{(1,2)+}_{mbc}{}^{b}\,\epsilon^{(2)c}\,, \\[1ex] \delta \Lambda^{(2)}_{m} ={}&\; 4\,E^{(2,2)-}_{ma}\,\epsilon^{(2)a} +4\,E^{(1,2)-}_{ma}\,\epsilon^{(1)a} -E^{(1,2)}\,\epsilon^{(1)}_{m} \nonumber\\ &\; -4\,C^{(1,1)+}_{ma}\,\epsilon^{(2)a} +4\,C^{(1,2)+}_{ma}\,\epsilon^{(1)a} -16\,K^{(1,2)+}_{mbc}{}^{b}\,\epsilon^{(1)c}\,, \\[1ex] \delta \overline\Lambda^{(1)} ={}&\; 4\,\overline E^{(1,1)-}_{ab}\,\overline\epsilon^{(1)-ab} +4\,\overline C^{(2,2)-}_{ab}\,\overline\epsilon^{(1)-ba} +4\,\overline E^{(1,2)-}_{ab}\,\overline\epsilon^{(2)-ab} -\overline E^{(1,2)}\,\overline\epsilon^{(2)} \nonumber\\ &\; +4\,\overline C^{(1,2)-}_{ab}\,\overline\epsilon^{(2)-ab} -16\,\overline K^{(1,2)-}_{abc}{}^{a}\, \overline\epsilon^{(2)-cb} +4\,\overline K^{(1,2)-}_{ab}{}^{ba}\, \overline\epsilon^{(2)}\,, \\[1ex] \delta \overline\Lambda^{(1)-}_{mn} ={}&\; 4\Bigl( \overline E^{(1,1)-}_{ma}\,\overline\epsilon^{(1)-}_{n}{}^{a} -\overline E^{(1,1)-}_{na}\,\overline\epsilon^{(1)-}_{m}{}^{a} +\frac12\,\varepsilon_{mnab}\, \overline E^{(1,1)-a}{}_{c}\, \overline\epsilon^{(1)-bc} \Bigr) \nonumber\\ &\; +\overline E^{(1,1)-}_{mn}\,\overline\epsilon^{(1)} -\overline C^{(2,2)-}_{mn}\,\overline\epsilon^{(1)} \nonumber\\ &\; +4\Bigl( \overline C^{(2,2)-}_{ma}\,\overline\epsilon^{(1)-}_{n}{}^{a} -\overline C^{(2,2)-}_{na}\,\overline\epsilon^{(1)-}_{m}{}^{a} -\frac12\,\varepsilon_{mnab}\, \overline C^{(2,2)-}_{c}{}^{a}\, \overline\epsilon^{(1)-bc} \Bigr) \nonumber\\ &\; +\overline E^{(1,2)-}_{mn}\,\overline\epsilon^{(2)} +2\Bigl( \overline E^{(1,2)-}_{ma}\,\overline\epsilon^{(2)-}_{n}{}^{a} -\overline E^{(1,2)-}_{na}\,\overline\epsilon^{(2)-}_{m}{}^{a} \Bigr) \nonumber\\ &\; +\overline E^{(1,2)}\,\overline\epsilon^{(2)-}_{mn} +\overline C^{(1,2)-}_{mn}\,\overline\epsilon^{(2)} \nonumber\\ &\; +2\Bigl( \overline C^{(1,2)-}_{ma}\,\overline\epsilon^{(2)-a}{}_{n} -\overline C^{(1,2)-}_{na}\,\overline\epsilon^{(2)-a}{}_{m} \Bigr) \nonumber\\ &\; -2\Bigl( \overline K^{(1,2)-}_{amn}{}^{a}\,\overline\epsilon^{(2)} -\overline K^{(1,2)-}_{anm}{}^{a}\,\overline\epsilon^{(2)} \Bigr) \nonumber\\ &\; +8\Bigl( \overline K^{(1,2)-}_{amb}{}^{a}\, \overline\epsilon^{(2)-b}{}_{n} -\overline K^{(1,2)-}_{anb}{}^{a}\, \overline\epsilon^{(2)-b}{}_{m} \Bigr) \nonumber\\ &\; +8\,\overline K^{(1,2)-}_{mnab}\, \overline\epsilon^{(2)-ab}\,, \\[1ex] \delta \overline\Lambda^{(2)} ={}&\; 4\,\overline E^{(2,2)-}_{ab}\,\overline\epsilon^{(2)-ab} +4\,\overline C^{(1,1)-}_{ab}\,\overline\epsilon^{(2)-ba} +4\,\overline E^{(1,2)-}_{ab}\,\overline\epsilon^{(1)-ab} \nonumber\\ &\; +\overline E^{(1,2)}\,\overline\epsilon^{(1)} +4\,\overline C^{(1,2)-}_{ab}\,\overline\epsilon^{(1)-ab} +16\,\overline K^{(1,2)-}_{abc}{}^{a}\, \overline\epsilon^{(1)-cb} \nonumber\\ &\; -4\,\overline K^{(1,2)-}_{ab}{}^{ba}\, \overline\epsilon^{(1)}\,, \\[1ex] \delta \overline\Lambda^{(2)-}_{mn} ={}&\; 4\Bigl( \overline E^{(2,2)-}_{ma}\,\overline\epsilon^{(2)-}_{n}{}^{a} -\overline E^{(2,2)-}_{na}\,\overline\epsilon^{(2)-}_{m}{}^{a} -\frac12\,\varepsilon_{mnab}\, \overline E^{(2,2)-}_{c}{}^{a}\, \overline\epsilon^{(2)-bc} \Bigr) \nonumber\\ &\; +\overline E^{(2,2)-}_{mn}\,\overline\epsilon^{(2)} -\overline C^{(1,1)-}_{mn}\,\overline\epsilon^{(2)} \nonumber\\ &\; +4\Bigl( \overline C^{(1,1)-}_{ma}\,\overline\epsilon^{(2)-}_{n}{}^{a} -\overline C^{(1,1)-}_{na}\,\overline\epsilon^{(2)-}_{m}{}^{a} -\frac12\,\varepsilon_{mnab}\, \overline C^{(1,1)-}_{c}{}^{a}\, \overline\epsilon^{(2)-bc} \Bigr) \nonumber\\ &\; +\overline E^{(1,2)-}_{mn}\,\overline\epsilon^{(1)} +2\Bigl( \overline E^{(1,2)-}_{ma}\,\overline\epsilon^{(1)-}_{n}{}^{a} -\overline E^{(1,2)-}_{na}\,\overline\epsilon^{(1)-}_{m}{}^{a} \Bigr) \nonumber\\ &\; -\overline E^{(1,2)}\,\overline\epsilon^{(1)-}_{mn} +\overline C^{(1,2)-}_{mn}\,\overline\epsilon^{(1)} \nonumber\\ &\; +2\Bigl( \overline C^{(1,2)-}_{m}{}^{a}\, \overline\epsilon^{(1)-}_{an} -\overline C^{(1,2)-}_{n}{}^{a}\, \overline\epsilon^{(1)-}_{am} \Bigr) \nonumber\\ &\; +2\Bigl( \overline K^{(1,2)-}_{amn}{}^{a} -\overline K^{(1,2)-}_{anm}{}^{a} \Bigr)\overline\epsilon^{(1)} \nonumber\\ &\; +8\Bigl( \overline K^{(1,2)-}_{amb}{}^{a}\, \overline\epsilon^{(1)-}_{n}{}^{b} -\overline K^{(1,2)-}_{anb}{}^{a}\, \overline\epsilon^{(1)-}_{m}{}^{b} \Bigr) \nonumber\\ &\; -8\,\overline K^{(1,2)-}_{mnab}\, \overline\epsilon^{(1)-ab}\,. \end{align} \end{subequations}
\paragraph{Twisted vector multiplet.}
\begin{table}[t] \centering \begin{tabular}{@{}cl@{}} \toprule Original field & Twisted components \\ \midrule
$A_m$ & $A_m$ \\ \addlinespace
$\lambda_\alpha{}^i$ & $\lambda_m^{(1)},\ \lambda_m^{(2)}$ \\ \addlinespace
$\overline\lambda^{\dot\alpha}{}_i$ & $\overline\lambda^{(1)},\ \overline\lambda_{mn}^{(1)-},\ \overline\lambda^{(2)},\ \overline{\lambda}^{(2)-}_{mn}$ \\ \addlinespace
$\phi^{ij}$ & $\phi^{(1,1)},\ \phi^{(1,2)},\ \phi^{(1,2)-}_{mn},\ \phi^{(2,2)}$ \\ \bottomrule \end{tabular} \caption{Twisted components of the Vafa-Witten vector multiplet.} \label{tab:vw-matter-fields} \end{table}
The fermions decompose in the same representation as the supersymmetry parameters, while the scalar field is embedded as
\begin{align}
    \phi^{ij} = \begin{pmatrix}
        \phi^{(1,1)} \varepsilon^{\dot \alpha \dot \beta} \, & \,\phi^{(1,2)} \varepsilon^{\dot \alpha \dot \beta} + \phi^{(1,2)-}_{mn}(\overline \sigma^{mn})^{\dot \alpha \dot \beta} \\
        -\phi^{(1,2)} \varepsilon^{\dot \beta \dot \alpha } - \phi^{(1,2)-}_{mn}(\overline \sigma^{mn})^{\dot \beta \dot \alpha}\, &\, \phi^{(2,2)} \varepsilon^{\dot \alpha \dot \beta}
    \end{pmatrix} \, .
\end{align}
The twisted equations are obtained with the following traces
\begin{subequations}
\begin{alignat}{2}
    \delta \lambda^{(1)}_m &:= -\frac12\delta \lambda^{(1)}_{\alpha}{}^{\dot \alpha} (\overline \sigma_m)_{\dot \alpha}{}^{\alpha} \, , &\qquad \delta \phi^{(1,1)} &:=\frac12 \delta \phi^{(1,1) \dot \alpha \dot \beta} \varepsilon_{ \dot \beta\dot \alpha} \, , \\
    \delta \overline \lambda^{(1)}&:= \frac12 \delta \overline \lambda^{(1)\dot \alpha}{}_{\dot \beta} \delta^{\dot \beta}{}_{\dot \alpha} \, , &\qquad  \delta \phi^{(1,2)}&:=\frac12 \delta \phi^{(1,2) \dot \alpha \dot \beta} \varepsilon_{\dot \beta \dot \alpha} \, ,\\
    \delta \overline \lambda^{(1)-}_{m n} &:= -\frac18 \delta \overline \lambda^{(1) \dot \alpha}{}_{\dot \beta} (\overline \sigma_{m n})^{\dot \beta}{}_{\dot \alpha} \, , &\qquad \delta \phi^{(1,2)-}_{m n} &:= \frac18 \delta \phi^{(1,2) \dot \alpha \dot \beta} (\overline \sigma_{m n})_{\dot \beta \dot \alpha} \, . 
\end{alignat}
\end{subequations}
Since the Vafa-Witten twist is completely symmetric under the exchange of the two $SU(2)$ blocks, the traces involving $(2)$ components are identical. The complete twisted variation of the matter multiplet in VW can be found in Appendix \ref{sec: VW matt}.

\subsection{Geometric Langlands twisted equations}\label{sec: twisted GL}
The GL twist consists of the following definitions of the $SU(4)$ gauge fields
\begin{align}\label{eq: SU(4) gauge fields GL}
    V_\mu{}^{i}{}_{j} = - \frac{1}{4} \omega_{\mu}{}^{ab} \begin{pmatrix}
        (\overline \sigma_{ab})^{\dot \alpha}{}_{\dot \beta} & 0 \\
        0 & -(\sigma_{ab})^{\alpha}{}_{\beta}
    \end{pmatrix}^{i}{}_{j} \, ; \quad \overline{V}_{\mu}{}_{i}{}^{j} = -(V_\mu ^{T})_{i}{}^{j} \, ,
\end{align}
corresponding to the following embedding
\begin{align}
    \begin{pmatrix}
        SU(2)_{r} & 0\\0&SU(2)_{l}
    \end{pmatrix}\, .
\end{align}
This embedding commutes with a $U(1)$, so that the full group structure is
\begin{align}
\left(SU(2)_l\times SU(2)_r\times U(1) \right)/\mathbb{Z}_2 \, .
\end{align}
We shall not keep track of the $U(1)$ factor here, the interested reader is referred to \cite{Kapustin:2006pk}. 
The $SU(4)$ indices split as $i \rightarrow \dot \alpha + \alpha$. We can twist the Weyl spinors as
\begin{align}
\epsilon_{\alpha}{}^{i} = \begin{pmatrix}
    \epsilon^{(1)}_{m} (\sigma^m)_{\alpha}{}^{\dot \beta} \, & \, \epsilon^{(2)+}_{mn} (\sigma^{mn})_{\alpha}{}^{\beta} + \delta_{\alpha}{}^{\beta} \epsilon^{(2)}
\end{pmatrix}  \, ; \quad \overline \epsilon^{\dot \alpha}{}_i = \begin{pmatrix}
    \overline \epsilon^{(1)-}_{mn} (\overline{\sigma}^{mn})^{\dot \alpha}{}_{\dot \beta} + \delta^{\dot \alpha}{}_{\dot \beta} \overline \epsilon^{(1)} \, & \, \overline \epsilon^{(2)}_{m} (\overline \sigma^m)^{\dot \alpha}{}_{\beta}
\end{pmatrix}  \, ,
\end{align}
while the bosonic fields of the Weyl multiplet are embedded as
\begin{align}
    T_{mn}^{+\,ij} = \begin{pmatrix}
        C^{(1,1)+}_{mn} \varepsilon^{\dot \alpha \dot \beta} & K^{(1,2)+}_{mn\, r} (\overline \sigma^r)^{\dot \alpha \beta}\\
        -K^{(1,2)+}_{mn\, r} (\overline \sigma^r)^{\beta \dot \alpha} &C^{(2,2)+}_{mn} \varepsilon^{\alpha \beta}
    \end{pmatrix} \, ; \quad \overline{T}^-_{mn \, ij} = \begin{pmatrix}
        \overline{C}^{(1,1)-}_{mn} \varepsilon_{\dot \alpha \dot \beta} & \overline{K
}^{(1,2)-}_{mn\, r} (\overline \sigma^r)_{\dot \alpha \beta}\\
-\overline{K
}^{(1,2)-}_{mn\, r} (\overline \sigma^r)_{\beta \dot \alpha} & \overline{C}^{(2,2)-}_{mn} \varepsilon_{\alpha \beta}
    \end{pmatrix} \, ,
\end{align}
\begin{align}
    E_{ij} = \begin{pmatrix}
        E^{(1,1)-}_{mn}(\overline \sigma^{mn})_{\dot \alpha \dot \beta} & E^{(1,2)}_{m} (\overline \sigma^m)_{\dot \alpha \beta}\\
        E^{(1,2)}_{m} (\overline \sigma^m)_{\beta \dot \alpha} &E^{(2,2)+}_{mn} (\sigma^{mn})_{\alpha \beta}
    \end{pmatrix} \, ; \quad \overline{E}^{ij} = \begin{pmatrix}
        \overline{E}^{(1,1)-}_{mn}(\overline \sigma^{mn})^{\dot \alpha \dot \beta} & \overline{E}^{(1,2)}_{m} (\overline \sigma^m)^{\dot \alpha \beta}\\
        \overline{E}^{(1,2)}_{m} (\overline \sigma^m)^{\beta \dot \alpha} & \overline{E}^{(2,2)+}_{mn} (\sigma^{mn})^{\alpha \beta}
    \end{pmatrix} \, .
\end{align}
\begin{table}[t]
\centering
\begin{tabular}{@{}cl@{}}
\toprule Original field & Twisted components \\ \midrule
$\epsilon_\alpha{}^i$ & $\epsilon_m^{(1)},\ \epsilon^{(2)+}_{mn},\ \epsilon^{(2)}$ \\ \addlinespace
$\overline\epsilon^{\dot\alpha}{}_i$ & $\overline\epsilon^{(1)},\ \overline\epsilon_{mn}^{(1)-},\ \overline\epsilon^{(2)}_m$ \\ \addlinespace $T_{mn}^{+\,ij}$ &  $C_{mn}^{(1,1)+},\ K^{(1,2)+}_{mn \, r} ,\ C_{mn}^{(2,2)+}$ \\ \addlinespace
$\overline T_{mn\,ij}^{-}$ & $\overline C_{mn}^{(1,1)-},\ \overline{K}^{(1,2)-}_{mn\,r} ,\ \overline C_{mn}^{(2,2)-}$ \\ \addlinespace
$E_{ij}$ & $E_{mn}^{(1,1)-},\ E_{m}^{(1,2)},\ E_{mn}^{(2,2)+}$ \\ \addlinespace
$\overline E^{ij}$ & \makecell[l]{ $\overline E_{mn}^{(1,1)-},\ \overline E^{(1,2)}_{m}\,,\, \overline E^{(2,2)+}_{mn}$} \\ \bottomrule 
\end{tabular} 
\caption{Twisted components of the supersymmetry parameters and bosonic Weyl-multiplet fields in the Geometric Langlands decomposition.}
\label{tab:gl-twisted-fields}
\end{table}

\paragraph{Twisted gravitino equations.}
The twisted equations can be obtained as the following traces of the supergravity equations
\begin{subequations}
\begin{alignat}{2}
    \delta \psi^{(1)}_{m n} &:= -\frac12 \delta \psi^{(1)}_{m \, \alpha}{}^{\dot \alpha} (\overline \sigma_n)_{\dot \alpha}{}^{\alpha} \, , &\qquad \delta \psi^{(2)}_m &:= \frac12 \delta \psi^{(2)}_{m \, \alpha}{}^{\beta} \delta_{\beta}{}^{\alpha} \, , \\
    \delta \overline \psi^{(1)}_{m} &:= \frac12 \delta \overline \psi^{(1)}_{m}{}^{\dot \alpha}{}_{\dot \beta} \delta^{\dot \beta}{}_{\dot \alpha} \, ,&\qquad \delta \psi^{(2)+}_{m\,  nr} &:= -\frac18\delta \psi^{(2)}_{m \, \alpha}{}^{\beta} (\sigma_{n r})_{\beta}{}^{\alpha} \, ,\\
     \delta \overline \psi^{(1)-}_{m \, nr}&:=-\frac18 \delta \overline \psi^{(1)}_{m}{}^{\dot \alpha}{}_{\dot \beta} (\overline \sigma_{n r})^{\dot \beta}{}_{\dot \alpha} \, , &\qquad \delta \overline \psi^{(2)}_{m n} &:= -\frac12 \delta \overline \psi^{(2)}_{m}{}^{\dot \alpha}{}_{\alpha} (\sigma_{n})^{\alpha}{}_{\dot \alpha} \, .
\end{alignat}
\end{subequations}
The twisted gravitino equations are therefore
\begin{subequations}
\begin{align}
\delta \psi^{(1)}_{mn} ={}& \nabla_{m}\epsilon^{(1)}_{n} +C^{(1,1)+}_{mn}\,\overline{\epsilon}^{(1)} +4\,C^{(1,1)+}_{m}{}^{a}\,\overline{\epsilon}^{(1)-}_{an} \nonumber\\ &+ K^{(1,2)+}_{am}{}^{a}\,\overline{\epsilon}^{(2)}_{n} +K^{(1,2)+}_{an}{}^{a}\,\overline{\epsilon}^{(2)}_{m} -\delta_{mn}\, K^{(1,2)+}_{ab}{}^{a}\,\overline{\epsilon}^{(2)}_{b} \nonumber\\ &+ \varepsilon_{mabc}\, K^{(1,2)+ab}{}_{n}\,\overline{\epsilon}^{(2)c} -\varepsilon_{mnab}\, K^{(1,2)+ca}{}_{c}\,\overline{\epsilon}^{(2)b} \nonumber\\ &- \frac14\Big( \nabla_{m}\epsilon^{(1)}_{n} -\nabla_{n}\epsilon^{(1)}_{m} +\delta_{mn}\,\nabla^{a}\epsilon^{(1)}_{a} -\varepsilon_{mnab}\,\nabla^{a}\epsilon^{(1)b} \Big) \, , \\ \delta \psi^{(2)}_{m} ={}& \frac34\,\nabla_{m}\epsilon^{(2)} +C^{(2,2)+}_{mb}\,\overline{\epsilon}^{(2)b} +4\,\overline{\epsilon}^{(1)-ab}\,K^{(1,2)+}_{amb} + \overline{\epsilon}^{(1)}\,K^{(1,2)+}_{am}{}^{a} +\nabla^{a}\epsilon^{(2)+}_{ma} \, , \\ \delta \psi^{(2)+}_{m\,nr} ={}& \nabla_{m}\epsilon^{(2)+}_{nr} -\frac14\Big( \delta_{mn}\,C^{(2,2)+}_{ra}\,\overline{\epsilon}^{(2)}_{a} -\delta_{mr}\,C^{(2,2)+}_{na}\,\overline{\epsilon}^{(2)a} \nonumber\\ &-C^{(2,2)+}_{rm}\,\overline{\epsilon}^{(2)}_{n} +C^{(2,2)+}_{nm}\,\overline{\epsilon}^{(2)}_{r} +C^{(2,2)+}_{nr}\,\overline{\epsilon}^{(2)}_{m} \Big) \nonumber\\ &+ 2\Big( K^{(1,2)+}_{mn}{}^{a}\,\overline{\epsilon}^{(1)-}_{ar} -K^{(1,2)+}_{mr}{}^{a}\,\overline{\epsilon}^{(1)-}_{an} \Big) \nonumber\\ &+ \Big( \delta_{mn}\,K^{(1,2)+}_{rab}\,\overline{\epsilon}^{(1)-ab} -\delta_{mr}\,K^{(1,2)+}_{nab}\,\overline{\epsilon}^{(1)-ab} \Big) \nonumber\\ &- \varepsilon_{mnr}{}^{a}\, K^{(1,2)+}_{abc}\,\overline{\epsilon}^{(1)-bc} -\frac12\,\overline{\epsilon}^{(1)} \Big( K^{(1,2)+}_{mnr} -K^{(1,2)+}_{mrn} \Big) \nonumber\\ &- \frac14\Big( \delta_{mn}\,\overline{\epsilon}^{(1)}\,K^{(1,2)+}_{ar}{}^{a} -\delta_{mr}\,\overline{\epsilon}^{(1)}\,K^{(1,2)+}_{an}{}^{a} \Big) +\frac14\,\varepsilon_{mnra}\, \overline{\epsilon}^{(1)}\,K^{(1,2)+}_{ba}{}^{b} \nonumber\\ &- \frac14\Big( \delta_{mn}\,\nabla^{a}\epsilon^{(2)+}_{ar} -\delta_{mr}\,\nabla^{a}\epsilon^{(2)+}_{an} +\nabla_{n}\epsilon^{(2)+}_{rm} -\nabla_{r}\epsilon^{(2)+}_{nm} +\nabla_{m}\epsilon^{(2)+}_{nr} \Big) \nonumber\\ &- \frac1{16}\Big( \delta_{mn}\,\nabla_{r}\epsilon^{(2)} -\delta_{mr}\,\nabla_{n}\epsilon^{(2)} -\varepsilon_{mnar}\,\nabla^{a}\epsilon^{(2)} \Big) \, , \\ \delta \overline{\psi}^{(1)}_{m} ={}& \frac34\,\nabla_{m}\overline{\epsilon}^{(1)} +\overline C^{(1,1)-}_{ma}\,\epsilon^{(1)a} +\overline K^{(1,2)-}_{am}{}^{a}\,\epsilon^{(2)} + 4\,\overline K^{(1,2)-}_{amb}\,\epsilon^{(2)+ab} +\nabla^{a}\overline{\epsilon}^{(1)-}_{ma} \, , \\ \delta \overline{\psi}^{(1)-}_{m\,nr} ={}& \nabla_{m}\overline\epsilon^{(1)-}_{nr} -\frac14\Big( \delta_{mn}\,\overline C^{(1,1)-}_{rl}\,\epsilon^{(1)l} -\delta_{mr}\,\overline C^{(1,1)-}_{nl}\,\epsilon^{(1)l} \nonumber\\ & -\overline C^{(1,1)-}_{rm}\,\epsilon^{(1)}_{n} +\overline C^{(1,1)-}_{nm}\,\epsilon^{(1)}_{r} +\epsilon^{(1)}_{m}\,\overline C^{(1,1)-}_{nr} \Big) \nonumber\\ &- \frac12\,\epsilon^{(2)}\Big( \overline K^{(1,2)-}_{mnr} -\overline K^{(1,2)-}_{mrn} \Big) \nonumber\\ &- \frac14\,\epsilon^{(2)}\Big( \delta_{mn}\,\overline K^{(1,2)-}_{ar}{}^{a} -\delta_{mr}\,\overline K^{(1,2)-}_{an}{}^{a} \Big) -\frac14\,\epsilon^{(2)}\varepsilon_{mnra}\, \overline K^{(1,2)-ba}{}_{b} \nonumber\\ &+ 2\Big( \overline K^{(1,2)-}_{mn}{}^{a}\,\epsilon^{(2)+}_{ar} -\overline K^{(1,2)-}_{mr}{}^{a}\,\epsilon^{(2)+}_{an} \Big) \nonumber\\ &+ \Big( \delta_{mn}\,\overline K^{(1,2)-}_{rab}\,\epsilon^{(2)+ab} -\delta_{mr}\,\overline K^{(1,2)-}_{nab}\,\epsilon^{(2)+ab} \Big) \nonumber\\ &+ \varepsilon_{mnra}\, \overline K^{(1,2)-abc}\,\epsilon^{(2)+}_{bc} \nonumber\\ &- \frac1{16}\Big( \delta_{mn}\,\nabla_{r}\overline{\epsilon}^{(1)} -\delta_{mr}\,\nabla_{n}\overline{\epsilon}^{(1)} -\varepsilon_{manr}\,\nabla^{a}\overline{\epsilon}^{(1)} \Big) \nonumber\\ &- \frac14\Big( \delta_{mn}\,\nabla^{a}\overline\epsilon^{(1)-}_{ar} -\delta_{mr}\,\nabla^{a}\overline\epsilon^{(1)-}_{an} +\nabla_{n}\overline\epsilon^{(1)-}_{rm} -\nabla_{r}\overline\epsilon^{(1)-}_{nm} +\nabla_{m}\overline\epsilon^{(1)-}_{nr} \Big) \, , \\ \delta \overline{\psi}^{(2)}_{mn} ={}& \nabla_{m}\overline\epsilon^{(2)}_{n} +\overline C^{(2,2)-}_{mn}\,\epsilon^{(2)} +4\,\overline C^{(2,2)-}_{m}{}^{a}\,\epsilon^{(2)+}_{an} \nonumber\\ &- \frac14\Big( \nabla_{m}\overline\epsilon^{(2)}_{n} -\nabla_{n}\overline\epsilon^{(2)}_{m} +\delta_{mn}\,\nabla^{a}\overline\epsilon^{(2)}_{a} +\varepsilon_{mnab}\,\nabla^{a}\overline\epsilon^{(2)b} \Big) \nonumber\\ &+ \overline K^{(1,2)-}_{am}{}^{a}\,\epsilon^{(1)}_{n} +\overline K^{(1,2)-}_{an}{}^{a}\,\epsilon^{(1)}_{m} -\delta_{mn}\,\overline K^{(1,2)-}_{a}{}^{ba}\,\epsilon^{(1)}_{b} \nonumber\\ &- \varepsilon_{mabc}\, \overline K^{(1,2)-ab}{}_{n}\,\epsilon^{(1)c} +\varepsilon_{mnab}\, \overline K^{(1,2)-}_{c}{}^{ac}\,\epsilon^{(1)b} \, .
\end{align}
\end{subequations}

\paragraph{Twisted gaugino equations.}
The twisted equations can be obtained as the following traces of the supergravity equations
\begin{subequations}
\begin{alignat}{2}
    \delta \Lambda^{(1)}_{m} &:= \frac12\delta \Lambda^{(1)}_{\alpha \dot \alpha} (\overline \sigma_m)^{\dot \alpha \alpha} \, , &\qquad \delta \Lambda^{(2)} &:= \frac12\delta \Lambda^{(2)}_{\alpha \beta} \varepsilon^{\beta \alpha} \, ,\\
    \delta \overline \Lambda^{(1)} &:= \frac12\delta \overline \Lambda^{(1) \dot \alpha \dot \beta} \varepsilon_{\dot \beta \dot \alpha} \, ,&\qquad \delta \Lambda^{(2)+}_{m n} &:= \frac18\delta \Lambda^{(2)}_{\alpha \beta} (\sigma_{m n})^{\beta \alpha} \, ,\\
    \quad \delta \overline \Lambda^{(1)-}_{m n} &:= \frac18\delta \overline \Lambda^{(1) \dot \alpha \dot \beta} (\overline \sigma_{m n})_{\dot \beta \dot \alpha} \, , &\qquad \delta \overline \Lambda^{(2)}_{m} &:= \frac12\delta \overline \Lambda^{(2) \dot \alpha \alpha} (\sigma_m)_{\alpha \dot \alpha} \, .
\end{alignat}
\end{subequations}
The twisted gaugino equations read
\begin{subequations} \begin{align} \delta \Lambda^{(1)}_{m} ={}& 4\,E^{(1,1)-}_{ma}\,\epsilon^{(1)a} -E^{(1,2)}_{m}\,\epsilon^{(2)} +4\,E^{(1,2)a}\,\epsilon^{(2)+}_{am} +4\,C^{(2,2)+}_{am}\,\epsilon^{(1)a} \nonumber\\ &- 4\,K^{(1,2)+}_{am}{}^{a}\,\epsilon^{(2)} -16\,K^{(1,2)+}_{amb}\,\epsilon^{(2)+ab}\,, \\ \delta \Lambda^{(2)} ={}& E^{(1,2)}_{a}\,\epsilon^{(1)a} +4\,E^{(2,2)+}_{ab}\,\epsilon^{(2)+ab} -4\,C^{(1,1)+}_{ab}\,\epsilon^{(2)+ab} \nonumber\\ &+ 4\,K^{(1,2)+}_{ba}{}^{b}\,\epsilon^{(1)a}\,, \\ \delta \Lambda^{(2)+}_{mn} ={}& E^{(2,2)+}_{mn}\,\epsilon^{(2)} -2\Bigl( E^{(2,2)+}_{ma}\,\epsilon^{(2)+a}{}_{n} -E^{(2,2)+}_{na}\,\epsilon^{(2)+a}{}_{m} \Bigr) \nonumber\\ &+ \frac14\Bigl( E^{(1,2)}_{m}\,\epsilon^{(1)}_{n} -E^{(1,2)}_{n}\,\epsilon^{(1)}_{m} +\varepsilon_{mnrs}\,E^{(1,2)r}\,\epsilon^{(1)s} \Bigr) \nonumber\\ &- 2\Bigl( C^{(1,1)+}_{ma}\,\epsilon^{(2)+a}{}_{n} -C^{(1,1)+}_{na}\,\epsilon^{(2)+a}{}_{m} \Bigr) -C^{(1,1)+}_{mn}\,\epsilon^{(2)} \nonumber\\ &- \Bigl( K^{(1,2)+}_{am}{}^{a}\,\epsilon^{(1)}_{n} -K^{(1,2)+}_{an}{}^{a}\,\epsilon^{(1)}_{m} \Bigr) -2\,K^{(1,2)+}_{mna}\,\epsilon^{(1)a} \nonumber\\ &- \varepsilon_{mnab}\,K^{(1,2)+ca}{}_{c}\,\epsilon^{(1)b}\,, \\ \delta \overline{\Lambda}^{(1)} ={}& 4\,\overline E^{(1,1)-}_{ab}\,\overline\epsilon^{(1)-ab} -\overline E^{(1,2)}_{a}\,\overline\epsilon^{(2)a} +4\,\overline K^{(1,2)-ab}{}_{a}\,\overline\epsilon^{(2)}_{b} \nonumber\\ &+ 4\,\overline C^{(2,2)-}_{ab}\,\overline\epsilon^{(1)-ba}\,, \\ \delta \overline{\Lambda}^{(1)-}_{mn} ={}& 2\Bigl( \overline E^{(1,1)-}_{ma}\,\overline\epsilon^{(1)-}_{n}{}^{a} -\overline E^{(1,1)-}_{na}\,\overline\epsilon^{(1)-}_{m}{}^{a} \Bigr) +\overline E^{(1,1)-}_{mn}\,\overline\epsilon^{(1)} \nonumber\\ &- \frac14\Bigl( \overline E^{(1,2)}_{m}\,\overline\epsilon^{(2)}_{n} -\overline E^{(1,2)}_{n}\,\overline\epsilon^{(2)}_{m} -\varepsilon_{mnab}\, \overline E^{(1,2)a}\,\overline\epsilon^{(2)b} \Bigr) \nonumber\\ &- \overline C^{(2,2)-}_{mn}\,\overline\epsilon^{(1)} +2\Bigl( \overline C^{(2,2)-}_{ma}\,\overline\epsilon^{(1)-}_{n}{}^{a} -\overline C^{(2,2)-}_{na}\,\overline\epsilon^{(1)-}_{m}{}^{a} \Bigr) \nonumber\\ &- \Bigl( \overline K^{(1,2)-}_{am}{}^{a}\,\overline\epsilon^{(2)}_{n} -\overline K^{(1,2)-}_{an}{}^{a}\,\overline\epsilon^{(2)}_{m} \Bigr) \nonumber\\ &- 2\Bigl( \overline K^{(1,2)-}_{man}\,\overline\epsilon^{(2)a} -\overline K^{(1,2)-}_{nam}\,\overline\epsilon^{(2)a} \Bigr) \nonumber\\ &- \varepsilon_{mnab}\, \overline K^{(1,2)-ca}{}_{c}\,\overline\epsilon^{(2)b}\,, \\ \delta \overline{\Lambda}^{(2)}_{m} ={}& 4\,\overline E^{(1,2)a}\,\overline\epsilon^{(1)-}_{ma} +\overline E^{(1,2)}_{m}\,\overline\epsilon^{(1)} +4\,\overline E^{(2,2)+}_{ma}\,\overline\epsilon^{(2)a} - 4\,\overline C^{(1,1)-}_{ma}\,\overline\epsilon^{(2)a}\nonumber\\ & -4\,\overline K^{(1,2)-}_{am}{}^{a}\,\overline\epsilon^{(1)} + 16\,\overline K^{(1,2)-ab}{}_{a}\, \overline\epsilon^{(1)-}_{mb} -8\,\overline K^{(1,2)-}_{abm}\, \overline\epsilon^{(1)-ab}\,. \end{align}
\end{subequations}
\paragraph{Twisted vector multiplet.}
\begin{table}[t]
\centering
\begin{tabular}{@{}cl@{}}
\toprule Original field & Twisted components \\ \midrule
$A_m$ & $A_m$ \\ \addlinespace
$\lambda_\alpha{}^i$ & $\lambda_m^{(1)},\ \lambda^{(2)},\ \lambda^{(2)+}_{mn}$ \\ \addlinespace
$\overline\lambda^{\dot\alpha}{}_i$ & $\overline\lambda^{(1)},\ \overline\lambda_{mn}^{(1)-},\ \overline\lambda^{(2)}_m$ \\ \addlinespace
$\phi^{ij}$ & $\phi^{(1,1)},\ \phi^{(1,2)}_m,\ \phi^{(2,2)}$ \\
\bottomrule
\end{tabular}
\caption{Twisted components of the Geometric Langlands vector multiplet.} \label{tab:gl-matter-fields}
\end{table}
The fermions are twisted in the same representation as the supersymmetry parameters and the scalar is twisted as
\begin{align}
    \phi^{ij} = \begin{pmatrix}
        \phi^{(1,1)} \varepsilon^{\dot \alpha \dot \beta} & \phi^{(1,2)}_m (\overline \sigma^m)^{\dot \alpha \beta}\\
        -\phi^{(1,2)}_m (\overline \sigma^m)^{\beta \dot \alpha} & \phi^{(2,2)} \varepsilon^{\alpha \beta}
    \end{pmatrix} \, .
\end{align}
The traces yielding the twisted matter equations are
\begin{subequations}
\begin{alignat}{2}
    \delta \lambda^{(1)}_{m} &:= -\frac12 \delta \lambda^{(1)}_{\alpha}{}^{\dot \beta} (\overline \sigma_m)_{\dot \beta}{}^{\alpha} \, , &\qquad \delta \lambda^{(2)+}_{m n} &:= -\frac18 \delta \lambda^{(2)}_{\alpha}{}^{\beta} (\sigma_{m n})_{\beta}{}^{\alpha} \, ,\\
    \delta \overline \lambda^{(1)-}_{m n} &:= -\frac18 \delta \overline \lambda^{(1) \dot \alpha}{}_{\dot \beta} (\overline \sigma_{m n})^{\dot \beta}{}_{\dot \alpha} \, , &\qquad \delta \lambda^{(2)}&:=\frac12 \delta \lambda^{(2)}_{\alpha}{}^{\beta} \delta_{\beta}{}^{\alpha} \, , \\
    \delta \overline \lambda^{(1)}&:=\frac12 \delta \overline \lambda^{(1) \dot \alpha}{}_{\dot \beta} \delta^{\dot \beta}{}_{\dot \alpha} \, , & \qquad \delta \overline \lambda^{(2)}_{m} &:= -\frac12 \delta \overline \lambda^{(2) \dot \alpha}{}_{\beta} (\sigma_m)^{\beta}{}_{\dot \alpha} \, , \\
    \delta \phi^{(1,1)}&:= \frac12 \delta \phi^{(1,1) \dot \alpha \dot \beta} \varepsilon_{\dot \beta \dot \alpha} \, , &\qquad \delta \phi^{(2,2)}&:= \frac12 \delta \phi^{(2,2) \alpha \beta} \varepsilon_{\beta \alpha} \, , \\
    \delta \phi^{(1,2)}_m&:= \frac12 \delta \phi^{(1,2) \dot \alpha \beta} (\sigma_{m})_{\beta \dot \alpha} \, .
\end{alignat}
\end{subequations}
The complete twisted variation of the matter multiplet in GL can be found in Appendix \ref{sec: GL matt}.

\section{Supersymmetric backgrounds}\label{sec: susy backgrounds}

\subsection{$\Omega$-like backgrounds}
We will call a background of $\Omega$-like if it preserves a supercharge obtained by turning on the scalar $\epsilon$ and one-form $\epsilon_\mu$ parts of the supersymmetry parameters, while the two-form parts are turned off. In particular, we suppose that vectors fields dual to the one-forms $\epsilon_\mu$ are conformal Killing (CKV).

Under these assumptions we show that $\Omega$-like deformations are supported by any of the three twists of $\mathcal{N}=4$. As a result, the deformed theories can be put on any manifold supporting the same isometries generated by the vector fields.

The DW $\Omega$-deformation preserves $N_T=1$ supercharges and consists of an embedding of Nekrasov's $\mathcal{N}=2$ deformation into the bigger $\mathcal{N}=4$ algebra. In the VW and GL deformations, on the other hand, $N_T=2$ supercharges are separately conserved. We still denote the number of conserved supercharges $N_T$, with the understanding that they are no longer topological because they are equivariantly deformed.

In the following subsections, we will display the relevant equations in form notation, which inevitably entails renaming of the vector multiplet and supersymmetry parameter fields. Our differential geometric conventions are collected in Appendix \ref{sec: diff geo}. We shall use the parameters $v,u$ for the scalars, while the one-form parts of the supersymmetry parameters are denoted $v V_1^\flat, uV_2^\flat$. The exception is given by the fuzzy sphere background, where there are no vector fields turned on and the scalars are denoted $w,u$. In renaming the vector multiplet twisted fields, we try to adhere as much as possible to standard nomenclature in order to make contact to the relevant original works. The dictionary between coordinate and form notation is summarised in tables that can be found in each subsection.

We solve the twisted gravitino and gaugino equations analytically for the background fields specified below. We then evaluate the dilatino equation on the explicit manifolds, orthonormal frames and vector fields given in Appendix \ref{sec: metrics}, solve the resulting component equations numerically for $D^{ij}{}_{kl}$, and record the corresponding solutions in each case. When computing the closure of the algebra, we further impose for the vector fields to be divergenceless so that they are Killing as opposed to confromal Killing. This is necessary to recover the wanted algebra \eqref{eq: algebra omega def}, otherwise CKVs would contribute with Weyl transformation factors.

\subsubsection{Donaldson-Witten $\Omega$-background}\label{sec: DW omega}
We will now show that a solution to the gravitino equations can be found upon setting all twisted components of $T, \overline T$ except the self-dual two-form $C^{(1,1)+}$ to zero. With these assumptions, all components of the gravitino vanish identically except for the two-form
\begin{align}
    \delta \psi^{(1)} =  \frac12 v(\dd V^\flat_1)^+ +v C^{(1,1)+} \, ,
\end{align}
which is set to zero by
\begin{align}
    C^{(1,1)+}=-\frac12 (\dd V^\flat_1)^+ \, .
\end{align}
Similarly, all twisted components of the gaugino vanish identically provided that we turn off $E, \overline{E}$.
In summary, the $\Omega$-background of the DW twist can be realised by
\begin{align}
    \begin{cases}
        C^{(1,1)+} = -\frac12 (\dd V^\flat_1)^+ \\
        \overline \epsilon^{(1)} = v \\
        V_1 \, \, \text{a CKV} 
    \end{cases} \, ,
\end{align}
and all other Weyl-multiplet fields set to zero.

The numerically verified solution to the dilatino is
\begin{align}
    D^{ij}{}_{kl}= -\frac{R}{2} \begin{pmatrix}
        \iu \varepsilon &0\\0&0
    \end{pmatrix}^{ij} \otimes \begin{pmatrix}
        \iu \varepsilon &0\\0&0
    \end{pmatrix}_{kl} - (\text{traces}) \, ,
\end{align}
where $R$ is the Ricci scalar and $\varepsilon=\varepsilon_{\alpha \beta}$ is the matrix used for lowering $SU(2)$ indices.

\paragraph{Closure of the algebra}

\begin{table}[t] \centering \begin{tabular}{@{}cc|cc@{}} \toprule Coordinate & Form & Coordinate & Form \\ \midrule $A_m$ & $A$ & $\frac{g \sqrt{- \iu}}{2}\Bigl( 8V_1^n\overline{\lambda}^{(1)-}_{nm} -2\overline{\lambda}^{(1)}V^\flat_{1m} -2\lambda^{(1)}_m \Bigr)$ & $\Psi$ \\ $-\iu g \left(\phi^{(1,1)} + \phi^{(2,2)}V_1^m V^\flat_{1m} \right)$ & $\Phi$ & $\sqrt{-\iu } \, \overline{\lambda}^{(1)}$ & $\eta$ \\ $\phi^{(2,2)}$ & $\overline{\Phi}$ & $\overline{\lambda}^{(1)-}_{mn}$ & $\chi^-$ \\ && $\overline{\epsilon}^{(1)}$ & $v$ \\ && $\epsilon^{(1)}_m$ & $vV^\flat_{1}$ \\ \bottomrule \end{tabular} \caption{Donaldson-Witten coordinate and differential-form notation for the bosonic fields, fermionic fields and supersymmetry parameters retained in the calculation. The one-form $\Psi$ is defined as the complete supersymmetry variation of $A$.}
\label{tab:DW redefined form dictionary}
\end{table}

We suppose that the vectors are now Killing instead of CKV. Let us define 
\begin{align}
Q=\frac{\sqrt{-\iu}}{2v} \delta\,\,.
\end{align}
In form notation (see table \ref{tab:DW redefined form dictionary}) the relevant twisted equations for checking the closure of the algebra read
\begin{subequations} \begin{align} Q A ={}& \,\Psi, \\[1ex]
Q\Psi ={}& \iu \,\iota_{V_1}F + \,D\Phi, \\[1ex]
Q\eta ={}& \iu \,\iota_{V_1}D\overline{\Phi} -\bigl[\Phi,\overline{\Phi}\bigr], \\[1ex]
Q\chi^- ={}& \sqrt{-\iu}\Bigl( -\frac{1}{4g}F^- -\frac12\bigl(D\overline{\Phi}\wedge V_1^\flat\bigr)^- -\frac{1}{4}\,\overline{\Phi} \bigl(\mathrm dV_1^\flat\bigr)^- -\frac{g}{8}\,\mathcal M^-\Bigr), \\[1ex]
Q\Phi ={}& \iu \,\iota_{V_1}\Psi, \\[1ex]
Q\overline{\Phi} ={}& \eta. \end{align} \end{subequations}
where 
\begin{align}
\mathcal{M}^-_{mn} = [\phi^{(1,2)\dot\alpha A},\phi^{(1,2)}_{\dot\beta A}](\overline{\sigma}_{mn})_{\dot\alpha}{}^{\dot\beta} \, .
\end{align}
The field redefinition in table \ref{tab:DW redefined form dictionary} makes it apparent that both the explicit dependence on $v$ and on the gauge coupling $g$ can be scaled away. Without loss of generality we could have defined the scalar part of the supercharge to be $\overline{\epsilon}^{(1)}=1$. Crucially, getting rid of the dependence on $g$ is possible since the gauge coupling is constant in this background. This will not be the case in more general solutions, such as the one discussed in Section \ref{sec: vw phi}.

The algebra closes as
\begin{equation}
\begin{aligned}
    Q^2 A = \iu \mathcal{L}_{V_1} A - D \varphi \, ,
\end{aligned}
\end{equation}
where we have used that $V_1$ is gauge neutral and defined
\begin{align}
    \varphi := \iu \iota_{V_1}A - \Phi  \, .
\end{align}
This is the expected algebra (cfr \eqref{eq: algebra omega def}). 
The $\mathcal{N}=4$ Donaldson-Witten twist is just the embedding of the $\mathcal{N}=2$ twist inside the bigger $\mathcal{N}=4$ algebra. If one disregards the additional $SU(2)_A$ flavour symmetry, the transformation of the matter fields of $\mathcal{N}=4$ can be exactly recast in terms of fields of $\mathcal{N}=2$ chiral multiplet which contains a vector $A$, a complex scalar $\Phi, \overline \Phi$ and two Weyl fermions reorganised into $\Psi, \eta$ and $\chi^-$. For the algebra to close off-shell one has to define an auxiliary field, which is essentially proportional to $\mathcal{M}^-$.

Notice that, defining an auxiliary field equal to the variation of $\chi^-$, one recovers the transformations in eq. (2.2) of \cite{Bonelli:2020xps}.\footnote{We believe in the reference there is a typo in the commutator term.} Due to opposite Clifford algebra conventions, our fields are anti-self-dual as opposed to self-dual.

\subsubsection{Vafa-Witten $\Omega$-background}\label{sec: vw omega}

\begin{table}[t] \centering \begin{tabular}{@{}cc|cc@{}} \toprule Coordinate & Form & Coordinate & Form \\ \midrule $A_m$ & $A$ & $\lambda^{(1)}_m$ & $\widetilde \psi$ \\ $\phi^{(1,1)}$ & $\sigma$ & $\lambda^{(2)}_m$ & $\psi$ \\ $\phi^{(2,2)}$ & $\overline{\sigma}$ & $\overline{\lambda}^{(1)}$ & $\widetilde \eta$ \\ $\phi^{(1,2)}$ & $C$ & $\overline{\lambda}^{(2)}$ & $\eta$ \\ $\phi^{(1,2)-}_{mn}$ & $B^-$ & $\overline{\lambda}^{(1)-}_{mn}$ & $\widetilde{\chi}^-$ \\ && $\overline{\lambda}^{(2)-}_{mn}$ & $\chi^-$ \\ && $\overline{\epsilon}^{(1)}$ & $v$ \\ && $\epsilon^{(1)}_m$ & $vV^\flat_1$ \\ && $\overline{\epsilon}^{(2)}$ & $u$ \\ && $\epsilon^{(2)}_m$ & $uV^\flat_2$ \\ \bottomrule \end{tabular} \caption{Vafa-Witten coordinate and differential-form notation for the bosonic fields (left) and fermionic fields and supersymmetry parameters (right).} \label{tab:VW form dictionary} \end{table}
If we now set all twisted components of $T, \overline T$ to zero except for the self-dual two-forms $C^{(1,1)+}$ and $C^{(2,2)+}$, all variations of components of the gravitino vanish identically except for the two-forms
\begin{subequations}
\begin{align}
    \delta \psi^{(1)} = \frac12v(\dd V^\flat_1)^+ + vC^{(1,1)+} \, , \\
    \delta \psi^{(2)} = \frac12u(\dd V^\flat_2)^+ + uC^{(2,2)+} \, .
\end{align}
\end{subequations}
These variations vanish provided that we turn on the following backgrounds
\begin{align}
    C^{(1,1)+} = -\frac12 (\dd V^\flat_1)^+ \, ; \quad C^{(2,2)+} = -\frac12 (\dd V^\flat_2)^+ \, .
\end{align}
In this background, some components of the $E, \overline E$ fields are required to be turn on for the variation of the gaugino to vanish. If we turn off all twisted components of $E, \overline E$ except for the anti-self-dual two-forms $E^{(1,1)-}$ and $E^{(2,2)-}$, all variations of the gaugino vanish except for the one-forms
\begin{subequations}
\begin{align}
    \delta \Lambda^{(1)} =& 4v\left(-\iota_{V_1} E^{(1,1)-} + \iota_{V_1} C^{(2,2)+} \right) \, ,\\
    \delta \Lambda^{(2)} =& 4u \left(-\iota_{V_2} E^{(2,2)-} + \iota_{V_2} C^{(1,1)+} \right) \, .
\end{align}
\end{subequations}
Upon setting
\begin{align}
    E^{(1,1)-}=\frac12 (\dd V^\flat_2)^- \, ; \quad E^{(2,2)-}=\frac12 (\dd V^\flat_1)^- \, ,
\end{align}
the equations rearrange as
\begin{subequations}
\begin{align}\label{eq: twisted +chir gaugino VW}
    \delta \Lambda^{(1)}=& -2v\iota_{V_1} \big( (\dd V^\flat_2)^- + (\dd V^\flat_2)^+ \big) = -2v \iota_{V_1} \dd V^\flat_2 \nonumber \\
    =& -2v\mathcal{L}_{V_1} V^\flat_2 + 2v\dd \big( \iota_{V_1} V^\flat_2 \big) \, ,\\
    \delta \Lambda^{(2)}=& -2u\mathcal{L}_{V_2} V^\flat_1 + 2u\dd \big( \iota_{V_2} V^\flat_1 \big) \, ,
\end{align}
\end{subequations}
which vanish if we consider commuting Killing vectors with constant inner product. In summary, the $\Omega$-background of the VW twist can be realised by
\begin{align}\label{eq: vw omega background}
    \begin{cases}
        C^{(1,1)+} = -\frac12 (\dd V^\flat_1)^+ \\
        E^{(1,1)-}=\frac12 (\dd V^\flat_2)^- \\
        \overline \epsilon^{(1)} = v \\
        V_1 \, \, \text{a KV} 
    \end{cases} \, , \quad  \begin{cases}
        C^{(2,2)+} = -\frac12 (\dd V^\flat_2)^+ \\
        E^{(2,2)-}=\frac12 (\dd V^\flat_1)^-\\
        \overline \epsilon^{(2)} = u \\
        V_2 \, \, \text{a KV} 
    \end{cases} \, ,
\end{align}
with $V_1,V_2$ such that
\begin{align}\label{eq: VW cond}
    [V_1, V_2]=0 \, , \quad \dd (\iota_{V_1} V^\flat_2)=0 \, ,
\end{align}
and all other background fields vanishing. Numerically verified solution of the dilatino is
\begin{align}
    D^{ij}{}_{kl} = -\frac{R}{6} \left( \begin{pmatrix}
        \iu \varepsilon&0\\0&\iu \varepsilon
    \end{pmatrix}^{ij} \otimes \begin{pmatrix}
        \iu \varepsilon&0\\0&\iu \varepsilon
    \end{pmatrix}_{kl}  + \varepsilon^{ij}{}_{kl} - (\iu \varepsilon \otimes \mathbb{1}_{2\times2})^{ij} \otimes (\iu \varepsilon \otimes \mathbb{1}_{2\times2})_{kl}  \right) \, ,
\end{align}
where $R$ is the Ricci scalar and $\varepsilon=\varepsilon_{\alpha \beta}$ is the matrix used for lowering $SU(2)$ indices.

\paragraph{Closure of the algebra}
We suppose that $V_1,V_2$ are Killing instead of conformal Killing. The transformations of the fields in this notation (see table \ref{tab:VW form dictionary}) are
\begin{subequations} \begin{align} \delta A ={}& -g\Big( 2vV^\flat_1\,\widetilde{\eta} -8v\,\iota_{V_1}\widetilde{\chi}^- +2uV^\flat_2\,\eta -8u\,\iota_{V_2}\chi^- +2v\,\widetilde{\psi} +2u\,\psi \Big), \\[2mm]
\delta\widetilde{\psi} ={}& \frac{2v}{g}\,\iota_{V_1}F^+ -2vD\sigma -2uDC +8uD^\ast B^- \nonumber\\ &\quad +2v\overline{\sigma}\, \iota_{V_1}\bigl(\dd(V^\flat_1)\bigr)^+ +2v\sigma\, \iota_{V_1}\bigl(\dd(V^\flat_2)\bigr)^+ \nonumber\\ &\quad +2vg[\sigma,\overline{\sigma}]\,V^\flat_1 -4ug[\sigma,C]\,V^\flat_2 \nonumber\\ &\quad +16vg\,\iota_{V_1}[C,B^-] -16ug\,\iota_{V_2}[\sigma,B^-] -32vg\,\mathcal N(B^-,V^\flat_1), \\[2mm]
\delta\psi ={}& \frac{2u}{g}\,\iota_{V_2}F^+ -2uD\overline{\sigma} -2vDC -8vD^\ast B^- \nonumber\\ &\quad +2u\sigma\, \iota_{V_2}\bigl(\dd(V^\flat_2)\bigr)^+ +2u\overline{\sigma}\, \iota_{V_2}\bigl(\dd(V^\flat_1)\bigr)^+ \nonumber\\ &\quad -4vg[\overline{\sigma},C]\,V^\flat_1 -2ug[\sigma,\overline{\sigma}]\,V^\flat_2 \nonumber\\ &\quad +16vg\,\iota_{V_1}[\overline{\sigma},B^-] -16ug\,\iota_{V_2}[C,B^-] -32ug\,\mathcal N(B^-,V^\flat_2), \\[2mm]
\delta\widetilde{\eta} ={}& -2v\,\iota_{V_1}D\overline{\sigma} +2u\,\iota_{V_2}DC +8u\,\iota_{V_2}D^\ast B^- \nonumber\\ &\quad +2gv[\overline{\sigma},\sigma] +4gu[\overline{\sigma},C] \, , \\[2mm] \delta\widetilde{\chi}^- ={}& -\frac{v}{2g}F^- -v\bigl(D\overline{\sigma}\wedge V^\flat_1\bigr)^- +u\bigl(DC\wedge V^\flat_2\bigr)^- \nonumber\\ &\quad +2u\mathcal P(B^-,V^\flat_2) -\frac{v}{2}\overline{\sigma}\, \bigl(\dd V^\flat_1\bigr)^- -\frac{v}{2}\sigma\, \bigl(\dd V^\flat_2\bigr)^- \nonumber\\ &\quad +u\mathcal Q(B^-,V^\flat_2) +4gv[C,B^-] +4gu[\overline{\sigma},B^-] +4gv\,\mathcal R(B^-), \\[2mm]
\delta\eta ={}& -2u\,\iota_{V_2}D\sigma +2v\,\iota_{V_1}DC -8v\,\iota_{V_1}D^\ast B^- \nonumber\\ &\quad +4gv[\sigma,C] -2gu[\overline{\sigma},\sigma] \, , \\[2mm]
\delta\chi^- ={}& -\frac{u}{2g}F^- +v\bigl(DC\wedge V^\flat_1\bigr)^- -u\bigl(D\sigma\wedge V^\flat_2\bigr)^- \nonumber\\ &\quad -2v\mathcal P(B^-,V^\flat_1) -\frac{u}{2}\overline{\sigma}\, \bigl(\dd V^\flat_1 \bigr)^- \nonumber\\ &\quad -v\mathcal Q(B^-,V^\flat_1) -4gv[\sigma,B^-] -4gu[C,B^-] +4gu\,\mathcal R(B^-) \nonumber\\ &\quad -\frac{u}{2}\sigma\, \bigl(\dd V^\flat_2\bigr)^-, \\[2mm]
\delta\sigma ={}& 2v\,\iota_{V_1}\widetilde{\psi} +2u\,\eta, \\ \delta\overline{\sigma} ={}& 2u\,\iota_{V_2}\psi +2v\,\widetilde{\eta}, \\ \delta C ={}& -u\,\widetilde{\eta} -v\,\eta +u\iota_{V_2}\widetilde{\psi} +v\iota_{V_1}\psi, \\ \delta B^- ={}& \frac{u}{2} \bigl(V^\flat_2\wedge\widetilde{\psi}\bigr)^- -\frac{v}{2} \bigl(V^\flat_1\wedge\psi\bigr)^- -u\,\widetilde{\chi}^- +v\,\chi^-. \end{align}
\end{subequations}
The contractions and auxiliary tensors are defined by
\begin{subequations}
\begin{align}
\left\langle X,Y\right\rangle :={}& \frac{1}{2}X_{mn}Y^{mn}, \\ \mathcal N(B^-,V)_m :={}& [B^-_{ms},B^{-rs}]V_r, \\ \mathcal P(B^-,V)_{mn} :={}& D_rB^-_{mn}V^r -D_mB^-_{nr}V^r +D_nB^-_{mr}V^r \nonumber\\ &\quad +D_rB^-_{n}{}^{r}V_m -D_rB^-_{m}{}^{r}V_n, \\ \mathcal Q(B^-,V)_{mn} :={}& (\dd V)^-_{mr}B^{-r}{}_{n} -(\dd V)^-_{nr}B^{-r}{}_{m} \nonumber\\ &\quad +B^-_{mn}\nabla_rV^r -B^-_{nr}\nabla_mV^r +B^-_{mr}\nabla_nV^r \nonumber\\ &\quad +B^-_{nr}\nabla^rV_m -B^-_{mr}\nabla^rV_n, \\ \mathcal R(B^-)_{mn} :={}& 2[B^-_{mr},B^-_{n}{}^{r}] \, , \\  (D^\ast B^-)_m :={}& -D^nB^-_{nm}, \qquad (\iota_VX)_m:=V^nX_{nm}. \end{align} \end{subequations}
Using the additional conditions \eqref{eq: VW cond} (and the identity \eqref{eq: myst identities}) we get
\begin{align} \delta^2 A ={}& -4\,\iota_{\,v^2V_1+u^2V_2}F -4D\varphi \, ,
\end{align}
where
\begin{align}\label{eq: VW omega gauge param}
\varphi ={}& -g\Big[ \bigl(v^2+u^2\iota_{V_2} V_2^\flat\bigr)\sigma +\bigl(u^2+v^2\iota_{V_1}V_1^\flat\bigr)\overline{\sigma} +2uv\bigl(1-\iota_{V_1}V^\flat_2\bigr)C -8uv\,\iota_{V_1}\iota_{V_2}B^-\Big] \, .
\end{align}
From the twisted matter equations in this background one can check that considering the pure twist with only one of the two parameters turned on, i.e. $V_1,V_2,v=0$ one recovers the BPS loci
\begin{subequations}
\begin{align}
\delta \chi^-&= -\frac{u}{2g} F^- -4gu[C,B^-] + 4g u \mathcal{R}(B^-) \, , \\
    \delta \widetilde{\psi} &= -2u DC + 8 u D^\ast B^- \, .
\end{align}
\end{subequations}
Up to the field redefinitions 
\begin{equation}
B^-\mapsto -\frac{\iu}{8g} B^- \, , \quad C \mapsto \frac{\iu}{2g} C \, , \quad \chi^- \mapsto -\frac{u}{2g} \chi^- \, , \quad \widetilde{\psi} \mapsto - \frac{\iu u}{g} \widetilde{\psi}
\end{equation}
these match the Vafa-Witten equations \cite{Vafa:1994tf}.\footnote{Notice that there appears to be a typo in eq. 2.56 of \cite{Vafa:1994tf}: the contraction of the covariant derivative with the $B$ field should be on the second index for it to be consistent with eq. 2.57. Compare for instance with eq. 3.3 in \cite{Manschot:2023rdh}.} The difference in self-duality is due to opposite Clifford algebra conventions.

\subsubsection{Geometric Langlands $\Omega$-background}\label{sec: GL omega} 
\begin{table}[t]
\centering
\begin{tabular}{@{}cc|cc@{}}
\toprule Coordinate & Form & Coordinate & Form \\ \midrule  $A_m$ &  $A$ & $\lambda^{(1)}_m$ & $\widetilde{\psi}$ \\
$\phi^{(1,1)}$ & $\sigma$&$\overline \lambda^{(1)}$ & $\widetilde{\eta}$\\
$\phi^{(2,2)}$ & $\overline \sigma$&$\overline \lambda^{(1)-}_{mn}$ & $\chi^-$\\
$\phi^{(1,2)}_m$ & $\phi$&$\lambda^{(2)}$ & $\eta$ \\
&&$\lambda^{(2)+}_{mn}$ & $\chi^+$\\
&&$\overline{\lambda}^{(2)}_m$ &$\psi$\\
&&$\overline \epsilon^{(1)}$ & $v$\\
&&$\epsilon^{(1)}_m$ & $v V^\flat_1$\\
&&$\epsilon^{(2)}$ & $u$\\
&&$\overline{\epsilon}^{(2)}_m$ & $u V^\flat_2$\\
\bottomrule
\end{tabular}
\caption{Geometric Langlands coordinate and form notation for the bosonic (on the left) and fermionic (on the right) fields of the twisted components of the matter multiplet and supersymmetry parameter.} \label{tab: GL dictionary}
\end{table}
Let us set to zero all components of $T,\overline T$ except for the self-dual two-form $C^{(1,1)+}$ and the anti-self-dual two-form $\overline C^{(2,2)-}$. All but the following variations of the components of the gravitino vanish:
\begin{align}
\delta\psi^{(1)} ={}&\frac12 v(\dd V^\flat_1)^+ +vC^{(1,1)+}\,, \\ \delta\overline{\psi}^{(2)} ={}&\frac12 u(\dd V^\flat_2)^-+u\overline C^{(2,2)-}\,.
\end{align}
These equations can therefore be solved by considering the following background fields: \begin{align}
C^{(1,1)+} =-\frac12(\dd V^\flat_1)^+\,; \qquad \overline C^{(2,2)-} =-\frac12(\dd V^\flat_2)^-\,. \end{align}
To find a solution to the gaugino equations, we turn off all components in $E,\overline E$ except for the self-dual two-form $E^{(2,2)+}$ and the anti-self-dual two-form $\overline E^{(1,1)-}$. With these assumptions, the only non-trivial variations of components of the gaugino are the two-forms \begin{subequations} \begin{align} \delta\Lambda^{(2)+} ={}&u\left(E^{(2,2)+}-C^{(1,1)+}\right)\,, \\ \delta\overline{\Lambda}^{(1)-} ={}&v\left(\overline E^{(1,1)-} -\overline C^{(2,2)-}\right)\,, \end{align} \end{subequations} and can be solved by setting \begin{align} E^{(2,2)+} =-\frac12(\dd V^\flat_1)^+\,; \qquad \overline E^{(1,1)-} =-\frac12(\dd V^\flat_2)^-\,. \end{align} In summary, the $\Omega$-background for the GL twist is realised by the following profiles: 
\begin{align} \begin{cases} C^{(1,1)+} =-\dfrac{1}{2}(\dd V^\flat_1)^+\,, \\[1ex] E^{(2,2)+} =-\dfrac{1}{2}(\dd V^\flat_1)^+\,, \\[1ex] \overline{\epsilon}^{(1)}=v\,, \\ V_1\ \text{a CKV}\,, \end{cases} \qquad \begin{cases} \overline C^{(2,2)-} =-\dfrac{1}{2}(\dd V^\flat_2)^-\,, \\[1ex] \overline E^{(1,1)-} =-\dfrac{1}{2}(\dd V^\flat_2)^-\,, \\[1ex] \epsilon^{(2)}=u\,, \\ V_2\ \text{a CKV}\,. \end{cases} \end{align}
Here $u$ and $v$ are non-zero constants, while all other background fields are vanishing. 
Importantly, we have to impose for the vector fields to be commuting
\begin{align}
    [V_1,V_2]=0 \, .
\end{align}
This condition is needed for the algebra to close correctly on the scalars. We also verified numerically that there is an obstruction to solving the dilatino equation if one considers non-commuting vector fields.
The numerically verified solution of the dilatino is
\begin{align}
D^{ij}{}_{kl} =-\frac{R}{6}\Bigg( &\begin{pmatrix} \iu \varepsilon&0\\ 0&\iu \varepsilon \end{pmatrix}^{ij} \otimes \begin{pmatrix} \iu \varepsilon&0\\ 0&\iu \varepsilon \end{pmatrix}_{kl} +\varepsilon^{ij}{}_{kl} - \begin{pmatrix} 0&\iu \varepsilon\\ \iu \varepsilon&0 \end{pmatrix}^{ij} \otimes \begin{pmatrix} 0&\iu \varepsilon\\ \iu \varepsilon&0 \end{pmatrix}_{kl} \nonumber\\ &+ 2\,R_{ab}V^a_1 V^b_2 \begin{pmatrix} \iu \varepsilon&0\\ 0&0 \end{pmatrix}^{ij} \otimes \begin{pmatrix} 0&0\\ 0&\iu \varepsilon \end{pmatrix}_{kl} \Bigg)\,,
\end{align}
where $R$ is the Ricci scalar, $R_{ab}$ the Ricci tensor and $\varepsilon=\varepsilon_{\alpha \beta}$ is the matrix used for lowering $SU(2)$ indices.

\paragraph{Closure of the algebra} We suppose that the vectors are Killing as opposed to conformal Killing. In the new notation, the relevant matter transformations are (see table \ref{tab: GL dictionary})
\begin{subequations}\label{eq: matter GL omega}
\begin{align}
\delta A ={}& g\Bigl( -2v\widetilde{\eta}V^\flat_1 +2u\eta V^\flat_2 +2u\psi -2v\widetilde{\psi} -8u\iota_{V_2}\chi^+ +8v\iota_{V_1}\chi^- \Bigr)\,, \\[1ex] \delta\widetilde{\psi} ={}& \frac{2v}{g}\iota_{V_1}F^+ -2vD\sigma -2uD\bigl(\iota_{V_2}\phi\bigr) -2uD_{V_2}\phi -2uV^\flat_2D^\ast\phi \nonumber\\ &\quad -2u\ast\bigl(V^\flat_2\wedge D\phi\bigr) +u\ast\bigl(\phi\wedge\dd V^\flat_2\bigr) \nonumber\\ &\quad -2v\overline{\sigma}\, \iota_{V_1}(\dd V^\flat_2)^- +2u\iota_{\phi^\sharp}(\dd V^\flat_2)^- +2v\overline{\sigma}\, \iota_{V_1}(\dd V^\flat_1)^+ \nonumber\\ &\quad -4gu[\sigma,\phi] +2vg[\sigma,\overline{\sigma}]V^\flat_1 -8vg\iota_{V_1}(\phi\wedge\phi)^-\,, \\[1ex] \delta\eta ={}& -2u\iota_{V_2}D\overline{\sigma} -2vD^\ast\phi -2gu[\sigma,\overline{\sigma}] +4vg\bigl[\overline{\sigma},\iota_{V_1}\phi\bigr]\,, \\[1ex] \delta\chi^+ ={}& -\frac{u}{2g}F^+ -u\bigl(D\overline{\sigma}\wedge V^\flat_2\bigr)^+ +v(D\phi)^+ -\frac{u}{2}\overline{\sigma}(\dd V^\flat_1)^+ \nonumber\\ &\quad -2gu(\phi\wedge\phi)^+ -2vg\bigl([\overline{\sigma},\phi]\wedge V^\flat_1\bigr)^+ -\frac{u}{2}\overline{\sigma}(\dd V^\flat_2)^+\,, \\[1ex] \delta\widetilde{\eta} ={}& -2uD^\ast\phi -2v\iota_{V_1}D\overline{\sigma} +2gv[\overline{\sigma},\sigma] +4ug\bigl[\overline{\sigma},\iota_{V_2}\phi\bigr]\,, \\[1ex] \delta\chi^- ={}& -\frac{v}{2g}F^- +u(D\phi)^- -v\bigl(D\overline{\sigma}\wedge V^\flat_1\bigr)^- -\frac{v}{2}\overline{\sigma}(\dd V^\flat_2)^- \nonumber\\ &\quad -2ug\bigl([\overline{\sigma},\phi]\wedge V^\flat_2\bigr)^- -2gv(\phi\wedge\phi)^- -\frac{v}{2}\overline{\sigma}(\dd V^\flat_1)^-\,, \\[1ex] \delta\psi ={}& \frac{2u}{g}\iota_{V_2}F^- -2uD\sigma -2vD\bigl(\iota_{V_1}\phi\bigr) -2vD_{V_1}\phi -2vV^\flat_1D^\ast\phi \nonumber\\ &\quad +2v\ast\bigl(V^\flat_1\wedge D\phi\bigr) -v\ast\bigl(\phi\wedge\dd V^\flat_1\bigr) +2v\iota_{\phi^\sharp}(\dd V^\flat_1)^+ \nonumber\\ &\quad -2u\overline{\sigma}\, \iota_{V_2}(\dd V^\flat_1)^+ +2u\overline{\sigma}\, \iota_{V_2}(\dd V^\flat_2)^- \nonumber\\ &\quad -4gv[\sigma,\phi] -2ug[\overline{\sigma},\sigma]V^\flat_2 -8ug\iota_{V_2}(\phi\wedge\phi)^+\,, \\[1ex] \delta\sigma ={}& -2u\iota_{V_2}\psi +2v\iota_{V_1}\widetilde{\psi}\,, \\[1ex] \delta\overline{\sigma} ={}& -2u\eta +2v\widetilde{\eta}\,, \\[1ex] \delta\phi ={}& -v\eta V^\flat_1 +u\widetilde{\eta}V^\flat_2 +v\psi -u\widetilde{\psi} -4v\iota_{V_1}\chi^+ +4u\iota_{V_2}\chi^-\,,
\end{align}
\end{subequations}
where
\begin{align}
    (D_{V} \phi)_m := V^n D_n \phi_m \, .
\end{align}
The algebra closes as
\begin{align}
\delta^2 A ={}& -4\,\iota_{\,v^2V_1-u^2V_2}F -4D\varphi\,,
\end{align}
where
\begin{equation}
    \begin{aligned}
        \varphi ={}& -g\Bigl[ (v^2-u^2)\sigma -2uv\,\iota_{V_1-V_2}\phi +\overline{\sigma}\Bigl( v^2\iota_{V_1}V_1^\flat -u^2\iota_{V_2}V_2^\flat \Bigr) \Bigr]\,.
    \end{aligned}
\end{equation}
Note that, for the algebra to close correctly, we have to consider commuting vector fields, otherwise a spurious term appears in the squared variation of $\sigma$
\begin{align}
        \delta^2 \sigma = -4 \iota_{v^2V_1 -u^2V_2} D \sigma + 4 [\varphi, \sigma] - 8uv \iota_{[V_1,V_2]}\phi \, .
    \end{align}
We suspect that such restriction on the vector fields of the GL $\Omega$-background is enforced by the dilatino equation, which has not been twisted in this paper. This is substantiated by numerical checks, confirming that while gravitino and gaugino can be solved for non-commuting vector fields, the dilatino equation admits no solution (not even a singular one).

From the set of equations \eqref{eq: matter GL omega}, in the limit of vanishing vector fields, we can recover the twisted equations for the pure twist appearing in the original work by Kapustin and Witten (equations 3.23-3.28 in \cite{Kapustin:2006pk}) using the following field rescaling
\begin{equation}
\begin{aligned}
    &\sigma \mapsto -\frac{\iu}{4g}\,\sigma \, , \quad \overline{\sigma} \mapsto -\frac{2\iu}{g}\,\overline{\sigma} \, , \quad \phi \mapsto \frac{\iu}{2g}\,\phi \, , \quad \psi \mapsto \frac{\iu}{2g}\,\psi \, , \quad \widetilde{\psi} \mapsto -\frac{1}{2g}\,\widetilde{\psi} \, ,  \\
    &\eta \mapsto -\frac{1}{g}\,\eta \, , \quad \widetilde{\eta} \mapsto -\frac{\iu}{g}\,\widetilde{\eta} \, , \quad \chi^+ \mapsto -\frac{1}{2g}\,\chi^+ \, , \quad \chi^- \mapsto -\frac{\iu}{2g}\,\chi^- \, , \quad v \mapsto \iu v \, .
\end{aligned}
\end{equation}
Notice also that KW use non-standard conventions for the gauge-covariant codifferential, whereby $D^\ast_{\text{KW}}=-D^\ast$. The relative factor of $\iu$ between the two supersymmetry parameters is a choice of relative normalisation of the two supercharges and cannot be avoided, while the factors of $\iu$ in the bosonic fields reflect the fact that our scalars are hermitian, whereas those of \cite{Kapustin:2006pk} are antihermitian.

\subsection{Background with non-constant gauge coupling}\label{sec: vw phi}
In the presence of the VW twist, we have had to impose a condition on the Killing vectors, in order to produce a supersymmetric $\Omega$-deformed background, of the form
\begin{equation}
\iota_{V_1} V^\flat_2=0\,.
\end{equation}
This constraint is of course symmetric in $V_{1,2}$. This type of constraint is relevant even for the $\Omega$-deformation on $\mathbb{R}^4$, but is not mentioned in previous work by Ito et al. which derive this deformation from a twisted dimensional reduction \cite{Ito:2011cr,Ito:2012hs,Ito:2013eva}. We believe this is due to the following effect: when the constraint is not satisfied, a spacetime varying coupling must be turned on in order to satisfy the supergravity equations. The presence of this varying coupling should be visible in the action for this background, in the form of terms which are not present in e.g. the standard two derivative action in \cite{deRoo:1984zyh} or \cite{Maxfield:2016lok}. Indeed, such higher derivative terms, which implicitly contain derivatives of the coupling as well as those of the dynamics fields, seem to be present in \cite{Ito:2013eva}.    

We now show how to construct the $\Omega$-deformation of the Vafa-Witten twisted theory when the constraint is not satisfied. In addition to the background discussed in Section \ref{sec: vw omega}, we turn on a non-constant field $\Phi$ in the Weyl multiplet, which can be interpreted as a varying gauge coupling, in such a way that enables us to relax the constraint.

We make the following choices for the supergravity fields
\begin{align}
    \overline{\Phi}^{1}=0 \, ,
\end{align}
and we choose $\overline{\Phi}^2=1$, such that the constraint \eqref{eq: constraint on Phi} forces $\Phi_2=1$ while $\Phi_1$ is unconstrained. The gauge field $a_\mu$ still vanishes for this background and so do all the terms in the supergravity equations involving $\overline \Phi^\alpha$, while allowing the term proportional to $\Phi_\alpha$ in $\delta \Lambda_i$ to be non trivial. Note that this choice of supergravity scalar field configurations is not in the gauge where one can simply identify the Yang-Mills coupling in the matter equations. 

We can now reconsider the same background fields considered in \eqref{eq: vw omega background}. The gravitino equations are unaffected and can be solved in the same way. On the other hand, the twisting of the term $2 \varepsilon^{\alpha \beta} \Phi_\alpha \mathcal{D}_\mu \Phi_\beta \gamma^\mu \overline \xi_i$ produces two additional terms in the non trivial gaugino equations
\begin{subequations}
\begin{align}
    \delta \Lambda^{(1)}=&  -2 v\dd \Phi_1-2v\mathcal{L}_{V_1} V^\flat_2 + 2v\dd \big( \iota_{V_1} V^\flat_2 \big) \, ,\\
    \delta \Lambda^{(2)}=& -2u \dd \Phi_1-2u\mathcal{L}_{V_2} V^\flat_1 + 2u\dd \big( \iota_{V_2} V^\flat_1 \big) \, ,
\end{align}
\end{subequations}
which prompts us to set
\begin{align}
    \Phi_1 = \iota_{V_1} V_2^\flat + c \, ,
\end{align}
with $c$ an arbitrary complex constant, which is the analogue of the ordinary Yang-Mills coupling. This allows for a relaxation of the constraint on the vector fields. This version of the $\Omega$-background is realised by
\begin{align}
    \begin{cases}
        C^{(1,1)+} = -\frac12 (\dd V_1^\flat)^+ \\
        E^{(1,1)-}=\frac12 (\dd V_2^\flat)^- \\
        \overline \epsilon^{(1)} = v \\
        V_1 \, \, \text{a KV} 
    \end{cases} \, , \quad  \begin{cases}
        C^{(2,2)+} = -\frac12 (\dd V_2^\flat)^+ \\
        E^{(2,2)-}_{mn}=\frac12 (\dd V_1^\flat)^-_{mn}\\
        \overline \epsilon^{(2)} = u \\
        V_2 \, \, \text{a KV} 
    \end{cases} \, ,
\end{align}
\begin{align}
    [V_1, V_2]=0 \, , \quad \Phi_1= \iota_{V_1} V_2^\flat +c \, ,
\end{align}
and all other background fields set to zero. The terms proportional to $\Phi_{\alpha}, \overline{\Phi}^{\alpha}$ in the dilatino equations vanish identically. The dilatino is thus solved by same configuration for $D^{ij}{}_{kl}$ as the one found in the constant coupling case.

$\mathcal{N}=4$ SYM with spacetime dependent couplings has previosuly appeared in similar setting: in F-theory constructions \cite{Harvey:2007ab,Martucci:2014ema} and in Janus constructions \cite{DHoker:2006qeo, Gaiotto:2008sd, Kim:2008dj, Kim:2009wv, Chen:2008tu}. Specifically, \cite{Martucci:2014ema} and its follow-ups consider a twist of $\mathcal{N}=4$ super Yang-Mills on a K\"{a}hler manifold with spacetime dependent $\tau$. However, as far as we know, our background is the first to incorporate the $\Omega$-deformation.

\paragraph{Closure of the algebra} In the background described in this section, we have adopted a different gauge fixing compared to the one used to write the matter equations in terms of the Yang-Mills coupling $g$ \eqref{eq: matter equations euclidean}. In order to check the closure of the algebra, one then has to reinstate the dependence on $\Phi_{\alpha}, \overline{\Phi}^\alpha$, which can be found in \cite{deRoo:1984zyh, deRoo:1985np}. We have checked that the algebra still closes in the same way as the one in Section \ref{sec: vw omega} with the gauge parameter
\begin{equation}
\begin{aligned}
    \varphi = - \Big[ (v^2(1+c+\iota_{V_1} V_2^\flat) &+ u^2 \iota_{V_2} V_2^\flat) \sigma + (u^2 (1+c+\iota_{V_1} V_2^\flat) +v^2 \iota_{V_1} V_1^\flat) \overline \sigma\\ &+2uv(1+c)C- 8uv \iota_{V_1} \iota_{V_2} B^-  \Big] \, .
\end{aligned}
\end{equation}

\subsection{Fuzzy sphere background}\label{sec: fuzzy sphere background}
In this section we describe a fuzzy sphere background \cite{Kato:2011yh}. This background can be obtained by turning on only the scalar components $\overline \epsilon^{(1)}, \overline \epsilon^{(2)}$ of the supersymmetry parameters in the Vafa-Witten twist, which we set to $w,u$ respectively. As for the Weyl-multiplet fields, we only turn on the four-form $\overline K^{(1,2)-}$ and two scalars $\overline E^{(1,2)}, E^{(1,2)}$. The four-form is anti-self-dual both in the first and in the second couple of indices and we set it to the following background value
\begin{align}
    \overline K^{(1,2)-}= t \, \overline \eta^a \otimes \overline{\eta}_a \, ,
\end{align}
where $\overline \eta^a$ are the anti-self-dual 't Hooft symbols \eqref{eq: 't Hooft symb}. With this choice of background fields, all gravitino variations vanish identically, while the scalar gaugino variations are
\begin{subequations}
\begin{align}
    \delta \overline \rho^{(1)} =& - u\overline E^{(1,2)} - 4u\overline{\mathcal{K}}^- \, ,\\
    \delta \overline \rho^{(2)} =&{}\  w \overline E^{(1,2)} + 4w\overline{\mathcal{K}}^- \, ,
\end{align}
\end{subequations}
where $\overline{\mathcal{K}}^-= \overline K^{(1,2)-}_{ab}{}^{ab}= \text{tr} \overline{K}^{(1,2)-}=12t$.
To show that the variations of all other components of the gaugino vanish, one has to use \eqref{eq: 't Hooft symb id}. Variations of the scalar gauginos vanish if we set $\overline{E}^{(1,2)}=-4\,\text{tr} \overline{K}^{(1,2)-}$. In summary, this background is realised by
\begin{align}
    \begin{cases}
        \overline K^{(1,2)-} = t \, \overline \eta^a \otimes \overline \eta_a \\
        \overline E^{(1,2)}= -48t  \\
        E^{(1,2)} = s \\
        \overline \epsilon^{(1)} = w \\ \overline \epsilon^{(2)} = u 
    \end{cases} \, .
\end{align}
We have set the extra field $E^{(1,2)}$ that we have turned on to a parameter $s$ for simplicity. At this stage this is simply an extra modulus of this background. In Appendix \ref{sec: fuzzy matching} we explicitly show that this background in flat space is the same as the one described in \cite{Kato:2011yh} by solving the dilatino equation on $\mathbb{R}^4$.

Since there is a leftover $SU(2)$ symmetry that rotates the $w,u$ parameters, these can be scaled away. Then it looks like this background depends on two deformation parameters $s,t$. An explicit calculation carried out in Appendix \ref{sec: fuzzy matching}, shows that reproducing the action in eq 2.35 of \cite{Kato:2011yh} requires $s \propto t \propto M$. To retain the full family of two-parameter solutions we may instead set $s \propto M \, , \, t \propto M-\widetilde m$ and, in addition, add a Q-exact term such as $\widetilde m \delta_{\overline{\epsilon}^{(2)}}(v \chi)$ to our action.

Note that this background is also available in the DW twist. It is essentially a deformation of the $\mathcal{N}=2$ theory extended to the VW twist. Such deformations are also known to exist for $\mathcal{N}=1$ theories. We have presented it only to emphasise that it fits naturally within the framework of rigid background supergravity and that this is the right context in which to generalise it. 

\paragraph{Closure of the algebra}
\begin{table}[t]
\centering \begin{tabular}{@{}cc|cc@{}} \toprule Coordinate & Form & Coordinate & Form \\
\midrule
$\phi^{(1,1)}$ & $\phi$ & $\lambda^{(1)}_m$ & $C$\\
$\phi^{(2,2)}$ & $\overline \phi$ & $\lambda^{(2)}_m$ & $\psi$ \\
$\phi^{(1,2)}$ & $v$ & $\overline \lambda^{(1)}$ & $ \lambda$\\
$\phi^{(1,2)-}_{mn}$ & $v^-$ & $\overline \lambda^{(2)}$ & $\chi$ \\
&& $\overline \lambda^{(1)-}_{mn}$ & $\lambda^-$\\
&& $\overline \lambda^{(2)-}_{mn}$ & $\chi^-$\\
&& $\overline \epsilon^{(1)}$ & $w$\\
&& $\overline{\epsilon}^{(2)}$ & $u$ \\
\bottomrule
\end{tabular} \caption{Dictionary between coordinate and form notation of the twisted matter fields adapted to the fuzzy sphere background. Bosonic fields are in the left column and fermionic fields and supersymmetry parameters in the right one.} \label{tab: fuzzy dictionary} \end{table}
We can write the matter equations in form notation using the dictionary in table \ref{tab: fuzzy dictionary}. The relevant equations are
\begin{subequations}
    \begin{align}
        \delta A ={}& -2g\left( w\,C +u\,\psi \right)\, , \\
        \delta C ={}& -2w\,D \phi -2u\,D v +8u\,D^\ast v^{-}\, , \displaybreak[0]\\ \delta \psi ={}& -2u\,D \overline \phi -2w\,D v - 8w\,D^\ast v^{-}\, .
    \end{align}
\end{subequations}
On the gauge field the algebra closes up to gauge transformations
\begin{align}
    \delta^2 A= 4g D \left( w^2\, \phi +u^2 \,\overline \phi + 2uw \, v  \right) \, .
\end{align}
Since the two-form field $\overline T$ of the Weyl multiplet has mass dimension one, the resulting theory is a topologically VW twisted theory with a mass deformation.

The BPS locus given by the scalar fermions reads
\begin{subequations}
\begin{align}
\delta\lambda &= s\,w\,v + s\,u\,\overline\phi + 2g\,w\,[\overline\phi,\phi] + 4g\,u\,[\overline\phi,v], \\
\delta\chi &= -s\,w\,\phi - s\,u\,v + 4g\,w\,[\phi,v] - 2g\,u\,[\overline\phi,\phi] \, .
\end{align}
\end{subequations}
This shows that this background supports BPS configurations where one of the deformation parameters $w,u$ is turned off and the scalar fields $\phi,\overline \phi, v$ form a representation of the $SU(2)$ algebra. The space spanned by scalars is known as a fuzzy sphere: a quantised version of a smooth two-sphere. This is a classical vacuum solution of the theory.

\section{Discussion}

\begin{table}[t]\centering\small
\begin{tabular}{@{}llllll@{}}\toprule
Background & Weyl fields & Parameters & Conditions & $N_T$  \\ \midrule
\makecell[l]{$\Omega$-deformed DW twist  (\S\ref{sec: DW omega})} & $C^{(1,1)+}$ & \makecell[c]{$V_1$} & $V_1$ Killing & \makecell[c]{1} \\ \addlinespace
\makecell[l]{$\Omega$-deformed VW twist  (\S\ref{sec: vw omega})} & \makecell[l]{$C^{(1,1)+},C^{(2,2)+}$,\\$E^{(1,1)-},E^{(2,2)-}$} & \makecell[c]{$V_1,V_2$} & \makecell[l]{$V_1,V_2$ Killing,\\ $[V_1,V_2]=0$,\\ $\dd(V_1\!\cdot\! V_2)=0$} & \makecell[c]{2}  \\ \addlinespace
\makecell[l]{$\Omega$-deformed VW twist - varying $\Phi$  (\S\ref{sec: vw phi})} & \makecell[l]{as above,\\ $\Phi_1=\iota_{V_1}V_2^\flat+c$} & \makecell[c]{$V_1,V_2$} & \makecell[l]{$V_1,V_2$ Killing,\\ $[V_1,V_2]=0$} & \makecell[c]{2} \\ \addlinespace
\makecell[l]{$\Omega$-deformed GL twist  (\S\ref{sec: GL omega})} & \makecell[l]{$C^{(1,1)+},\overline C^{(2,2)-}$,\\$E^{(2,2)+},\overline E^{(1,1)-}$} & \makecell[c]{$V_1,V_2$} & \makecell[l]{$V_1,V_2$ Killing,\\ $[V_1,V_2]=0$} & \makecell[c]{2} \\ \addlinespace
\makecell[l]{Fuzzy sphere (\S\ref{sec: fuzzy sphere background})} & \makecell[l]{$\overline K^{(1,2)-},\overline E^{(1,2)}$,\\$E^{(1,2)}$} & \makecell[c]{$s,t$} &  & \makecell[c]{2}  \\
\bottomrule\end{tabular}
\caption{Summary of the supersymmetric backgrounds found in this paper. All other Weyl multiplet fields vanish, except $D^{ij}{}_{kl}$, which is fixed by the dilatino equation.}
\label{tab: summary backgrounds}
\end{table}

We have derived rigid supersymmetric backgrounds using the 4d $\mathcal{N}=4$ Weyl multiplet. The backgrounds generalise the $\Omega$-deformation, viewed as a non-topological deformation of the Donaldson-Witten twist, to the three inequivalent topological twists of  $\mathcal{N}=4$ SYM. In the process, we have derived the twisted supersymmetry equations for some of the Weyl multiplet fermions, and for the fields of  $\mathcal{N}=4$ SYM in a rigid bosonic background. We briefly discussed also a deformation leading to fuzzy sphere vacua. The backgrounds are summarised in table \ref{tab: summary backgrounds}. 

The backgrounds presented here are applicable to $\mathcal{N}=4$ gauge theory on any oriented Euclidean spin manifold. The backgrounds for the VW and GL twists can be consistently defined also on non-spin Euclidean four-manifolds. The backgrounds generalise previous work on the $\Omega$-deformation of $\mathcal{N}=4$ SYM in flat space presented in \cite{Ito:2011cr,Ito:2013eva} and the flat space fuzzy sphere background derived in \cite{Kato:2011yh}. While most of the backgrounds incorporate the standard gauge coupling constant of $\mathcal{N}=4$ SYM, we identified a specific subset of backgrounds which required a complex, and spacetime-dependent, coupling. Consideration of an observable defined by such a background as worthy of study is very much in the spirit of the backgrounds introduced in \cite{Festuccia:2011ws}.  

The generalised $\Omega$-backgrounds depend on a number of parameters which determine the specific Killing vectors used. We count parameters of the $\Omega$-deformation by considering the space of admissible vector fields on a fixed Euclidean manifold. Ostensibly, the action of the deformed theory depends separately on all of these parameters. It would be useful to determine whether this dependence of the action survives in the values of observables such as the partition function. This is not always the case. For instance, the supergravity fields associated with the original $\Omega$ background of the DW twist in $\mathbb{R}^4$ depend on only one linear combination of the commuting Killing vectors in the Cartan subalgebra of the $SO(4)$ rotation group. Moreover, some deformations may be exact with respect to conserved supersymmetries, which would imply that the values of all observables in the twisted theory are independent of the values of the deformation. 

In comparing with the results of \cite{Ito:2011cr,Ito:2013eva}, we have identified two mismatches in terms of parameter counting. Our GL backgrounds depend on twice as many parameters, which we attribute to a more general procedure for identifying deformations. Our most general VW background depends on the same number of parameters as in \cite{Ito:2013eva}, but requires a spacetime-dependent coupling when a certain constraint is not satisfied. The number of free parameters in the presence of the constraint is halved. We have argued that these subtleties are implicitly taken into account also in \cite{Ito:2013eva}.

The parameter counting for the fuzzy sphere solution is somewhat difficult to compare with \cite{Kato:2011yh}. We believe that our background corresponds to one out of two possible deformation parameters presented in \cite{Kato:2011yh}, and that the additional parameter deforms the theory in a $\delta$-exact way and hence has no effect on the value of observables in the twisted theory. There are of course many such exact deformations. 

The original $\Omega$-deformation was very successfully used to study various aspects of $\mathcal{N}=2$ gauge theories \cite{Nekrasov:2002qd,Nekrasov:2003rj,Nekrasov:2009rc,Nekrasov:2010ka}. It is our hope that the generalised backgrounds we introduced prove to be useful for the study of $\mathcal{N}=4$ SYM. The increased number of parameters entering our backgrounds make it plausible that they can be used to study the theory in ways which are not possible by simply regarding it as an $\mathcal{N}=2$ theory. 

We have taken here some modest steps towards a classification of supersymmetric backgrounds for $\mathcal{N}=4$ SYM, of the type discussed for $\mathcal{N}=1$ in \cite{Closset:2013vra} and for $\mathcal{N}=2$ in \cite{Klare:2013dka}. There are likely additional interesting backgrounds, which are not available with a lower amount of supersymmetry, and which support interesting observables as is for instance the case for the 3d $\mathcal{N}=4$ gauge theories studied in \cite{Dedushenko:2016jxl,Dedushenko:2018icp,Dedushenko:2017avn}. There also exist additional rigid backgrounds compatible only with the twists of $\mathcal{N}=4$ SYM, some of which take advantage of the non-perturbative S-duality group. We hope to report on these possibilities in the near future.  

It would be interesting to study our backgrounds as deformations of higher dimensional theories on spacetimes which contain a four manifold factor. Such setups have been used to generalise the original $\Omega$-deformation and the Nekrasov partition function. For instance, see recent work on computing refined Vafa-Witten invariants \cite{Khlaif:2026rnb}. 

\section*{Acknowledgments} This work was supported by the UKRI STFC consolidated grant ST/X000583/1, “New
Frontiers in Particle Physics, Cosmology and Gravity”. EM would like to acknowledge SISSA and the organisers of the conference \textit{Quivers, Symmetries and SCFTs 2026} for hospitality during the final stages of this project. The authors would like to thank Benjamin Suzzoni and Pietro Capuozzo for many interesting discussions and for collaboration on related projects. The authors would also like to thank Fri\dh rik Freyr Gautason for many useful comments. EM also thanks Sara Pasquetti for her encouragement.

\clearpage
\appendix

\section{Conventions}\label{sec: conventions}
\subsection{Spinor and Clifford algebra conventions}\label{sec: clifford}

We denote positive (or left-handed $SU(2)_l$) chirality indices and negative (or right-handed $SU(2)_r$) chirality ones respectively with undotted/dotted Greek letters from the beginning of the alphabet. We contract undotted indices as $\searrow$ and dotted ones as $\nearrow$. We follow the Euclidean sigma matrices convention
\begin{align}\label{eq: Euclidean sigmas}
    (\sigma_{\mu})_{\alpha \dot \alpha} = ( \tau_{1},  \tau_{2}, \tau_{3}, - \iu \mathbb{1}_{2}) \, &; \quad \, (\overline \sigma_{\mu})^{\dot \alpha \alpha} = (\tau_{1}, \tau_{2},\tau_{3}, \iu \mathbb{1}_{2}) \, ; \nonumber \\
    (\sigma_{\mu \nu})_{\alpha}{}^{\beta} = \frac{1}{2}(\sigma_\mu \overline \sigma_\nu - \sigma_\nu \overline \sigma_\mu)_{\alpha}{}^{\beta}\, &; \quad (\overline \sigma_{\mu \nu})^{\dot \alpha}{}_{\dot \beta} = \frac{1}{2}(\overline \sigma_\mu \sigma_\nu - \overline \sigma_\nu \sigma_\mu)^{\dot \alpha}{}_{\dot \beta} \, ,
\end{align}
where $\tau_i$ are the Pauli matrices.
The gamma matrices are constructed as
\begin{align}
    \gamma_\mu = \begin{pmatrix}
        0 & \sigma_\mu \\
        \overline \sigma_\mu & 0
    \end{pmatrix} \, ; \quad \gamma_{\mu \nu} =\gamma_{[\mu}\gamma_{\nu]}=\begin{pmatrix}
        \sigma_{\mu \nu}&0\\0&\overline \sigma_{\mu \nu}
    \end{pmatrix} \, .
\end{align}
We define the following matrices
\begin{align}
    \varepsilon^{\alpha \beta} = \varepsilon^{\dot \alpha \dot \beta} = \begin{pmatrix}
        0&1\\-1&0
    \end{pmatrix} \, ; \quad \varepsilon_{\alpha \beta}=\varepsilon_{\dot \alpha \dot \beta} = \begin{pmatrix}
        0&-1\\1&0
    \end{pmatrix} \, ,
\end{align}
used to raise and lower $SU(2)$ indices as
\begin{align}
    \chi^{\beta} = \varepsilon^{\beta \alpha}\chi_{\alpha} \, &; \quad \overline{\chi}_{\dot \beta} = \varepsilon_{\dot \beta \dot \alpha} \overline{\chi}^{\dot \alpha} \, ;\\ \chi_{\beta} = \varepsilon_{\beta \alpha}\chi^{\alpha} \, &; \quad \overline{\chi}^{\dot \beta} = \varepsilon^{\dot \beta \dot \alpha} \overline{\chi}_{\dot \alpha} \, ;\\ \quad \chi^\alpha \psi_{\alpha} = \chi_\alpha \varepsilon^{\beta \alpha } \psi_\beta \, &; \quad  \overline \chi_{\dot \alpha} \overline \psi^{\dot \alpha}=\overline \chi^{\dot \alpha} \varepsilon_{\dot \beta \dot \alpha } \overline \psi^{\dot \beta} \, .
\end{align}
We have a charge conjugation matrix
\begin{align}
    C = \begin{pmatrix}
        \varepsilon^{\alpha \beta} &0\\
        0 & \varepsilon_{\dot \alpha \dot \beta}
    \end{pmatrix} \,,
\end{align}
satisfying
\begin{align}
    C^T = -C \, ; \quad C \gamma_a C^{-1} = + \gamma_a^T \, .
\end{align}
This choice ensures that the chiral components of the bilinears defined below are contracted according to the conventions stated above. It differs from the charge conjugation matrix of \cite{Maxfield:2016lok}, which satisfies $C\gamma_a C^{-1}=-\gamma_a^T$, by the sign of the negative-chirality block; the consequences for the matter equations are discussed below equation \eqref{eq: gauge conventions}.
In Euclidean signature the $\mathcal{N}=4$ supersymmetry parameters are independent Weyl spinors $\epsilon_{\alpha}{}^{i}$ and $\overline \epsilon^{\dot \alpha}{}_{i}$ transforming in the $\mathbf{4}$ and $\mathbf{\overline 4}$ of $SU(4)$. We split the $SU(4)$ index into the two blocks $i=\tilde i \oplus \hat i=(1,2)\oplus(3,4)$ and write
\begin{align}
    \epsilon_{\alpha}{}^{i} \longrightarrow  \begin{pmatrix}
        \epsilon_{\alpha}^{(1) \tilde i} \, & \, \epsilon_{\alpha}^{(2) \hat i}
    \end{pmatrix}  \, ; \quad \overline \epsilon^{\dot \alpha}{}_{i} \longrightarrow  \begin{pmatrix}
        \overline \epsilon^{\dot \alpha (1)}{}_{\tilde i} \, & \, \overline \epsilon^{\dot \alpha (2)}{}_{\hat i}
    \end{pmatrix} \, .
\end{align}
Here and below, the superscripts $(1)$ and $(2)$ refer to the first and second $SU(2)$ blocks, respectively.
Following the chiral notation for the $SU(4)$ $i$ indices \cite{Bergshoeff:1980is}, upper and lower indices transform in the defining and conjugate representations respectively.

Although the supergravity equations are chiral, it is convenient for us to write the equations in terms of projected Dirac spinors which ultimately depend only on one chirality. The projected Dirac spinors for the supersymmetry parameters are defined as
\begin{align}
    \xi^{i} = \begin{pmatrix}
        \epsilon_{\alpha}{}^{i} \\0
    \end{pmatrix} = \begin{pmatrix}
        \epsilon_{\alpha}^{(1) \tilde i} & \epsilon_{\alpha}^{(2) \hat i} \\0 & 0
    \end{pmatrix} \, ; \quad \overline \xi_{i} = \begin{pmatrix}
        0 \\
        \overline \epsilon^{\dot \alpha}{}_{i}
    \end{pmatrix} = \begin{pmatrix}
        0 & 0 \\
        \overline \epsilon^{\dot \alpha (1)}{}_{\tilde i} & \overline \epsilon^{\dot \alpha (2)}{}_{\hat i}
    \end{pmatrix} \, .
\end{align}
Similarly, the fermions of the Euclidean $\mathcal N=4$ vector multiplet are defined in terms of projected Dirac spinors as
\begin{align}
    \lambda^{i} = \begin{pmatrix}
        \lambda_{\alpha}{}^{i} \\0
    \end{pmatrix} \, ; \quad \overline \lambda_{i} = \begin{pmatrix}
        0 \\
        \overline \lambda^{\dot \alpha}{}_{i}
    \end{pmatrix} \, .
\end{align}
For any pair of barred or unbarred Dirac spinors $\chi$ and $\psi$, we define the bilinears as
\begin{align} \chi\gamma_{\mu_1\ldots\mu_k}\psi := \chi \bigl(C^T\gamma_{\mu_1\ldots\mu_k}\bigr) \psi \, ,
\end{align}
where
\begin{align}
    (C^T \gamma_{\mu_1,...,\mu_k})^T = - (-1)^{\frac{k(k-1)}{2}} C^T \gamma_{\mu_1,...,\mu_k} \, .
\end{align}
Therefore, the chiral components of a Dirac spinor bilinear are
\begin{align}
    \chi \psi := \chi^\alpha \psi_\alpha \, ; \quad \overline \chi \overline \psi := \overline \chi_{\dot \alpha} \overline \psi^{\dot \alpha} \, .
\end{align}
In our conventions, the supersymmetry parameters $\xi^i, \overline \xi_i$ are taken to be commuting, while dynamical fermions are Grassmann-odd. Consequently, 
\begin{align}
    \xi_1\gamma_{\mu_1,...,\mu_k} \xi_2 = + \xi_2 (C^T\gamma_{\mu_1,...,\mu_k})^T \xi_1 \, .
\end{align}
and the supersymmetry algebra is evaluated as $\{ \delta_{\xi_1},\delta_{\xi_2} \}$ rather than $[ \delta_{\xi_1},\delta_{\xi_2} ]$.

\subsection{Differential geometric and twisting conventions}\label{sec: diff geo}
Greek letters from the middle of the alphabet denote coordinate indices, while Latin letters from the middle of the alphabet denote frame indices. Upper indices label vector components, whereas lower indices label covector or differential-form components. If an object naturally has vector indices, using the musical isomorphism we shall denote its metric dual with $\flat$, whereas if an object naturally has covector indices we shall denote its metric dual with a $\sharp$. Thus, a vector field is written as $V = V^\mu \frac{\partial}{\partial x^\mu}$ and its metric-dual one-form is \begin{align} V^\flat = V^\flat_\mu \dd x^\mu \, , \qquad V^\flat_\mu = g_{\mu\nu}V^\nu \, , \end{align} while a one-form is written as $\omega = \omega_\mu \dd x^\mu$ and its metric-dual vector field is
\begin{align}
    \omega^\sharp = \omega^{\sharp \mu} \frac{\partial}{\partial x^\mu} \, , \quad \omega^{\sharp \mu} = g^{\mu \nu} \omega_\nu \, .
\end{align}
Symmetrisation and antisymmetrisation of indices involve a factor of a half
\begin{align}
    \nabla_{[m}V^\flat_{n]} := \frac12 \big(  \nabla_m V^\flat_n - \nabla_n V^\flat_m \big) \, ; \quad \nabla_{(m}V^\flat_{n)} := \frac12 \big(  \nabla_m V^\flat_n + \nabla_n V^\flat_m \big) \, .
\end{align}
The exterior derivative of a 1-form is defined as
\begin{align}
    (\dd V^\flat)_{mn} =  2\nabla_{[m}V^\flat_{n]} \, .
\end{align}
Forms are normalised as $\omega=\frac{1}{p!}\omega_{m_1\ldots m_p} \dd x^{m_1}\wedge\ldots\wedge \dd x^{m_p}$, so that
\begin{align}
(\alpha\wedge\beta)_{m_1\ldots m_{p+q}}=\frac{(p+q)!}{p!\,q!}\alpha_{[m_1\ldots m_p}\beta_{m_{p+1}\ldots m_{p+q}]} \, .
\end{align}
The Hodge dual in four dimensions is
\begin{align}
(\ast\omega)_{m_1\ldots m_{4-p}}=\frac{1}{p!}\,\omega^{n_1\ldots n_p}\varepsilon_{n_1\ldots n_p m_1\ldots m_{4-p}}\,,
\end{align}
which for a two-form reduces to $(\ast A)_{mn}=\frac12\varepsilon_{mnrs}A^{rs}$.
Self-dual and anti-self-dual projections of a two-form $A$ are defined as
\begin{align}
    A^{\pm}_{mn} = \frac{1}{2} \big(A_{mn} \pm (\ast A)_{mn} \big) \, ,
\end{align}
where the Levi-Civita symbol is such that
\begin{align}
    \varepsilon_{1234}=1 \, .
\end{align}
The product of a self-dual and an anti-self-dual two-form is symmetric in its free indices
\begin{align}\label{eq: myst identities}
    A^+_{m}{}^r B^-_{rn} = A^+_{n}{}^rB^-_{rm} \, .
\end{align}
In our conventions, $\sigma_{ab}$ is self-dual and $\overline \sigma_{ab}$ is anti-self-dual
\begin{align}
    \sigma_{ab} =\frac12 \varepsilon^{abcd} \sigma_{cd} \, ; \quad \overline \sigma_{ab} =-\frac12 \varepsilon^{abcd} \overline \sigma_{cd} \, .
\end{align}
We define the anti-self-dual 't Hooft symbols
\begin{align}\label{eq: 't Hooft symb}
    \overline\eta^1_{mn} = \left(
\begin{array}{cccc}
 0 & 0 & 0 & -1 \\
 0 & 0 & 1 & 0 \\
 0 & -1 & 0 & 0 \\
 1 & 0 & 0 & 0 \\
\end{array}
\right) \, , \quad \overline\eta^2_{mn}=\left(
\begin{array}{cccc}
 0 & 0 & -1 & 0 \\
 0 & 0 & 0 & -1 \\
 1 & 0 & 0 & 0 \\
 0 & 1 & 0 & 0 \\
\end{array}
\right) \, , \quad \overline \eta^3_{mn}= \left(
\begin{array}{cccc}
 0 & 1 & 0 & 0 \\
 -1 & 0 & 0 & 0 \\
 0 & 0 & 0 & -1 \\
 0 & 0 & 1 & 0 \\
\end{array} \right) \, ,
\end{align}
which satisfy
\begin{align}\label{eq: 't Hooft symb id}
    \overline \eta^a_{mr} \overline \eta_{a\, n}{}^r= 3 \delta_{mn} \, .
\end{align}
We denote $\mathcal{L}_{V}$, $\iota_{V}$ the Lie derivative and interior product with respect to the vector field $V$. We may also use the commutator to denote a Lie derivative: $\mathcal{L}_V W = [V,W]$. The Lie derivative, interior product and exterior derivative are related by Cartan's formula
\begin{align}
    \mathcal{L}_{V} = \iota_V \circ \dd + \dd \circ \iota_V \, .
\end{align}
A vector field $V^\mu$ is Killing if it satisfies
\begin{align}
    \mathcal{L}_V g_{\mu\nu} = \nabla_\mu V^\flat_\nu+\nabla_\nu V^\flat_\mu = 0 \, ,
\end{align}
which implies that the covariant divergence vanishes identically $\nabla_\mu V^\mu =0$.
A vector field $V^\mu$ is conformally Killing if it satisfies
\begin{align}
    \mathcal{L}_V g_{\mu \nu} - \frac2d \text{div}(V) g_{\mu \nu} =0 \, ,
\end{align}
where $d$ is the number of spacetime dimensions. It follows that a Killing vector is automatically also a conformal Killing vector.

When twisting the fields we shall make use of tensors which turn $\text{Spin}(4) \simeq SU(2) \times SU(2)$ indices into chiral $SU(2)$ indices. The $SU(2)$ invariant tensors
\begin{align}\label{eq: SU(2) inv tens}
    \varepsilon^{\alpha \beta} (\delta^{\alpha}{}_{\beta}) \, , \, \varepsilon^{\dot \alpha \dot \beta} (\delta^{\dot \alpha}{}_{\dot \beta}) &\longrightarrow (\mathbf 1,\mathbf 1) \, , \nonumber\\
    (\sigma_\mu)_{\alpha \dot \alpha}, (\overline \sigma_\mu)^{ \dot \alpha \alpha} &\longrightarrow (\mathbf 2, \mathbf 2) \, , \\
    (\sigma_{\mu \nu})_{\alpha}{}^{\beta} \, , (\overline \sigma_{\mu \nu})^{\dot \alpha}{}_{\dot \beta} &\longrightarrow (\mathbf 3,\mathbf 1) \, , (\mathbf 1, \mathbf 3) \, , \nonumber
\end{align}
where $(\bullet,\bullet)$ denotes the dimensions $SU(2)\times SU(2)$ representation.

For each twist, the appropriate choice of $SU(4)_R$ connection reduces the full covariant derivative of the supersymmetry parameters to the Levi-Civita covariant derivative for the twisted fields.
After decomposing $\phi^{ij}$ into its twisted components, the $SU(4)_R$ connection reproduces the Levi-Civita connection appropriate to their tensorial type. We denote the resulting gauge-covariant derivative by \begin{align} D_\mu \coloneqq \nabla_\mu+[A_\mu,\,\cdot\,]. \end{align} Thus, on any twisted component of $\phi^{ij}$, $\mathcal D_\mu\phi^{ij}=D_\mu\phi^{ij}$. Since the supersymmetry parameters are gauge neutral, $D_\mu\epsilon=\nabla_\mu\epsilon$. We will also make use of the gauge-covariant codifferential, which in form notation is defined in terms of the gauge-covariant derivative as
\begin{align}
    D^\ast:= -\ast D \ast \, ,
\end{align}
where $\ast$ is the Hodge dual in four dimensions.

\subsection{Background geometries and Killing vector fields}\label{sec: metrics}
In this subsection we collect specific geometries on which we numerically solved the dilatino equation. Notice that the vector fields depend on complex constants $c_i$. In flat space these are what in most papers is denoted $\epsilon_i$: the deformation parameters associated to rotations of the $x_1,x_2$ and $x_3,x_4$ planes.
\subsubsection{$\mathbb{R}^4$}
We first consider $\mathbb{R}^4$ in Cartesian coordinates $x^\mu$, with metric
\begin{align} \dd s^2 =   (\dd x^1)^2 + (\dd x^2)^2 + (\dd x^3)^2 + (\dd x^4)^2\, .
\end{align}
In these coordinates the vielbeins read
\begin{align}
    e^m{}_\mu = \delta^m{}_\mu \, .
\end{align}
We consider two commuting Killing vectors
\begin{align}
    V_1 &= c_{1} x^2 \frac{\partial}{\partial x^1} - c_1 x^1 \frac{\partial}{\partial x^2} + c_2 x^4 \frac{\partial}{\partial x^3} - c_2 x^3 \frac{\partial}{\partial x^4} \, , \nonumber \\
    V_2 &= c_{3} x^2 \frac{\partial}{\partial x^1} - c_3 x^1 \frac{\partial}{\partial x^2} + c_4 x^4 \frac{\partial}{\partial x^3} - c_4 x^3 \frac{\partial}{\partial x^4} \, , \nonumber \\
    &[V_1, V_2] = 0 \, .
\end{align}
Together, these vector fields generate the standard commuting $U(1)$ rotations in the $(x^1,x^2)$ and $(x^3,x^4)$ planes.
The following set of Killing vectors will also be useful
\begin{align}
    V_1 &= c_{1} x^2 \frac{\partial}{\partial x^1} - c_1 x^1 \frac{\partial}{\partial x^2} \, ; \quad  V_2 = c_2 x^4 \frac{\partial}{\partial x^3} - c_2 x^3 \frac{\partial}{\partial x^4} \, , \nonumber \\
    &[V_1, V_2] = 0 \, ; \qquad \dd (V_1 \cdot V_2)=0 \, .
\end{align}

\subsubsection{$S^3 \times \mathbb{R}$}
This background can be written in toroidal coordinates as
\begin{align}
    \dd s^2 = \dd  \theta^2 + \sin^2{\theta} \dd \varphi^2 + \cos^2{\theta} \dd \tau^2 +  \dd t^2 \, , \nonumber \\
    \theta \in \left[ 0, \frac{\pi}{2} \right) \, , \, \varphi \in \left[ 0, 2 \pi \right) \, , \, \tau \in \left[ 0,2 \pi \right) \,  .
\end{align}
In these coordinates the orthonormal frame reads
\begin{align}
e^1 &= e^1{}_\mu \dd x^\mu \nonumber \\ &\quad = \sin(\tau+\varphi)\,\dd\theta +\cos\theta\sin\theta\cos(\tau+\varphi)\,\dd\varphi  -\cos\theta\sin\theta\cos(\tau+\varphi)\,\dd\tau, \\ e^2 &= e^2{}_\mu \dd x^\mu \nonumber\\ &\quad = -\cos(\tau+\varphi)\,\dd\theta +\cos\theta\sin\theta\sin(\tau+\varphi)\,\dd\varphi -\cos\theta\sin\theta\sin(\tau+\varphi)\,\dd\tau, \\ e^3 &= e^3{}_\mu \dd x^\mu \nonumber\\ &\quad = \sin^2\theta\,\dd\varphi +\cos^2\theta\,\dd\tau, \\ e^4 &= e^4{}_\mu \dd x^\mu \nonumber\\ &\quad = \dd t \, .
\end{align}
We will consider two independent Killing vectors, each generating a $U(1)$ rotation and a translation along the fourth direction
\begin{align}
    V_1 = c_{1} \frac{\partial}{\partial \varphi} + c_2\frac{\partial}{\partial \tau} + c_3 \frac{\partial}{\partial t} \, , \quad V_2 =c_{4} \frac{\partial}{\partial \varphi} + c_5\frac{\partial}{\partial \tau} + c_6 \frac{\partial}{\partial t} \, , \quad \left[V_1, V_2 \right] =0\, .
\end{align}
The following set of Killing vectors will also be useful
\begin{align}
    V_1 &= c_{1} \frac{\partial}{\partial \varphi} + c_1\frac{\partial}{\partial \tau} + c_2 \frac{\partial}{\partial t} \, , \quad V_2 =c_{3} \frac{\partial}{\partial \varphi} + c_3\frac{\partial}{\partial \tau} + c_4 \frac{\partial}{\partial t} \, , \nonumber \\
     &[V_1, V_2] = 0 \, ; \qquad \dd (V_1 \cdot V_2)=0 \, .
\end{align}

\section{Twisted vector multiplet transformations}\label{sec: vec mult transf}
For completeness, we collect here the Donaldson-Witten, Vafa-Witten and Geometric Langlands vector multiplet transformations in a general bosonic Weyl multiplet background. The notation and projection conventions are those introduced in Sections \ref{sec: twisted DW}, \ref{sec: twisted VW} and \ref{sec: twisted GL}. We list the variations of the gauge field and fermions, followed by those of the twisted scalars.
\subsection{Donaldson-Witten twist}\label{app: DW vector multiplet}
\paragraph{Gauge field.}
\begin{align}
    \delta A_m
    =&\ -g\big(
    8 \overline \lambda^{(1)-}_{mn} \epsilon^{(1)n}
    + 2 \overline \lambda^{(1)} \epsilon^{(1)}_m
    + 8 \lambda^{(1) n} \overline \epsilon^{(1)-}_{mn}
    + 2 \lambda^{(1)}_m \overline \epsilon^{(1)}
    \nonumber \\
    &\qquad
    +\epsilon^{(2)}_{\alpha}{}^{A}
    (\sigma_m)^{\alpha}{}_{\dot \alpha}
    \overline \lambda^{(2) \dot \alpha}{}_A
    - \overline \epsilon^{(2) \dot \alpha}{}_{A}
    (\overline \sigma_m)_{\dot \alpha}{}^{\alpha}
    \lambda^{(2)}_{\alpha}{}^A
    \big) \, .
\end{align}
\paragraph{Weyl Fermions.}
\begin{align}
\delta \lambda^{(1)}_{m} ={}& -\frac{1}{g}\left( F_{m}{}^{n}\epsilon^{(1)}_{n} +\frac{1}{2}\varepsilon_{mnrs}F^{rs}\epsilon^{(1)n} \right) -2\,D_m\phi^{(1,1)}\,\overline{\epsilon}^{(1)} +8\,D^{n}\phi^{(1,1)}\,\overline{\epsilon}^{(1)-}_{mn} \nonumber\\ &+D_n\phi^{(1,2)\dot\alpha A} (\sigma^n\overline{\epsilon}^{(2)})_{\alpha A} (\overline{\sigma}_m)_{\dot\alpha}{}^{\alpha} \nonumber\\[3pt] &-4\,\overline{E}^{(1,1)-}_{m}{}^{n} \phi^{(2,2)}\epsilon^{(1)}_{n} +\frac12\,\overline{E}^{(1,1)-}_{rs} (\overline{\sigma}^{rs}\phi^{(1,2)})^{\dot\alpha}{}_{A} \epsilon^{(2)}_{\alpha}{}^{A} (\overline{\sigma}_m)_{\dot\alpha}{}^{\alpha} \nonumber\\ &+\frac12\,\overline{E}^{(1,2)\dot\alpha A} \phi^{(1,2)}_{\dot\beta A}\epsilon^{(1)}_{n} (\overline{\sigma}_m\sigma^n)_{\dot\alpha}{}^{\dot\beta} -\frac12\,\overline{E}^{(1,2)\dot\alpha A} \phi^{(1,1)}\epsilon^{(2)}_{\alpha A} (\overline{\sigma}_m)_{\dot\alpha}{}^{\alpha} \nonumber\\[3pt] &+4\,\epsilon^{(1)n}C^{(1,1)+}_{mn}\phi^{(2,2)} +4\,\epsilon^{(1)n}C^{(2,2)+}_{mn}\phi^{(1,1)} -4\,\epsilon^{(1)n}T^{(1,2)+}_{mn}{}^{\dot\alpha A} \phi^{(1,2)}_{\dot\alpha A} \nonumber\\[3pt] &+2g\,\epsilon^{(1)}_{m} [\phi^{(1,1)},\phi^{(2,2)}] -2g\,[\phi^{(1,1)},\phi^{(1,2)\dot\alpha}{}_{A}] \epsilon^{(2)}_{\alpha}{}^{A} (\overline{\sigma}_m)_{\dot\alpha}{}^{\alpha} \nonumber\\ &+g\,[\phi^{(1,2)\dot\alpha A},\phi^{(1,2)}_{\dot\beta A}] \epsilon^{(1)n} (\overline{\sigma}_{mn})_{\dot\alpha}{}^{\dot\beta} \nonumber\\[3pt] &-\phi^{(1,1)}\nabla_m\overline{\epsilon}^{(1)} +4\,\phi^{(1,1)}\nabla^{n}\overline{\epsilon}^{(1)-}_{mn} +\frac12\,\phi^{(1,2)\dot\alpha A} (\overline{\sigma}_{m}\sigma^n)_{\dot\alpha\dot\beta} \nabla_n\overline{\epsilon}^{(2)\dot\beta}{}_{A}\, , \displaybreak[0]\\ \delta \lambda^{(2)}_{\alpha}{}^{A} ={}& -\frac{1}{2g}F_{mn} (\sigma^{mn}\epsilon^{(2)})_{\alpha}{}^{A} +2\,D_m\phi^{(1,2)\dot\gamma A} (\sigma^m\overline{\sigma}^{rs})_{\alpha\dot\gamma} \overline{\epsilon}^{(1)-}_{rs} \nonumber\\ &+2\,D_m\phi^{(1,2)\dot\gamma A} (\sigma^m)_{\alpha\dot\gamma}\overline{\epsilon}^{(1)} -2\,D_m\phi^{(2,2)} (\sigma^m\overline{\epsilon}^{(2)})_{\alpha}{}^{A} \nonumber\\ &+\overline{E}^{(1,2)\dot\beta A}\phi^{(2,2)} \epsilon^{(1)}_{m}(\sigma^m)_{\alpha\dot\beta} +\overline{E}^{(1,2)\dot\beta A} \phi^{(1,2)}{}_{\dot\beta B} \epsilon^{(2)}_{\alpha}{}^{B} \nonumber\\ &+\overline{E}^{(2,2)AB} \phi^{(1,2)\dot\beta}{}_{B} \epsilon^{(1)}_{m}(\sigma^m)_{\alpha\dot\beta} +\overline{E}^{(2,2)AB}\phi^{(1,1)} \epsilon^{(2)}_{\alpha B} \nonumber\\ &+\phi^{(2,2)}C^{(1,1)+}_{mn} (\sigma^{mn}\epsilon^{(2)})_{\alpha}{}^{A} +\phi^{(1,1)}C^{(2,2)+}_{mn} (\sigma^{mn}\epsilon^{(2)})_{\alpha}{}^{A} \nonumber\\ &-\phi^{(1,2)}_{\dot\gamma B} T^{(1,2)+}_{mn}{}^{\dot\gamma B} (\sigma^{mn}\epsilon^{(2)})_{\alpha}{}^{A} \nonumber\\ &+2g\,[\phi^{(2,2)},\phi^{(1,1)}] \epsilon^{(2)}_{\alpha}{}^{A} +4g\,[\phi^{(2,2)},\phi^{(1,2)\dot\alpha A}] \epsilon^{(1)m}(\sigma_m)_{\alpha\dot\alpha} \nonumber\\ &-2g\,[\phi^{(1,2)\dot\beta A}, \phi^{(1,2)}_{\dot\beta B}] \epsilon^{(2)}_{\alpha}{}^{B} \nonumber\\ &+\phi^{(1,2)\dot\beta A} (\sigma^m)_{\alpha\dot\beta} \nabla_m\overline{\epsilon}^{(1)} +\phi^{(1,2)\dot\beta A} (\sigma^m)_{\alpha\dot\gamma} \nabla_m\overline{\epsilon}^{(1)-}_{rs} (\overline{\sigma}^{rs})^{\dot\gamma}{}_{\dot\beta} \nonumber\\ &-\phi^{(2,2)} (\sigma^m)_{\alpha\dot\gamma} \nabla_m\overline{\epsilon}^{(2)\dot\gamma A}\, , \displaybreak[0]\\  \delta \overline{\lambda}^{(1)} ={}& \frac{2}{g}F^{mn}\overline{\epsilon}^{(1)-}_{mn} -2\,D_m\phi^{(2,2)}\,\epsilon^{(1)m} -D_m\phi^{(1,2)}_{\dot\alpha A} (\overline{\sigma}^{m}\epsilon^{(2)})^{\dot\alpha A} \nonumber\\ &+4\,E^{(1,1)-mn}\phi^{(1,1)} \overline{\epsilon}^{(1)-}_{mn} -\frac12\,E^{(1,1)-}_{mn} \phi^{(1,2)\dot\gamma}{}_{A} (\overline{\sigma}^{mn})_{\dot\alpha\dot\gamma} \overline{\epsilon}^{(2)\dot\alpha A} \nonumber\\ &+\frac12\,E^{(1,2)}_{\dot\alpha A}\phi^{(2,2)} \overline{\epsilon}^{(2)\dot\alpha A} -\frac12\,E^{(1,2)}_{\dot\alpha A} \phi^{(1,2)\dot\alpha A}\overline{\epsilon}^{(1)} \nonumber\\ &-\frac12\,E^{(1,2)}_{\dot\alpha A} \phi^{(1,2)\dot\gamma A} (\overline{\sigma}^{mn})^{\dot\alpha}{}_{\dot\gamma} \overline{\epsilon}^{(1)-}_{mn} \nonumber\\ &-4\,\overline{C}^{(1,1)-mn}\phi^{(1,1)} \overline{\epsilon}^{(1)-}_{mn} -4\,\overline{C}^{(2,2)-mn}\phi^{(2,2)} \overline{\epsilon}^{(1)-}_{mn} \nonumber\\ &+4\,\overline{T}^{(1,2)-mn}{}_{\dot\gamma A} \phi^{(1,2)\dot\gamma A} \overline{\epsilon}^{(1)-}_{mn} \nonumber\\ &-\phi^{(2,2)}\nabla_m\epsilon^{(1)m} -\frac12\,\phi^{(1,2)}_{\dot\alpha A} (\overline{\sigma}^{m})^{\dot\alpha\beta} \nabla_m\epsilon^{(2)}_{\beta}{}^{A} \nonumber\\ &+2g\,[\phi^{(2,2)},\phi^{(1,1)}] \overline{\epsilon}^{(1)} -2g\,[\phi^{(2,2)},\phi^{(1,2)}_{\dot\beta A}] \overline{\epsilon}^{(2)\dot\alpha A} \delta^{\dot\beta}{}_{\dot\alpha} \nonumber\\ &-g\,[\phi^{(1,2)}_{\dot\beta A}, \phi^{(1,2)\dot\gamma A}] \overline{\epsilon}^{(1)-}_{mn} (\overline{\sigma}^{mn})^{\dot\alpha}{}_{\dot\gamma} \delta^{\dot\beta}{}_{\dot\alpha}\, , \displaybreak[0]\\ \delta \overline{\lambda}^{(1)-}_{mn} ={}& -\frac{1}{4g}F_{mn}\overline{\epsilon}^{(1)} +\frac{1}{8g}\varepsilon_{mnrs}F^{rs}\overline{\epsilon}^{(1)} +\frac{1}{g}\varepsilon_{mnrs}F^{r}{}_{t} \overline{\epsilon}^{(1)-ts} \nonumber\\[3pt] &+\frac12\,\varepsilon_{mnrs}D^{r}\phi^{(2,2)} \epsilon^{(1)s} -\frac12\,D_m\phi^{(2,2)}\epsilon^{(1)}_{n} +\frac12\,D_n\phi^{(2,2)}\epsilon^{(1)}_{m} \nonumber\\ &+\frac14\,D_r\phi^{(1,2)}_{\dot\beta A} (\overline{\sigma}^{r}\epsilon^{(2)})^{\dot\alpha A} (\overline{\sigma}_{mn})^{\dot\beta}{}_{\dot\alpha} \nonumber\\[3pt] &-E^{(1,1)-}_{mn}\phi^{(1,1)}\overline{\epsilon}^{(1)} +2\,E^{(1,1)-}_{m}{}^{t}\phi^{(1,1)} \overline{\epsilon}^{(1)-}_{tn} -2\,E^{(1,1)-}_{n}{}^{t}\phi^{(1,1)} \overline{\epsilon}^{(1)-}_{tm} \nonumber\\ &-\frac18\,E^{(1,1)-}_{rs} (\overline{\sigma}^{rs})_{\dot\beta\dot\gamma} \phi^{(1,2)\dot\gamma A} \overline{\epsilon}^{(2)\dot\alpha}{}_{A} (\overline{\sigma}_{mn})^{\dot\beta}{}_{\dot\alpha} \nonumber\\ &-\frac18\,\phi^{(2,2)}E^{(1,2)}_{\dot\beta A} \overline{\epsilon}^{(2)\dot\alpha A} (\overline{\sigma}_{mn})^{\dot\beta}{}_{\dot\alpha} \nonumber\\ &+\frac18\,E^{(1,2)}_{\dot\beta A} (\overline{\sigma}^{rs})^{\dot\alpha}{}_{\dot\gamma} \phi^{(1,2)\dot\gamma A} \overline{\epsilon}^{(1)-}_{rs} (\overline{\sigma}_{mn})^{\dot\beta}{}_{\dot\alpha} \nonumber\\ &+\frac18\,E^{(1,2)}_{\dot\beta A} \phi^{(1,2)\dot\alpha A}\overline{\epsilon}^{(1)} (\overline{\sigma}_{mn})^{\dot\beta}{}_{\dot\alpha} \nonumber\\[3pt] &+\overline{C}^{(1,1)-}_{mn}\phi^{(1,1)} \overline{\epsilon}^{(1)} +2\,\overline{C}^{(1,1)-}_{m}{}^{r}\phi^{(1,1)} \overline{\epsilon}^{(1)-}_{rn} -2\,\overline{C}^{(1,1)-}_{n}{}^{r}\phi^{(1,1)} \overline{\epsilon}^{(1)-}_{rm} \nonumber\\ &+\overline{C}^{(2,2)-}_{mn}\phi^{(2,2)} \overline{\epsilon}^{(1)} +2\,\overline{C}^{(2,2)-}_{m}{}^{r}\phi^{(2,2)} \overline{\epsilon}^{(1)-}_{rn} -2\,\overline{C}^{(2,2)-}_{n}{}^{r}\phi^{(2,2)} \overline{\epsilon}^{(1)-}_{rm} \nonumber\\ &-\overline{T}^{(1,2)-}_{mn\dot\alpha A} \phi^{(1,2)\dot\alpha A}\overline{\epsilon}^{(1)} -2\,\overline{T}^{(1,2)-}_{mr\dot\alpha A} \phi^{(1,2)\dot\alpha A} \overline{\epsilon}^{(1)-r}{}_{n} \nonumber\\ &+2\,\overline{T}^{(1,2)-}_{nr\dot\alpha A} \phi^{(1,2)\dot\alpha A} \overline{\epsilon}^{(1)-r}{}_{m} \nonumber\\[3pt] &-\frac14\,\phi^{(2,2)} \left(\nabla_m\epsilon^{(1)}_{n} -\nabla_n\epsilon^{(1)}_{m}\right) +\frac14\,\phi^{(2,2)} \varepsilon_{mnrs}\nabla^{r}\epsilon^{(1)s} \nonumber\\ &+\frac18\,\phi^{(1,2)}_{\dot\beta A} \nabla_r\epsilon^{(2)}_{\beta}{}^{A} (\overline{\sigma}_{mn}\overline{\sigma}^{r})^{\dot\beta\beta} \nonumber\\[3pt] &+2g\,[\phi^{(2,2)},\phi^{(1,1)}] \overline{\epsilon}^{(1)-}_{mn} +\frac{g}{2}\,[\phi^{(2,2)},\phi^{(1,2)}_{\dot\beta A}] \overline{\epsilon}^{(2)\dot\alpha A} (\overline{\sigma}_{mn})^{\dot\beta}{}_{\dot\alpha} \nonumber\\ &+\frac{g}{4}\,[\phi^{(1,2)}_{\dot\beta A}, \phi^{(1,2)\dot\alpha A}] \overline{\epsilon}^{(1)} (\overline{\sigma}_{mn})^{\dot\beta}{}_{\dot\alpha} \nonumber\\ &+\frac{g}{4}\,[\phi^{(1,2)}_{\dot\beta A}, \phi^{(1,2)\dot\gamma A}] \overline{\epsilon}^{(1)-}_{rs} (\overline{\sigma}_{mn}\overline{\sigma}^{rs}) ^{\dot\beta}{}_{\dot\gamma}\, , \displaybreak[0]\\  \delta \overline{\lambda}^{(2)\dot\alpha}{}_{A} ={}& -\frac{1}{2g}F_{mn} (\overline{\sigma}^{mn}\overline{\epsilon}^{(2)}) ^{\dot\alpha}{}_{A} +2\,D_m\phi^{(1,2)}_{\dot\beta A} (\overline{\sigma}^{m}\sigma^{n})^{\dot\alpha\dot\beta} \epsilon^{(1)}_{n} \nonumber\\ &-2\,D_m\phi^{(1,1)} (\overline{\sigma}^{m}\epsilon^{(2)})^{\dot\alpha}{}_{A} \nonumber\\[3pt] &-E^{(1,2)\dot\alpha}{}_{A}\phi^{(1,1)} \overline{\epsilon}^{(1)} -E^{(1,2)\dot\beta}{}_{A}\phi^{(1,1)} (\overline{\sigma}^{mn})^{\dot\alpha}{}_{\dot\beta} \overline{\epsilon}^{(1)-}_{mn} \nonumber\\ &+E^{(1,2)}_{\dot\beta A}\phi^{(1,2)\dot\beta B} \overline{\epsilon}^{(2)\dot\alpha}{}_{B} -\phi^{(1,2)\dot\alpha B}E^{(2,2)}_{AB} \overline{\epsilon}^{(1)} \nonumber\\ &-\phi^{(1,2)\dot\beta B}E^{(2,2)}_{AB} (\overline{\sigma}^{mn})^{\dot\alpha}{}_{\dot\beta} \overline{\epsilon}^{(1)-}_{mn} +\phi^{(2,2)}E^{(2,2)}_{AB} \overline{\epsilon}^{(2)\dot\alpha B} \nonumber\\[3pt] &+\left( \overline{C}^{(1,1)-}_{mn}\phi^{(1,1)} +\overline{C}^{(2,2)-}_{mn}\phi^{(2,2)} -\overline{T}^{(1,2)-}_{mn\dot\beta B} \varepsilon^{\dot\beta\dot\gamma} \phi^{(1,2)}_{\dot\gamma C}\varepsilon^{BC} \right) \nonumber\\ &\hspace{3cm}\times (\overline{\sigma}^{mn}\overline{\epsilon}^{(2)}) ^{\dot\alpha}{}_{A} \nonumber\\[3pt] &+\phi^{(1,2)}_{\dot\beta A} (\overline{\sigma}^{m}\sigma^{n})^{\dot\alpha\dot\beta} \nabla_m\epsilon^{(1)}_{n} -\phi^{(1,1)} (\overline{\sigma}^{m})^{\dot\alpha\beta} \nabla_m\epsilon^{(2)}_{\beta A} \nonumber\\[3pt] &+2g\,[\phi^{(1,1)},\phi^{(2,2)}] \overline{\epsilon}^{(2)\dot\alpha}{}_{A} +2g\,[\phi^{(1,2)\dot\beta}{}_{A}, \phi^{(1,2)}_{\dot\beta}{}^{B}] \overline{\epsilon}^{(2)\dot\alpha}{}_{B} \nonumber\\ &-4g\,[\phi^{(1,1)},\phi^{(1,2)\dot\alpha}{}_{A}] \overline{\epsilon}^{(1)} -4g\,[\phi^{(1,1)},\phi^{(1,2)\dot\beta}{}_{A}] \overline{\epsilon}^{(1)-}_{mn} (\overline{\sigma}^{mn})^{\dot\alpha}{}_{\dot\beta}\, . \end{align}

\paragraph{Scalars.}

\begin{align}
\delta \phi^{(1,1)}
={}&\ 2\,\epsilon^{(1)}_{m}\lambda^{(1)m}
-\overline{\epsilon}^{(2)\dot\alpha}{}_{A}
 \overline{\lambda}^{(2)}_{\dot\alpha}{}^{A} \, , \displaybreak[0] \\[4pt]
\delta \phi^{(2,2)AB}
={}&\ -\epsilon^{(2)}_{\alpha}{}^{A}
       \lambda^{(2)\alpha B}
+\epsilon^{(2)}_{\alpha}{}^{B}
 \lambda^{(2)\alpha A}
+2\,\overline{\epsilon}^{(1)}
   \overline{\lambda}^{(1)}\varepsilon^{AB}
\nonumber\\
&\quad
+8\,\overline{\epsilon}^{(1)-}_{mn}
   \overline{\lambda}^{(1)-mn}\varepsilon^{AB} \, ,
\displaybreak[0]\\[4pt]
\delta \phi^{(1,2)\dot\alpha A}
={}&\ -\epsilon^{(1)}_{m}
(\sigma^{m})_{\alpha}{}^{\dot\alpha}
\lambda^{(2)\alpha A}
-\lambda^{(1)}_{m}
(\sigma^{m})_{\alpha}{}^{\dot\alpha}
\epsilon^{(2)\alpha A}
\nonumber\\
&\quad
+\overline{\epsilon}^{(1)}
 \overline{\lambda}^{(2)\dot\alpha A}
+\overline{\lambda}^{(1)}
 \overline{\epsilon}^{(2)\dot\alpha A}
\nonumber\\
&\quad
-\overline{\lambda}^{(1)-}_{mn}
 \bigl(
 \overline{\sigma}^{mn}\overline{\epsilon}^{(2)}
 \bigr)^{\dot\alpha A}
-\overline{\epsilon}^{(1)-}_{mn}
 \bigl(
 \overline{\sigma}^{mn}\overline{\lambda}^{(2)}
 \bigr)^{\dot\alpha A}\, .
\end{align}

\subsection{Vafa-Witten twist}\label{sec: VW matt}
\paragraph{Gauge field.}
\begin{align}
\delta A_m
&=-g \big(
2 \epsilon^{(1)}_m \overline{\lambda}^{(1)}
+8 \epsilon^{(1)n}\overline{\lambda}^{(1)-}_{mn}
+2 \epsilon^{(2)}_m \overline{\lambda}^{(2)}
+8 \epsilon^{(2)n}\overline{\lambda}^{(2)-}_{mn} 
\nonumber\\
&\quad
+2 \overline{\epsilon}^{(1)} \lambda^{(1)}_m
+8 \overline{\epsilon}^{(1)-}_{mn}\lambda^{(1)n}
+2 \overline{\epsilon}^{(2)} \lambda^{(2)}_m
+8 \overline{\epsilon}^{(2)-}_{mn}\lambda^{(2)n} \big)\, .
\end{align}

\paragraph{Weyl fermions.}

\begin{align}
\delta \lambda^{(1)}_m &= -\frac{1}{g} \left( F_{mn}\epsilon^{(1)n} +\frac{1}{2}\varepsilon_{mnrs}F^{rs}\epsilon^{(1)n} \right) \nonumber\\ &\quad -2 D_m\phi^{(1,1)}\,\overline{\epsilon}^{(1)} +8 D_n\phi^{(1,1)}\,\overline{\epsilon}^{(1)-}_{m}{}^{n} -2 D_m\phi^{(1,2)}\,\overline{\epsilon}^{(2)} +8 D_n\phi^{(1,2)}\,\overline{\epsilon}^{(2)-}_{m}{}^{n} \nonumber\\ &\quad +8 D_n\phi^{(1,2)-}_{m}{}^{n}\,\overline{\epsilon}^{(2)} -16 \varepsilon^{nrst} D_n\phi^{(1,2)-}_{mr}\, \overline{\epsilon}^{(2)-}_{st} \nonumber\\ &\quad -4 \overline{E}^{(1,1)-}_{mn}\phi^{(2,2)}\epsilon^{(1)n} +4 \overline{E}^{(1,1)-}_{mn}\phi^{(1,2)}\epsilon^{(2)n} -4 \overline{E}^{(1,2)-}_{mn}\phi^{(1,1)}\epsilon^{(2)n} \nonumber\\ &\quad +4 \overline{E}^{(1,2)-}_{mn}\phi^{(1,2)}\epsilon^{(1)n} -\overline{E}^{(1,2)}\phi^{(1,2)}\epsilon^{(1)}_m +\overline{E}^{(1,2)}\phi^{(1,1)}\epsilon^{(2)}_m \nonumber\\ &\quad -4 \overline{E}^{(1,2)}\phi^{(1,2)-}_{mn}\epsilon^{(1)n} -8 \epsilon^{(1)}_m \overline{E}^{(1,2)-}_{nr}\phi^{(1,2)-nr} \nonumber\\ &\quad -8 \varepsilon^{nrst} \epsilon^{(1)}_{n} \phi^{(1,2)-}_{mr} \overline{E}^{(1,2)-}_{st} +8 \epsilon^{(2)}_m \overline{E}^{(1,1)-}_{nr}\phi^{(1,2)-nr} \nonumber\\ &\quad +8 \varepsilon^{nrst} \epsilon^{(2)}_{n} \phi^{(1,2)-}_{mr} \overline{E}^{(1,1)-}_{st} \nonumber\\ &\quad +4 C^{(1,1)+}_{mn}\phi^{(2,2)}\epsilon^{(1)n} +4 C^{(2,2)+}_{mn}\phi^{(1,1)}\epsilon^{(1)n} -8 C^{(1,2)+}_{mn}\phi^{(1,2)}\epsilon^{(1)n} \nonumber\\ &\quad -32 K^{(1,2)+}_{mnrs}\phi^{(1,2)-rs}\epsilon^{(1)n} \nonumber\\ &\quad +2g[\phi^{(1,1)} \, , \, \phi^{(2,2)}]\epsilon^{(1)}_m -4g[\phi^{(1,1)}\, , \, \phi^{(1,2)}]\epsilon^{(2)}_m -16g[\phi^{(1,2)} \, , \, \phi^{(1,2)-}_{ms}]\epsilon^{(1)s} \nonumber\\ &\quad +16g[\phi^{(1,1)} \, , \, \phi^{(1,2)-}_{ms}]\epsilon^{(2)s} -32g [\phi^{(1,2)-}_{ms} \, , \, \phi^{(1,2)-rs}] \epsilon^{(1)}_{r} \nonumber\\ &\quad -\phi^{(1,1)}\nabla_m\overline{\epsilon}^{(1)} +4 \phi^{(1,1)}\nabla_n\overline{\epsilon}^{(1)-}_{m}{}^{n} -\phi^{(1,2)}\nabla_m\overline{\epsilon}^{(2)} +4 \phi^{(1,2)}\nabla_n\overline{\epsilon}^{(2)-}_{m}{}^{n} \nonumber\\ &\quad +4 \phi^{(1,2)-}_{m}{}^{n}\nabla_n\overline{\epsilon}^{(2)} -8 \varepsilon^{nrst} \phi^{(1,2)-}_{mr} \nabla_n\overline{\epsilon}^{(2)-}_{st} \, , \displaybreak[0]\\ \delta \lambda^{(2)}_m &= -\frac{1}{g} \left( F_{mn}\epsilon^{(2)n} +\frac{1}{2}\varepsilon_{mnrs}F^{rs}\epsilon^{(2)n} \right) \nonumber\\ &\quad -2 D_m\phi^{(2,2)}\,\overline{\epsilon}^{(2)} +8 D_n\phi^{(2,2)}\,\overline{\epsilon}^{(2)-}_{m}{}^{n} -2 D_m\phi^{(1,2)}\,\overline{\epsilon}^{(1)} +8 D_n\phi^{(1,2)}\,\overline{\epsilon}^{(1)-}_{m}{}^{n} \nonumber\\ &\quad -8 D_n\phi^{(1,2)-}_{m}{}^{n}\,\overline{\epsilon}^{(1)} +16 \varepsilon^{nrst} D_n\phi^{(1,2)-}_{mr}\, \overline{\epsilon}^{(1)-}_{st} \nonumber\\ &\quad -4 \overline{E}^{(2,2)-}_{mn}\phi^{(1,1)}\epsilon^{(2)n} +4 \overline{E}^{(2,2)-}_{mn}\phi^{(1,2)}\epsilon^{(1)n} -4 \overline{E}^{(1,2)-}_{mn}\phi^{(2,2)}\epsilon^{(1)n} \nonumber\\ &\quad +4 \overline{E}^{(1,2)-}_{mn}\phi^{(1,2)}\epsilon^{(2)n} +\overline{E}^{(1,2)}\phi^{(1,2)}\epsilon^{(2)}_m -\overline{E}^{(1,2)}\phi^{(2,2)}\epsilon^{(1)}_m \nonumber\\ &\quad -4 \overline{E}^{(1,2)}\phi^{(1,2)-}_{mn}\epsilon^{(2)n} +8 \epsilon^{(2)}_m \overline{E}^{(1,2)-}_{nr}\phi^{(1,2)-nr} \nonumber\\ &\quad +8 \varepsilon^{nrst} \epsilon^{(2)}_{n} \phi^{(1,2)-}_{mr} \overline{E}^{(1,2)-}_{st} -8 \epsilon^{(1)}_m \overline{E}^{(2,2)-}_{nr}\phi^{(1,2)-nr} \nonumber\\ &\quad -8 \varepsilon^{nrst} \epsilon^{(1)}_{n} \phi^{(1,2)-}_{mr} \overline{E}^{(2,2)-}_{st} \nonumber\\ &\quad +4 C^{(2,2)+}_{mn}\phi^{(1,1)}\epsilon^{(2)n} +4 C^{(1,1)+}_{mn}\phi^{(2,2)}\epsilon^{(2)n} -8 C^{(1,2)+}_{mn}\phi^{(1,2)}\epsilon^{(2)n} \nonumber\\ &\quad -32 K^{(1,2)+}_{mnrs}\phi^{(1,2)-rs}\epsilon^{(2)n} \nonumber\\ &\quad -4g[\phi^{(2,2)} \, , \,\phi^{(1,2)}]\epsilon^{(1)}_m -2g[\phi^{(1,1)} \, , \,\phi^{(2,2)}]\epsilon^{(2)}_m -16g[\phi^{(2,2)}\, , \,\phi^{(1,2)-}_{ms}]\epsilon^{(1)s} \nonumber\\ &\quad +16g[\phi^{(1,2)}\, , \,\phi^{(1,2)-}_{ms}]\epsilon^{(2)s} -32g[\phi^{(1,2)-}_{ms} \, , \, \phi^{(1,2)-rs}] \epsilon^{(2)}_{r} \nonumber\\ &\quad -\phi^{(2,2)}\nabla_m\overline{\epsilon}^{(2)} +4 \phi^{(2,2)}\nabla_n\overline{\epsilon}^{(2)-}_{m}{}^{n} -\phi^{(1,2)}\nabla_m\overline{\epsilon}^{(1)} +4 \phi^{(1,2)}\nabla_n\overline{\epsilon}^{(1)-}_{m}{}^{n} \nonumber\\ &\quad -4 \phi^{(1,2)-}_{m}{}^{n}\nabla_n\overline{\epsilon}^{(1)} +8 \varepsilon^{nrst} \phi^{(1,2)-}_{mr} \nabla_n\overline{\epsilon}^{(1)-}_{st} \, , \displaybreak[0]\\ \delta \overline{\lambda}^{(1)} &= \frac{2}{g}F_{mn}\overline{\epsilon}^{(1)-mn} -2 D_m\phi^{(2,2)}\,\epsilon^{(1)m} +2 D_m\phi^{(1,2)}\,\epsilon^{(2)m} -8 D_m\phi^{(1,2)-mn}\epsilon^{(2)}_{n} \nonumber\\ &\quad +4 E^{(1,1)-}_{mn}\phi^{(1,1)} \overline{\epsilon}^{(1)-mn} +\phi^{(2,2)}E^{(1,2)}\overline{\epsilon}^{(2)} +\phi^{(1,2)}E^{(1,2)}\overline{\epsilon}^{(1)} \nonumber\\ &\quad +4 \phi^{(2,2)}E^{(1,2)-}_{mn} \overline{\epsilon}^{(2)-mn} +4 \phi^{(1,2)}E^{(1,2)-}_{mn} \overline{\epsilon}^{(1)-mn} +4 \phi^{(1,2)}E^{(1,1)-}_{mn} \overline{\epsilon}^{(2)-mn} \nonumber\\ &\quad -16 \phi^{(1,2)-}_{mn}E^{(1,1)-rn} \overline{\epsilon}^{(2)-m}{}_{r} +16 \phi^{(1,2)-}_{mn}E^{(1,2)-rn} \overline{\epsilon}^{(1)-m}{}_{r} \nonumber\\ &\quad +4 \phi^{(1,2)-}_{mn}E^{(1,2)} \overline{\epsilon}^{(1)-mn} -4 \phi^{(1,2)-}_{mn}E^{(1,2)-mn} \overline{\epsilon}^{(1)} \nonumber\\ &\quad +4 \phi^{(1,2)-}_{mn}E^{(1,1)-mn} \overline{\epsilon}^{(2)} -4 \overline{C}^{(1,1)-}_{mn}\phi^{(1,1)} \overline{\epsilon}^{(1)-mn} \nonumber\\ &\quad -4 \overline{C}^{(2,2)-}_{mn}\phi^{(2,2)} \overline{\epsilon}^{(1)-mn} -8 \overline{C}^{(1,2)-}_{mn}\phi^{(1,2)} \overline{\epsilon}^{(1)-mn} \nonumber\\ &\quad +32 \overline{K}^{(1,2)-}_{rsmn}\phi^{(1,2)-mn} \overline{\epsilon}^{(1)-rs} \nonumber\\ &\quad +2g\,\overline{\epsilon}^{(1)} \bigl[\phi^{(2,2)},\phi^{(1,1)}\bigr] +4g\,\overline{\epsilon}^{(2)} \bigl[\phi^{(2,2)},\phi^{(1,2)}\bigr] \nonumber\\ &\quad -16g\,\bigl[\phi^{(1,2)},\phi^{(1,2)-}_{mn}\bigr] \overline{\epsilon}^{(1)-mn} -16g\,\bigl[\phi^{(2,2)},\phi^{(1,2)-}_{mn}\bigr] \overline{\epsilon}^{(2)-mn} \nonumber\\ &\quad -32g\,\bigl[\phi^{(1,2)-}_{mr},\phi^{(1,2)-m}{}_{s}\bigr] \overline{\epsilon}^{(1)-rs} \nonumber\\ &\quad -\phi^{(2,2)}\nabla_m\epsilon^{(1)m} +\phi^{(1,2)}\nabla_m\epsilon^{(2)m} -4 \phi^{(1,2)-}_{mn}\nabla^{m}\epsilon^{(2)n} \, , \end{align}
\begin{align}
\delta \overline{\lambda}^{(1)-}_{mn} &= -\frac{1}{g} \left( F_{mr}\overline{\epsilon}^{(1)-r}{}_{n} -F_{nr}\overline{\epsilon}^{(1)-r}{}_{m} \right) -\frac{1}{4g} \left( \overline{\epsilon}^{(1)}F_{mn} -\frac{1}{2}\varepsilon_{mnrs} \overline{\epsilon}^{(1)}F^{rs} \right) \nonumber\\ &\quad -\frac{1}{2} \left( D_m\phi^{(2,2)}\,\epsilon^{(1)}_{n} -D_n\phi^{(2,2)}\,\epsilon^{(1)}_{m} -\varepsilon_{mnrs}D^{r}\phi^{(2,2)}\,\epsilon^{(1)s} \right) \nonumber\\ &\quad +\frac{1}{2} \left( D_m\phi^{(1,2)}\,\epsilon^{(2)}_{n} -D_n\phi^{(1,2)}\,\epsilon^{(2)}_{m} -\varepsilon_{mnrs}D^{r}\phi^{(1,2)}\,\epsilon^{(2)s} \right) \nonumber\\ &\quad +2\left( D_r\phi^{(1,2)-}_{mn}\epsilon^{(2)r} -D_m\phi^{(1,2)-}_{nr}\epsilon^{(2)r} +D_n\phi^{(1,2)-}_{mr}\epsilon^{(2)r} \right) \nonumber\\ &\quad +2\left( D_r\phi^{(1,2)-}_{n}{}^{r}\epsilon^{(2)}_{m} -D_r\phi^{(1,2)-}_{m}{}^{r}\epsilon^{(2)}_{n} \right) \nonumber\\ &\quad -\frac{1}{4}\phi^{(2,2)} \left( \nabla_m\epsilon^{(1)}_{n} -\nabla_n\epsilon^{(1)}_{m} -\varepsilon_{mnrs}\nabla^{r}\epsilon^{(1)s} \right) \nonumber\\ &\quad +\frac{1}{4}\phi^{(1,2)} \left( \nabla_m\epsilon^{(2)}_{n} -\nabla_n\epsilon^{(2)}_{m} -\varepsilon_{mnrs}\nabla^{r}\epsilon^{(2)s} \right) \nonumber\\ &\quad +E^{(1,2)}\phi^{(1,2)} \overline{\epsilon}^{(1)-}_{mn} -E^{(1,2)}\phi^{(1,2)-}_{mn} \overline{\epsilon}^{(1)} +E^{(1,2)}\phi^{(2,2)} \overline{\epsilon}^{(2)-}_{mn} \nonumber\\ &\quad +2E^{(1,2)} \left( \phi^{(1,2)-}_{mr}\overline{\epsilon}^{(1)-r}{}_{n} -\phi^{(1,2)-}_{nr}\overline{\epsilon}^{(1)-r}{}_{m} \right) \nonumber\\ &\quad -E^{(1,1)-}_{mn}\phi^{(1,2)} \overline{\epsilon}^{(2)} -E^{(1,1)-}_{mn}\phi^{(1,1)} \overline{\epsilon}^{(1)} \nonumber\\ &\quad +2\left( E^{(1,1)-}_{mr}\phi^{(1,2)-r}{}_{n} -E^{(1,1)-}_{nr}\phi^{(1,2)-r}{}_{m} \right)\overline{\epsilon}^{(2)} \nonumber\\ &\quad +2\phi^{(1,2)} \left( E^{(1,1)-}_{mr}\overline{\epsilon}^{(2)-r}{}_{n} -E^{(1,1)-}_{nr}\overline{\epsilon}^{(2)-r}{}_{m} \right) \nonumber\\ &\quad +2\phi^{(1,1)} \left( E^{(1,1)-}_{mr}\overline{\epsilon}^{(1)-r}{}_{n} -E^{(1,1)-}_{nr}\overline{\epsilon}^{(1)-r}{}_{m} \right) \nonumber\\ &\quad -8\left( E^{(1,1)-}_{mr}\phi^{(1,2)-r}{}_{s} \overline{\epsilon}^{(2)-s}{}_{n} -E^{(1,1)-}_{nr}\phi^{(1,2)-r}{}_{s} \overline{\epsilon}^{(2)-s}{}_{m} \right) \nonumber\\ &\quad -E^{(1,2)-}_{mn}\phi^{(2,2)} \overline{\epsilon}^{(2)} -E^{(1,2)-}_{mn}\phi^{(1,2)} \overline{\epsilon}^{(1)} \nonumber\\ &\quad -2\left( E^{(1,2)-}_{mr}\phi^{(1,2)-r}{}_{n} -E^{(1,2)-}_{nr}\phi^{(1,2)-r}{}_{m} \right)\overline{\epsilon}^{(1)} \nonumber\\ &\quad +2\phi^{(1,2)} \left( E^{(1,2)-}_{mr}\overline{\epsilon}^{(1)-r}{}_{n} -E^{(1,2)-}_{nr}\overline{\epsilon}^{(1)-r}{}_{m} \right) \nonumber\\ &\quad +2\phi^{(2,2)} \left( E^{(1,2)-}_{mr}\overline{\epsilon}^{(2)-r}{}_{n} -E^{(1,2)-}_{nr}\overline{\epsilon}^{(2)-r}{}_{m} \right) \nonumber\\ &\quad +8\left( E^{(1,2)-}_{mr}\phi^{(1,2)-r}{}_{s} \overline{\epsilon}^{(1)-s}{}_{n} -E^{(1,2)-}_{nr}\phi^{(1,2)-r}{}_{s} \overline{\epsilon}^{(1)-s}{}_{m} \right) \nonumber\\ &\quad +2\overline{C}^{(1,2)-}_{mn} \phi^{(1,2)}\overline{\epsilon}^{(1)} +\overline{C}^{(1,1)-}_{mn} \phi^{(1,1)}\overline{\epsilon}^{(1)} +\overline{C}^{(2,2)-}_{mn} \phi^{(2,2)}\overline{\epsilon}^{(1)} \nonumber\\ &\quad +4\phi^{(1,2)} \left( \overline{C}^{(1,2)-}_{mr} \overline{\epsilon}^{(1)-r}{}_{n} -\overline{C}^{(1,2)-}_{nr} \overline{\epsilon}^{(1)-r}{}_{m} \right) \nonumber\\ &\quad +2\phi^{(1,1)} \left( \overline{C}^{(1,1)-}_{mr} \overline{\epsilon}^{(1)-r}{}_{n} -\overline{C}^{(1,1)-}_{nr} \overline{\epsilon}^{(1)-r}{}_{m} \right) \nonumber\\ &\quad +2\phi^{(2,2)} \left( \overline{C}^{(2,2)-}_{mr} \overline{\epsilon}^{(1)-r}{}_{n} -\overline{C}^{(2,2)-}_{nr} \overline{\epsilon}^{(1)-r}{}_{m} \right) \nonumber\\ &\quad -8\overline{K}^{(1,2)-}_{mnrs} \phi^{(1,2)-rs}\overline{\epsilon}^{(1)} \nonumber\\ &\quad +16\left( \overline{\epsilon}^{(1)-}_{mt} \overline{K}^{(1,2)-t}{}_{nrs} \phi^{(1,2)-rs} -\overline{\epsilon}^{(1)-}_{nt} \overline{K}^{(1,2)-t}{}_{mrs} \phi^{(1,2)-rs} \right) \nonumber\\ &\quad +2g\, \bigl[\phi^{(2,2)},\phi^{(1,1)}\bigr] \overline{\epsilon}^{(1)-}_{mn} +4g\, \bigl[\phi^{(2,2)},\phi^{(1,2)}\bigr] \overline{\epsilon}^{(2)-}_{mn} \nonumber\\ &\quad +4g\, \bigl[\phi^{(1,2)},\phi^{(1,2)-}_{mn}\bigr] \overline{\epsilon}^{(1)} +4g\, \bigl[\phi^{(2,2)},\phi^{(1,2)-}_{mn}\bigr] \overline{\epsilon}^{(2)} \nonumber\\ &\quad +8g\, \bigl[\phi^{(1,2)-}_{mr},\phi^{(1,2)-}_{n}{}^r\bigr] \overline{\epsilon}^{(1)} \nonumber\\ &\quad -8g\, \bigl[\phi^{(1,2)},\phi^{(1,2)-}_{mr}\bigr] \overline{\epsilon}^{(1)-r}{}_{n} +8g\, \bigl[\phi^{(1,2)},\phi^{(1,2)-}_{nr}\bigr] \overline{\epsilon}^{(1)-r}{}_{m} \nonumber\\ &\quad -8g\, \bigl[\phi^{(2,2)},\phi^{(1,2)-}_{mr}\bigr] \overline{\epsilon}^{(2)-r}{}_{n} +8g\, \bigl[\phi^{(2,2)},\phi^{(1,2)-}_{nr}\bigr] \overline{\epsilon}^{(2)-r}{}_{m} \nonumber\\ &\quad -16g\, \bigl[\phi^{(1,2)-}_{mn},\phi^{(1,2)-}_{rs}\bigr] \overline{\epsilon}^{(1)-rs} \nonumber\\ &\quad +\left( \phi^{(1,2)-}_{mn}\nabla_r\epsilon^{(2)r} -\phi^{(1,2)-}_{nr}\nabla_m\epsilon^{(2)r} +\phi^{(1,2)-}_{mr}\nabla_n\epsilon^{(2)r} \right) \nonumber\\ &\quad +\left( \phi^{(1,2)-}_{nr}\nabla^{r}\epsilon^{(2)}_{m} -\phi^{(1,2)-}_{mr}\nabla^{r}\epsilon^{(2)}_{n} \right) \, , \displaybreak[0]\\
\delta \overline{\lambda}^{(2)} &= \frac{2}{g}F_{mn}\overline{\epsilon}^{(2)-mn} -2D_m\phi^{(1,1)}\,\epsilon^{(2)m} +2D_m\phi^{(1,2)}\,\epsilon^{(1)m} +8D_m\phi^{(1,2)-mn}\epsilon^{(1)}_{n} \nonumber\\ &\quad +4E^{(2,2)-}_{mn}\phi^{(2,2)} \overline{\epsilon}^{(2)-mn} -\phi^{(1,1)}E^{(1,2)} \overline{\epsilon}^{(1)} -\phi^{(1,2)}E^{(1,2)} \overline{\epsilon}^{(2)} \nonumber\\ &\quad +4\phi^{(1,1)} E^{(1,2)-}_{mn}\overline{\epsilon}^{(1)-mn} +4\phi^{(1,2)} E^{(1,2)-}_{mn}\overline{\epsilon}^{(2)-mn} +4\phi^{(1,2)} E^{(2,2)-}_{mn}\overline{\epsilon}^{(1)-mn} \nonumber\\ &\quad +16\phi^{(1,2)-}_{mn} E^{(2,2)-rn}\overline{\epsilon}^{(1)-m}{}_{r} -16\phi^{(1,2)-}_{mn} E^{(1,2)-rn}\overline{\epsilon}^{(2)-m}{}_{r} \nonumber\\ &\quad +4\phi^{(1,2)-}_{mn} E^{(1,2)} \overline{\epsilon}^{(2)-mn} +4\phi^{(1,2)-}_{mn} E^{(1,2)-mn} \overline{\epsilon}^{(2)} \nonumber\\ &\quad -4\phi^{(1,2)-}_{mn} E^{(2,2)-mn} \overline{\epsilon}^{(1)} \nonumber\\ &\quad -4\overline{C}^{(2,2)-}_{mn} \phi^{(2,2)} \overline{\epsilon}^{(2)-mn} -4\overline{C}^{(1,1)-}_{mn} \phi^{(1,1)} \overline{\epsilon}^{(2)-mn} \nonumber\\ &\quad -8\overline{C}^{(1,2)-}_{mn} \phi^{(1,2)} \overline{\epsilon}^{(2)-mn} +32\overline{K}^{(1,2)-}_{rsmn} \phi^{(1,2)-mn} \overline{\epsilon}^{(2)-rs} \nonumber\\ &\quad +4g\,\overline{\epsilon}^{(1)} \bigl[\phi^{(1,1)},\phi^{(1,2)}\bigr] -2g\,\overline{\epsilon}^{(2)} \bigl[\phi^{(2,2)},\phi^{(1,1)}\bigr] \nonumber\\ &\quad +16g\, \bigl[\phi^{(1,1)},\phi^{(1,2)-}_{mn}\bigr] \overline{\epsilon}^{(1)-mn} +16g\, \bigl[\phi^{(1,2)},\phi^{(1,2)-}_{mn}\bigr] \overline{\epsilon}^{(2)-mn} \nonumber\\ &\quad -32g\, \bigl[\phi^{(1,2)-}_{mr}, \phi^{(1,2)-m}{}_{s}\bigr] \overline{\epsilon}^{(2)-rs} \nonumber\\ &\quad -\phi^{(1,1)}\nabla_m\epsilon^{(2)m} +\phi^{(1,2)}\nabla_m\epsilon^{(1)m} +4\phi^{(1,2)-}_{mn}\nabla^{m}\epsilon^{(1)n} \, , \displaybreak[0]\\
\delta \overline{\lambda}^{(2)-}_{mn} &= -\frac{1}{4g}\overline{\epsilon}^{(2)} \left( F_{mn} -\frac{1}{2}\varepsilon_{mnrs}F^{rs} \right) -\frac{1}{g} \left( F_{mr}\overline{\epsilon}^{(2)-r}{}_{n} -F_{nr}\overline{\epsilon}^{(2)-r}{}_{m} \right) \nonumber\\ &\quad +\frac{1}{2} \left( D_m\phi^{(1,2)}\,\epsilon^{(1)}_{n} -D_n\phi^{(1,2)}\,\epsilon^{(1)}_{m} -\varepsilon_{mnrs}D^{r}\phi^{(1,2)}\,\epsilon^{(1)s} \right) \nonumber\\ &\quad -\frac{1}{2} \left( D_m\phi^{(1,1)}\,\epsilon^{(2)}_{n} -D_n\phi^{(1,1)}\,\epsilon^{(2)}_{m} -\varepsilon_{mnrs}D^{r}\phi^{(1,1)}\,\epsilon^{(2)s} \right) \nonumber\\ &\quad -2\left( D_r\phi^{(1,2)-}_{mn}\epsilon^{(1)r} -D_m\phi^{(1,2)-}_{nr}\epsilon^{(1)r} +D_n\phi^{(1,2)-}_{mr}\epsilon^{(1)r} \right) \nonumber\\ &\quad -2\left( D_r\phi^{(1,2)-}_{n}{}^{r}\epsilon^{(1)}_{m} -D_r\phi^{(1,2)-}_{m}{}^{r}\epsilon^{(1)}_{n} \right) \nonumber\\ &\quad -\overline{\epsilon}^{(1)} \phi^{(1,2)}E^{(2,2)-}_{mn} -\overline{\epsilon}^{(1)} \phi^{(1,1)}E^{(1,2)-}_{mn} \nonumber\\ &\quad -2\overline{\epsilon}^{(1)} \left( E^{(2,2)-}_{mr}\phi^{(1,2)-r}{}_{n} -E^{(2,2)-}_{nr}\phi^{(1,2)-r}{}_{m} \right) \nonumber\\ &\quad -E^{(1,2)}\phi^{(1,1)} \overline{\epsilon}^{(1)-}_{mn} -8\overline{\epsilon}^{(1)-}_{mn} E^{(2,2)-}_{rs}\phi^{(1,2)-rs} \nonumber\\ &\quad -\overline{\epsilon}^{(2)} \phi^{(2,2)}E^{(2,2)-}_{mn} -\overline{\epsilon}^{(2)} \phi^{(1,2)}E^{(1,2)-}_{mn} -\overline{\epsilon}^{(2)}E^{(1,2)} \phi^{(1,2)-}_{mn} \nonumber\\ &\quad +2\overline{\epsilon}^{(2)} \left( E^{(1,2)-}_{mr}\phi^{(1,2)-r}{}_{n} -E^{(1,2)-}_{nr}\phi^{(1,2)-r}{}_{m} \right) \nonumber\\ &\quad +2\phi^{(2,2)} \left( E^{(2,2)-}_{mr}\overline{\epsilon}^{(2)-r}{}_{n} -E^{(2,2)-}_{nr}\overline{\epsilon}^{(2)-r}{}_{m} \right) \nonumber\\ &\quad +2\phi^{(1,2)} \left( E^{(1,2)-}_{mr}\overline{\epsilon}^{(2)-r}{}_{n} -E^{(1,2)-}_{nr}\overline{\epsilon}^{(2)-r}{}_{m} \right) \nonumber\\ &\quad +2\phi^{(1,2)} \left( E^{(2,2)-}_{mr}\overline{\epsilon}^{(1)-r}{}_{n} -E^{(2,2)-}_{nr}\overline{\epsilon}^{(1)-r}{}_{m} \right) \nonumber\\ &\quad +2\phi^{(1,1)} \left( E^{(1,2)-}_{mr}\overline{\epsilon}^{(1)-r}{}_{n} -E^{(1,2)-}_{nr}\overline{\epsilon}^{(1)-r}{}_{m} \right) \nonumber\\ &\quad +2E^{(1,2)} \left( \phi^{(1,2)-}_{mr}\overline{\epsilon}^{(2)-r}{}_{n} -\phi^{(1,2)-}_{nr}\overline{\epsilon}^{(2)-r}{}_{m} \right) \nonumber\\ &\quad -8\left( E^{(2,2)-}_{mr}\phi^{(1,2)-r}{}_{s} \overline{\epsilon}^{(1)-s}{}_{n} -E^{(2,2)-}_{nr}\phi^{(1,2)-r}{}_{s} \overline{\epsilon}^{(1)-s}{}_{m} \right) \nonumber\\ &\quad -8E^{(2,2)-}_{mn} \overline{\epsilon}^{(1)-}_{rs}\phi^{(1,2)-rs} +8\phi^{(1,2)-}_{mn} E^{(2,2)-}_{rs}\overline{\epsilon}^{(1)-rs} \nonumber\\ &\quad +8\left( E^{(1,2)-}_{mr}\phi^{(1,2)-r}{}_{s} \overline{\epsilon}^{(2)-s}{}_{n} -E^{(1,2)-}_{nr}\phi^{(1,2)-r}{}_{s} \overline{\epsilon}^{(2)-s}{}_{m} \right) \nonumber\\ &\quad +8E^{(1,2)-}_{mn} \overline{\epsilon}^{(2)-}_{rs}\phi^{(1,2)-rs} +8\overline{\epsilon}^{(2)-}_{mn} E^{(1,2)-}_{rs}\phi^{(1,2)-rs} \nonumber\\ &\quad -8\phi^{(1,2)-}_{mn} E^{(1,2)-}_{rs}\overline{\epsilon}^{(2)-rs} \nonumber\\ &\quad +\overline{\epsilon}^{(2)} \phi^{(1,1)}\overline{C}^{(1,1)-}_{mn} +2\overline{\epsilon}^{(2)} \phi^{(1,2)}\overline{C}^{(1,2)-}_{mn} +\overline{\epsilon}^{(2)} \phi^{(2,2)}\overline{C}^{(2,2)-}_{mn} \nonumber\\ &\quad -8\overline{\epsilon}^{(2)} \overline{K}^{(1,2)-}_{mnrs}\phi^{(1,2)-rs} \nonumber\\ &\quad +2\phi^{(1,1)} \left( \overline{C}^{(1,1)-}_{mr} \overline{\epsilon}^{(2)-r}{}_{n} -\overline{C}^{(1,1)-}_{nr} \overline{\epsilon}^{(2)-r}{}_{m} \right) \nonumber\\ &\quad +4\phi^{(1,2)} \left( \overline{C}^{(1,2)-}_{mr} \overline{\epsilon}^{(2)-r}{}_{n} -\overline{C}^{(1,2)-}_{nr} \overline{\epsilon}^{(2)-r}{}_{m} \right) \nonumber\\ &\quad +2\phi^{(2,2)} \left( \overline{C}^{(2,2)-}_{mr} \overline{\epsilon}^{(2)-r}{}_{n} -\overline{C}^{(2,2)-}_{nr} \overline{\epsilon}^{(2)-r}{}_{m} \right) \nonumber\\ &\quad +16\left( \overline{\epsilon}^{(2)-}_{mt} \overline{K}^{(1,2)-t}{}_{nrs} \phi^{(1,2)-rs} -\overline{\epsilon}^{(2)-}_{nt} \overline{K}^{(1,2)-t}{}_{mrs} \phi^{(1,2)-rs} \right) \nonumber\\ &\quad -E^{(1,2)}\phi^{(1,2)} \overline{\epsilon}^{(2)-}_{mn} \nonumber\\ &\quad -4g\, \bigl[\phi^{(1,1)},\phi^{(1,2)-}_{mn}\bigr] \overline{\epsilon}^{(1)} -4g\, \bigl[\phi^{(1,2)},\phi^{(1,2)-}_{mn}\bigr] \overline{\epsilon}^{(2)} \nonumber\\ &\quad +8g\, \bigl[\phi^{(1,2)-}_{mr},\phi^{(1,2)-}_{n}{}^r\bigr] \overline{\epsilon}^{(2)} \nonumber\\ &\quad -2g\, \bigl[\phi^{(2,2)},\phi^{(1,1)}\bigr] \overline{\epsilon}^{(2)-}_{mn} +4g\, \bigl[\phi^{(1,1)},\phi^{(1,2)}\bigr] \overline{\epsilon}^{(1)-}_{mn} \nonumber\\ &\quad +8g\, \bigl[\phi^{(1,1)},\phi^{(1,2)-}_{mr}\bigr] \overline{\epsilon}^{(1)-r}{}_{n} -8g\, \bigl[\phi^{(1,1)},\phi^{(1,2)-}_{nr}\bigr] \overline{\epsilon}^{(1)-r}{}_{m} \nonumber\\ &\quad +8g\, \bigl[\phi^{(1,2)},\phi^{(1,2)-}_{mr}\bigr] \overline{\epsilon}^{(2)-r}{}_{n} -8g\, \bigl[\phi^{(1,2)},\phi^{(1,2)-}_{nr}\bigr] \overline{\epsilon}^{(2)-r}{}_{m} \nonumber\\ &\quad -16g\, \bigl[\phi^{(1,2)-}_{mn},\phi^{(1,2)-}_{rs}\bigr] \overline{\epsilon}^{(2)-rs} \nonumber\\ &\quad +\frac{1}{4}\phi^{(1,2)} \left( \nabla_m\epsilon^{(1)}_{n} -\nabla_n\epsilon^{(1)}_{m} -\varepsilon_{mnrs}\nabla^{r}\epsilon^{(1)s} \right) \nonumber\\ &\quad -\frac{1}{4}\phi^{(1,1)} \left( \nabla_m\epsilon^{(2)}_{n} -\nabla_n\epsilon^{(2)}_{m} -\varepsilon_{mnrs}\nabla^{r}\epsilon^{(2)s} \right) \nonumber\\ &\quad -\left( \phi^{(1,2)-}_{mn}\nabla_r\epsilon^{(1)r} -\phi^{(1,2)-}_{nr}\nabla_m\epsilon^{(1)r} +\phi^{(1,2)-}_{mr}\nabla_n\epsilon^{(1)r} \right) \nonumber\\ &\quad -\left( \phi^{(1,2)-}_{nr}\nabla^{r}\epsilon^{(1)}_{m} -\phi^{(1,2)-}_{mr}\nabla^{r}\epsilon^{(1)}_{n} \right) \, .
\end{align}

\paragraph{Scalars.}

\begin{align}
\delta \phi^{(1,1)}
={}&\
2\,\epsilon^{(1)m}\lambda^{(1)}_m
+2\,\overline{\epsilon}^{(2)}\overline{\lambda}^{(2)}
+8\,\overline{\epsilon}^{(2)-}_{mn}
\overline{\lambda}^{(2)-mn}\, , \displaybreak[0]\\
\delta \phi^{(2,2)}
={}&\
2\,\epsilon^{(2)m}\lambda^{(2)}_m
+2\,\overline{\epsilon}^{(1)}\overline{\lambda}^{(1)}
+8\,\overline{\epsilon}^{(1)-}_{mn}
\overline{\lambda}^{(1)-mn}\, , \displaybreak[0]\\
\delta \phi^{(1,2)}
={}&\
-\overline{\epsilon}^{(2)}\overline{\lambda}^{(1)}
-\overline{\epsilon}^{(1)}\overline{\lambda}^{(2)}
+\epsilon^{(2)m}\lambda^{(1)}_m
+\epsilon^{(1)m}\lambda^{(2)}_m
\nonumber\\
&\quad
-4\,\overline{\epsilon}^{(2)-}_{mn}
\overline{\lambda}^{(1)-mn}
-4\,\overline{\epsilon}^{(1)-}_{mn}
\overline{\lambda}^{(2)-mn}\, , \displaybreak[0]\\
\delta \phi^{(1,2)-}_{mn} ={}& +\frac{1}{4}\left(\epsilon^{(2)}_m\lambda^{(1)}_n-\epsilon^{(2)}_n\lambda^{(1)}_m-\varepsilon_{mnrs}\epsilon^{(2)r}\lambda^{(1)s}\right)
   -\frac{1}{4}\left(\epsilon^{(1)}_m\lambda^{(2)}_n-\epsilon^{(1)}_n\lambda^{(2)}_m-\varepsilon_{mnrs}\epsilon^{(1)r}\lambda^{(2)s}\right) \nonumber\\
   &-\overline{\lambda}^{(2)}\overline{\epsilon}^{(1)-}_{mn}+\overline{\lambda}^{(1)}\overline{\epsilon}^{(2)-}_{mn}
   -\overline{\epsilon}^{(2)}\overline{\lambda}^{(1)-}_{mn}+\overline{\epsilon}^{(1)}\overline{\lambda}^{(2)-}_{mn} \nonumber\\
   &+2\left(\overline{\epsilon}^{(2)-}_{mr}\overline{\lambda}^{(1)-r}{}_{n}-\overline{\epsilon}^{(2)-}_{nr}\overline{\lambda}^{(1)-r}{}_{m}\right)
   -2\left(\overline{\epsilon}^{(1)-}_{mr}\overline{\lambda}^{(2)-r}{}_{n}-\overline{\epsilon}^{(1)-}_{nr}\overline{\lambda}^{(2)-r}{}_{m}\right) \, .
\end{align}

\subsection{Geometric Langlands twist}\label{sec: GL matt}
\paragraph{Gauge field.}
\begin{align}
\delta A_m
&= -g \big(
2\overline{\lambda}^{(1)}\epsilon^{(1)}_m
-2\lambda^{(2)}\overline{\epsilon}^{(2)}_m
-2\epsilon^{(2)}\overline{\lambda}^{(2)}_m
+2\overline{\epsilon}^{(1)}\lambda^{(1)}_m
\nonumber\\
&\quad
-8\overline{\epsilon}^{(2)n}\lambda^{(2)+}_{mn}
-8\epsilon^{(2)+}_{mn}\overline{\lambda}^{(2)n}
+8\overline{\epsilon}^{(1)-}_{mn}\lambda^{(1)n}
+8\epsilon^{(1)n}\overline{\lambda}^{(1)-}_{mn} \big) .
\end{align}

\paragraph{Weyl fermions.}
\begin{align} \delta \lambda^{(1)}_m &= -\frac{1}{g} \left( F_{mn} +\frac{1}{2}\varepsilon_{mnrs}F^{rs} \right)\epsilon^{(1)n} \nonumber\\ &\quad -2D_m\phi^{(1,1)}\,\overline{\epsilon}^{(1)} +8D_n\phi^{(1,1)}\,\overline{\epsilon}^{(1)-}_{m}{}^{n} -2D_m\phi^{(1,2)}_{n}\,\overline{\epsilon}^{(2)n} +2\overline{\epsilon}^{(2)}_{m}D_n\phi^{(1,2)n} \nonumber\\ &\quad -2D_n\phi^{(1,2)}_{m}\,\overline{\epsilon}^{(2)n} -2\varepsilon_{mnrs}D^n\phi^{(1,2)s}\overline{\epsilon}^{(2)r} \nonumber\\ &\quad -\phi^{(1,1)}\nabla_m\overline{\epsilon}^{(1)} +4\phi^{(1,1)}\nabla_n\overline{\epsilon}^{(1)-}_{m}{}^{n} -\phi^{(1,2)}_{n}\nabla_m\overline{\epsilon}^{(2)n} \nonumber\\ &\quad +\phi^{(1,2)n}\nabla_n\overline{\epsilon}^{(2)}_{m} -\phi^{(1,2)}_{m}\nabla_n\overline{\epsilon}^{(2)n} -\varepsilon_{mnrs}\phi^{(1,2)s}\nabla^n\overline{\epsilon}^{(2)r} \nonumber\\ &\quad +\phi^{(1,1)}\epsilon^{(2)}\overline{E}^{(1,2)}_{m} +4\phi^{(1,1)}\overline{E}^{(1,2)}_{n} \epsilon^{(2)+}_{m}{}^{n} +\overline{E}^{(1,2)}_{m}\epsilon^{(1)n}\phi^{(1,2)}_{n} \nonumber\\ &\quad +\epsilon^{(1)}_{m}\overline{E}^{(1,2)}_{n}\phi^{(1,2)n} -\phi^{(1,2)}_{m}\overline{E}^{(1,2)}_{n}\epsilon^{(1)n} +\varepsilon_{mnrs}\overline{E}^{(1,2)n} \epsilon^{(1)r}\phi^{(1,2)s} \nonumber\\ &\quad -4\phi^{(2,2)}\overline{E}^{(1,1)-}_{mn}\epsilon^{(1)n} +4\epsilon^{(2)}\overline{E}^{(1,1)-}_{mn}\phi^{(1,2)n} \nonumber\\ &\quad +16\overline{E}^{(1,1)-}_{mr} \epsilon^{(2)+rs}\phi^{(1,2)}_{s} \nonumber\\ &\quad +4\phi^{(2,2)}C^{(1,1)+}_{mn}\epsilon^{(1)n} +4\phi^{(1,1)}C^{(2,2)+}_{mn}\epsilon^{(1)n} +8K^{(1,2)+}_{mnr}\epsilon^{(1)n}\phi^{(1,2)r} \nonumber\\ &\quad -4g\,\epsilon^{(2)} \big[\phi^{(1,1)},\phi^{(1,2)}_{m}\big] -16g\,\epsilon^{(2)+}_{mn} \big[\phi^{(1,1)},\phi^{(1,2)n}\big] \nonumber\\ &\quad +2g\,\epsilon^{(1)}_{m} \big[\phi^{(1,1)},\phi^{(2,2)}\big] +4g\,\epsilon^{(1)n} \big[\phi^{(1,2)}_{m},\phi^{(1,2)}_{n}\big] \nonumber\\ &\quad -2g\,\varepsilon_{mnrs}\epsilon^{(1)s} \big[\phi^{(1,2)n},\phi^{(1,2)r}\big] \, , \displaybreak[0]\\ \delta \lambda^{(2)} &= \frac{2}{g}F_{mn}\epsilon^{(2)+mn} \nonumber\\ &\quad -2D_m\phi^{(2,2)}\,\overline{\epsilon}^{(2)m} +2\overline{\epsilon}^{(1)}D_m\phi^{(1,2)m} +8\overline{\epsilon}^{(1)-}_{mn}D^m\phi^{(1,2)n} \nonumber\\ &\quad -\epsilon^{(2)}\overline{E}^{(1,2)}_{m}\phi^{(1,2)m} +\phi^{(2,2)}\overline{E}^{(1,2)}_{m}\epsilon^{(1)m} -4\overline{E}^{(1,2)}_{m}\epsilon^{(2)+mn}\phi^{(1,2)}_{n} \nonumber\\ &\quad +4\phi^{(1,1)}\epsilon^{(2)+}_{mn}\overline{E}^{(2,2)+mn} -4\overline{E}^{(2,2)+}_{mn}\epsilon^{(1)n}\phi^{(1,2)m} \nonumber\\ &\quad -4\phi^{(2,2)}C^{(1,1)+}_{mn}\epsilon^{(2)+mn} -4\phi^{(1,1)}C^{(2,2)+}_{mn}\epsilon^{(2)+mn} -8K^{(1,2)+}_{mnr}\epsilon^{(2)+mn}\phi^{(1,2)r} \nonumber\\ &\quad -2g\,\epsilon^{(2)} \big[\phi^{(1,1)},\phi^{(2,2)}\big] +4g\,\epsilon^{(1)m} \big[\phi^{(2,2)},\phi^{(1,2)}_{m}\big] \nonumber\\ &\quad +8g\,\epsilon^{(2)+mn} \big[\phi^{(1,2)}_{m},\phi^{(1,2)}_{n}\big] \nonumber\\ &\quad -\phi^{(2,2)}\nabla_m\overline{\epsilon}^{(2)m} +\phi^{(1,2)m}\nabla_m\overline{\epsilon}^{(1)} +4\phi^{(1,2)n} \nabla_m\overline{\epsilon}^{(1)-m}{}_{n} \, , \displaybreak[0]\\ \delta \lambda^{(2)+}_{mn} &= -\frac{1}{4g}\epsilon^{(2)} \left( F_{mn} +\frac{1}{2}\varepsilon_{mnrs}F^{rs} \right) -\frac{1}{g} \left( F_{mr}\epsilon^{(2)+r}{}_{n} - F_{nr}\epsilon^{(2)+r}{}_{m} \right) \nonumber\\ &\quad -\frac{1}{2}\left( D_m\phi^{(2,2)}\,\overline{\epsilon}^{(2)}_{n} -D_n\phi^{(2,2)}\,\overline{\epsilon}^{(2)}_{m} +\varepsilon_{mnrs}D^r\phi^{(2,2)}\,\overline{\epsilon}^{(2)s} \right) \nonumber\\ &\quad +\frac{1}{2}\overline{\epsilon}^{(1)} \left( D_m\phi^{(1,2)}_{n} -D_n\phi^{(1,2)}_{m} +\varepsilon_{mnrs}D^r\phi^{(1,2)s} \right) \nonumber\\ &\quad +2\overline{\epsilon}^{(1)-}_{mn}D_r\phi^{(1,2)r} +2\left( \overline{\epsilon}^{(1)-}_{mr}D^r\phi^{(1,2)}_{n} -\overline{\epsilon}^{(1)-}_{nr}D^r\phi^{(1,2)}_{m} \right) \nonumber\\ &\quad +2\left( \overline{\epsilon}^{(1)-}_{nr}D_m\phi^{(1,2)r} -\overline{\epsilon}^{(1)-}_{mr}D_n\phi^{(1,2)r} \right) \nonumber\\ &\quad +4\left( \overline{\epsilon}^{(1)-}_{nr}D^r\phi^{(1,2)}_{m} -\overline{\epsilon}^{(1)-}_{mr}D^r\phi^{(1,2)}_{n} \right) \nonumber\\ &\quad +\frac{1}{4}\phi^{(2,2)} \left( \overline{E}^{(1,2)}_{n}\epsilon^{(1)}_{m} -\overline{E}^{(1,2)}_{m}\epsilon^{(1)}_{n} +\varepsilon_{mnrs}\overline{E}^{(1,2)s}\epsilon^{(1)r} \right) \nonumber\\ &\quad +\frac{1}{4}\epsilon^{(2)} \left( \overline{E}^{(1,2)}_{m}\phi^{(1,2)}_{n} -\overline{E}^{(1,2)}_{n}\phi^{(1,2)}_{m} +\varepsilon_{mnrs}\overline{E}^{(1,2)r}\phi^{(1,2)s} \right) \nonumber\\ &\quad -\epsilon^{(2)+}_{mn}\overline{E}^{(1,2)}_{r}\phi^{(1,2)r} +\left( \overline{E}^{(1,2)}_{m}\epsilon^{(2)+}_{nr}\phi^{(1,2)r} -\overline{E}^{(1,2)}_{n}\epsilon^{(2)+}_{mr}\phi^{(1,2)r} \right) \nonumber\\ &\quad -\left( \phi^{(1,2)}_{m}\epsilon^{(2)+}_{nr}\overline{E}^{(1,2)r} -\phi^{(1,2)}_{n}\epsilon^{(2)+}_{mr}\overline{E}^{(1,2)r} \right) \nonumber\\ &\quad -\epsilon^{(2)}\phi^{(1,1)}\overline{E}^{(2,2)+}_{mn} +2\phi^{(1,1)} \left( \overline{E}^{(2,2)+}_{mr}\epsilon^{(2)+r}{}_{n} -\overline{E}^{(2,2)+}_{nr}\epsilon^{(2)+r}{}_{m} \right) \nonumber\\ &\quad -\overline{E}^{(2,2)+}_{mn}\epsilon^{(1)r}\phi^{(1,2)}_{r} +\left( \epsilon^{(1)}_{m}\overline{E}^{(2,2)+}_{nr}\phi^{(1,2)r} -\epsilon^{(1)}_{n}\overline{E}^{(2,2)+}_{mr}\phi^{(1,2)r} \right) \nonumber\\ &\quad -\left( \phi^{(1,2)}_{m}\overline{E}^{(2,2)+}_{nr}\epsilon^{(1)r} -\phi^{(1,2)}_{n}\overline{E}^{(2,2)+}_{mr}\epsilon^{(1)r} \right) \nonumber\\ &\quad +\epsilon^{(2)}\phi^{(2,2)}C^{(1,1)+}_{mn} +\epsilon^{(2)}\phi^{(1,1)}C^{(2,2)+}_{mn} +2\epsilon^{(2)}K^{(1,2)+}_{mnr}\phi^{(1,2)r} \nonumber\\ &\quad +2\phi^{(2,2)} \left( C^{(1,1)+}_{mr}\epsilon^{(2)+r}{}_{n} -C^{(1,1)+}_{nr}\epsilon^{(2)+r}{}_{m} \right) \nonumber\\ &\quad +2\phi^{(1,1)} \left( C^{(2,2)+}_{mr}\epsilon^{(2)+r}{}_{n} -C^{(2,2)+}_{nr}\epsilon^{(2)+r}{}_{m} \right) \nonumber\\ &\quad +4\left( K^{(1,2)+}_{mrs}\epsilon^{(2)+r}{}_{n}\phi^{(1,2)s} -K^{(1,2)+}_{nrs}\epsilon^{(2)+r}{}_{m}\phi^{(1,2)s} \right) \nonumber\\ &\quad -2g\big[\phi^{(1,1)},\phi^{(2,2)}\big]\epsilon^{(2)+}_{mn} \nonumber\\ &\quad -4g\,\epsilon^{(2)+}_{mr} \big[\phi^{(1,2)r},\phi^{(1,2)}_{n}\big] +4g\,\epsilon^{(2)+}_{nr} \big[\phi^{(1,2)r},\phi^{(1,2)}_{m}\big] \nonumber\\ &\quad -g\,\epsilon^{(2)} \big[\phi^{(1,2)}_{m},\phi^{(1,2)}_{n}\big] -\frac{g}{2}\varepsilon_{mnrs}\epsilon^{(2)} \big[\phi^{(1,2)r},\phi^{(1,2)s}\big] \nonumber\\ &\quad -g\big[\phi^{(2,2)},\phi^{(1,2)}_{m}\big]\epsilon^{(1)}_{n} +g\big[\phi^{(2,2)},\phi^{(1,2)}_{n}\big]\epsilon^{(1)}_{m} \nonumber\\ &\quad -g\varepsilon_{mnrs} \big[\phi^{(2,2)},\phi^{(1,2)r}\big]\epsilon^{(1)s} \nonumber\\ &\quad -\frac{1}{4}\phi^{(2,2)} \left( \nabla_m\overline{\epsilon}^{(2)}_{n} -\nabla_n\overline{\epsilon}^{(2)}_{m} +\varepsilon_{mnrs}\nabla^r\overline{\epsilon}^{(2)s} \right) \nonumber\\ &\quad +\frac{1}{4}\phi^{(1,2)}_{n}\nabla_m\overline{\epsilon}^{(1)} -\frac{1}{4}\phi^{(1,2)}_{m}\nabla_n\overline{\epsilon}^{(1)} +\frac{1}{4}\varepsilon_{mnrs}\phi^{(1,2)s} \nabla^r\overline{\epsilon}^{(1)} \nonumber\\ &\quad +\phi^{(1,2)r}\nabla_r\overline{\epsilon}^{(1)-}_{mn} +\left( \phi^{(1,2)}_{n}\nabla^r\overline{\epsilon}^{(1)-}_{mr} -\phi^{(1,2)}_{m}\nabla^r\overline{\epsilon}^{(1)-}_{nr} \right) \nonumber\\ &\quad +\left( \phi^{(1,2)r}\nabla_m\overline{\epsilon}^{(1)-}_{nr} -\phi^{(1,2)r}\nabla_n\overline{\epsilon}^{(1)-}_{mr} \right) \nonumber\\ &\quad +2\left( \phi^{(1,2)}_{m}\nabla^r\overline{\epsilon}^{(1)-}_{nr} -\phi^{(1,2)}_{n}\nabla^r\overline{\epsilon}^{(1)-}_{mr} \right) \, , \displaybreak[0]\\ \delta \overline{\lambda}^{(1)} &= \frac{2}{g}F_{mn}\overline{\epsilon}^{(1)-mn} \nonumber\\ &\quad +2\epsilon^{(2)}D_m\phi^{(1,2)m} +8\epsilon^{(2)+}_{mn}D^m\phi^{(1,2)n} -2D_m\phi^{(2,2)}\,\epsilon^{(1)m} \nonumber\\ &\quad -\phi^{(2,2)}E^{(1,2)}_m\overline{\epsilon}^{(2)m} +4E^{(1,1)-}_{mn}\overline{\epsilon}^{(2)m}\phi^{(1,2)n} +4\phi^{(1,1)}E^{(1,1)-}_{mn}\overline{\epsilon}^{(1)-mn} \nonumber\\ &\quad +4E^{(1,2)}_m\phi^{(1,2)}_n\overline{\epsilon}^{(1)-mn} +\overline{\epsilon}^{(1)}E^{(1,2)}_m\phi^{(1,2)m} \nonumber\\ &\quad -4\phi^{(1,1)}\overline{C}^{(1,1)-}_{mn} \overline{\epsilon}^{(1)-mn} -4\phi^{(2,2)}\overline{C}^{(2,2)-}_{mn} \overline{\epsilon}^{(1)-mn} \nonumber\\ &\quad -8\overline{K}^{(1,2)-}_{mnr} \overline{\epsilon}^{(1)-mn}\phi^{(1,2)r} \nonumber\\ &\quad +2g\,\overline{\epsilon}^{(1)} \big[\phi^{(2,2)},\phi^{(1,1)}\big] +4g\,\overline{\epsilon}^{(2)m} \big[\phi^{(2,2)},\phi^{(1,2)}_m\big] \nonumber\\ &\quad +8g\,\overline{\epsilon}^{(1)-mn} \big[\phi^{(1,2)}_m,\phi^{(1,2)}_n\big] \nonumber\\ &\quad -\phi^{(2,2)}\nabla_m\epsilon^{(1)m} +\phi^{(1,2)m}\nabla_m\epsilon^{(2)} +4\phi^{(1,2)m}\nabla_n\epsilon^{(2)+n}{}_{m} \, , \displaybreak[0]\\ \delta \overline{\lambda}^{(1)-}_{mn} &= -\frac{1}{4g}\overline{\epsilon}^{(1)} \left( F_{mn}-\frac{1}{2}\varepsilon_{mnrs}F^{rs} \right) -\frac{1}{g} \left( F_{mr}\overline{\epsilon}^{(1)-r}{}_{n} -F_{nr}\overline{\epsilon}^{(1)-r}{}_{m} \right) \nonumber\\ &\quad +\frac{1}{2}\epsilon^{(2)} \left( D_m\phi^{(1,2)}_{n} -D_n\phi^{(1,2)}_{m} -\varepsilon_{mnrs}D^r\phi^{(1,2)s} \right) \nonumber\\ &\quad +2\epsilon^{(2)+}_{mn}D_r\phi^{(1,2)r} +2\left( \epsilon^{(2)+}_{mr}D^r\phi^{(1,2)}_{n} -\epsilon^{(2)+}_{nr}D^r\phi^{(1,2)}_{m} \right) \nonumber\\ &\quad +2\left( \epsilon^{(2)+}_{nr}D_m\phi^{(1,2)r} -\epsilon^{(2)+}_{mr}D_n\phi^{(1,2)r} \right) \nonumber\\ &\quad +4\left( \epsilon^{(2)+}_{nr}D^r\phi^{(1,2)}_{m} -\epsilon^{(2)+}_{mr}D^r\phi^{(1,2)}_{n} \right) \nonumber\\ &\quad -\frac{1}{2} \left( D_m\phi^{(2,2)}\epsilon^{(1)}_{n} -D_n\phi^{(2,2)}\epsilon^{(1)}_{m} -\varepsilon_{mnrs}D^r\phi^{(2,2)}\epsilon^{(1)s} \right) \nonumber\\ &\quad +\frac{1}{4}\phi^{(2,2)} \left( E^{(1,2)}_{m}\overline{\epsilon}^{(2)}_{n} -E^{(1,2)}_{n}\overline{\epsilon}^{(2)}_{m} -\varepsilon_{mnrs}E^{(1,2)r}\overline{\epsilon}^{(2)s} \right) \nonumber\\ &\quad -\left( E^{(1,1)-}_{mr}\overline{\epsilon}^{(2)r}\phi^{(1,2)}_{n} -E^{(1,1)-}_{nr}\overline{\epsilon}^{(2)r}\phi^{(1,2)}_{m} \right) \nonumber\\ &\quad -2\left( \phi^{(1,2)}_{m}E^{(1,1)-}_{nr}\overline{\epsilon}^{(2)r} -\phi^{(1,2)}_{n}E^{(1,1)-}_{mr}\overline{\epsilon}^{(2)r} \right) \nonumber\\ &\quad -E^{(1,1)-}_{mn}\overline{\epsilon}^{(2)}_{r}\phi^{(1,2)r} +\left( \overline{\epsilon}^{(2)}_{m}E^{(1,1)-}_{nr}\phi^{(1,2)r} -\overline{\epsilon}^{(2)}_{n}E^{(1,1)-}_{mr}\phi^{(1,2)r} \right) \nonumber\\ &\quad -\overline{\epsilon}^{(1)}\phi^{(1,1)}E^{(1,1)-}_{mn} +\frac{1}{4}\overline{\epsilon}^{(1)} \left( E^{(1,2)}_{n}\phi^{(1,2)}_{m} -E^{(1,2)}_{m}\phi^{(1,2)}_{n} +\varepsilon_{mnrs}E^{(1,2)r}\phi^{(1,2)s} \right) \nonumber\\ &\quad +2\phi^{(1,1)} \left( E^{(1,1)-}_{mr}\overline{\epsilon}^{(1)-r}{}_{n} -E^{(1,1)-}_{nr}\overline{\epsilon}^{(1)-r}{}_{m} \right) \nonumber\\ &\quad +\overline{\epsilon}^{(1)-}_{mn}E^{(1,2)}_{r}\phi^{(1,2)r} +\left( E^{(1,2)}_{m}\overline{\epsilon}^{(1)-}_{nr}\phi^{(1,2)r} -E^{(1,2)}_{n}\overline{\epsilon}^{(1)-}_{mr}\phi^{(1,2)r} \right) \nonumber\\ &\quad +2\left( \overline{\epsilon}^{(1)-}_{mr}\phi^{(1,2)r}E^{(1,2)}_{n} -\overline{\epsilon}^{(1)-}_{nr}\phi^{(1,2)r}E^{(1,2)}_{m} \right) \nonumber\\ &\quad -2\left( \overline{\epsilon}^{(1)-}_{mr}E^{(1,2)r}\phi^{(1,2)}_{n} -\overline{\epsilon}^{(1)-}_{nr}E^{(1,2)r}\phi^{(1,2)}_{m} \right) \nonumber\\ &\quad -\left( \phi^{(1,2)}_{m}\overline{\epsilon}^{(1)-}_{nr}E^{(1,2)r} -\phi^{(1,2)}_{n}\overline{\epsilon}^{(1)-}_{mr}E^{(1,2)r} \right) \nonumber\\ &\quad +\overline{\epsilon}^{(1)}\phi^{(1,1)} \overline{C}^{(1,1)-}_{mn} +\overline{\epsilon}^{(1)}\phi^{(2,2)} \overline{C}^{(2,2)-}_{mn} +2\overline{\epsilon}^{(1)} \overline{K}^{(1,2)-}_{mnr}\phi^{(1,2)r} \nonumber\\ &\quad +2\phi^{(1,1)} \left( \overline{C}^{(1,1)-}_{mr}\overline{\epsilon}^{(1)-r}{}_{n} -\overline{C}^{(1,1)-}_{nr}\overline{\epsilon}^{(1)-r}{}_{m} \right) \nonumber\\ &\quad +2\phi^{(2,2)} \left( \overline{C}^{(2,2)-}_{mr}\overline{\epsilon}^{(1)-r}{}_{n} -\overline{C}^{(2,2)-}_{nr}\overline{\epsilon}^{(1)-r}{}_{m} \right) \nonumber\\ &\quad +4\left( \overline{K}^{(1,2)-}_{mrs} \overline{\epsilon}^{(1)-r}{}_{n}\phi^{(1,2)s} -\overline{K}^{(1,2)-}_{nrs} \overline{\epsilon}^{(1)-r}{}_{m}\phi^{(1,2)s} \right) \nonumber\\ &\quad +2g\big[\phi^{(2,2)},\phi^{(1,1)}\big] \overline{\epsilon}^{(1)-}_{mn} \nonumber\\ &\quad -g\big[\phi^{(2,2)},\phi^{(1,2)}_{m}\big] \overline{\epsilon}^{(2)}_{n} +g\big[\phi^{(2,2)},\phi^{(1,2)}_{n}\big] \overline{\epsilon}^{(2)}_{m} \nonumber\\ &\quad +g\,\varepsilon_{mnrs} \big[\phi^{(2,2)},\phi^{(1,2)r}\big]\overline{\epsilon}^{(2)s} \nonumber\\ &\quad -g\,\overline{\epsilon}^{(1)} \big[\phi^{(1,2)}_{m},\phi^{(1,2)}_{n}\big] +\frac{g}{2}\varepsilon_{mnrs}\overline{\epsilon}^{(1)} \big[\phi^{(1,2)r},\phi^{(1,2)s}\big] \nonumber\\ &\quad -4g\,\overline{\epsilon}^{(1)-}_{mr} \big[\phi^{(1,2)r},\phi^{(1,2)}_{n}\big] +4g\,\overline{\epsilon}^{(1)-}_{nr} \big[\phi^{(1,2)r},\phi^{(1,2)}_{m}\big] \nonumber\\ &\quad +\frac{1}{4} \left( \phi^{(1,2)}_{n}\nabla_m\epsilon^{(2)} -\phi^{(1,2)}_{m}\nabla_n\epsilon^{(2)} -\varepsilon_{mnrs}\phi^{(1,2)s}\nabla^r\epsilon^{(2)} \right) \nonumber\\ &\quad +\phi^{(1,2)r}\nabla_r\epsilon^{(2)+}_{mn} +\left( \phi^{(1,2)}_{n}\nabla^r\epsilon^{(2)+}_{mr} -\phi^{(1,2)}_{m}\nabla^r\epsilon^{(2)+}_{nr} \right) \nonumber\\ &\quad +\left( \phi^{(1,2)r}\nabla_m\epsilon^{(2)+}_{nr} -\phi^{(1,2)r}\nabla_n\epsilon^{(2)+}_{mr} \right) \nonumber\\ &\quad +2\left( \phi^{(1,2)}_{m}\nabla^r\epsilon^{(2)+}_{nr} -\phi^{(1,2)}_{n}\nabla^r\epsilon^{(2)+}_{mr} \right) \nonumber\\ &\quad -\frac{1}{4}\phi^{(2,2)} \left( \nabla_m\epsilon^{(1)}_{n} -\nabla_n\epsilon^{(1)}_{m} -\varepsilon_{mnrs}\nabla^r\epsilon^{(1)s} \right) \, , \displaybreak[0]\\
\delta \overline{\lambda}^{(2)}_m &= -\frac{1}{g} \left( F_{mn}-\frac{1}{2}\varepsilon_{mnrs}F^{rs} \right)\overline{\epsilon}^{(2)n} \nonumber\\ &\quad -2D_m\phi^{(1,1)}\,\epsilon^{(2)} +8D_n\phi^{(1,1)}\,\epsilon^{(2)+}_{m}{}^{n} -2D_m\phi^{(1,2)}_{n}\epsilon^{(1)n} \nonumber\\ &\quad +2\epsilon^{(1)}_{m}D_n\phi^{(1,2)n} -2\epsilon^{(1)n}D_n\phi^{(1,2)}_{m} +2\varepsilon_{mnrs}D^n\phi^{(1,2)s}\epsilon^{(1)r} \nonumber\\ &\quad -E^{(1,2)}_m\phi^{(1,1)}\overline{\epsilon}^{(1)} -4\phi^{(1,1)}E^{(1,2)}_n \overline{\epsilon}^{(1)-}_{m}{}^{n} \nonumber\\ &\quad -E^{(1,2)}_m\overline{\epsilon}^{(2)n}\phi^{(1,2)}_{n} -\overline{\epsilon}^{(2)}_mE^{(1,2)}_n\phi^{(1,2)n} +\phi^{(1,2)}_mE^{(1,2)}_n\overline{\epsilon}^{(2)n} \nonumber\\ &\quad +\varepsilon_{mnrs}E^{(1,2)n} \overline{\epsilon}^{(2)r}\phi^{(1,2)s} \nonumber\\ &\quad +4\overline{\epsilon}^{(1)}E^{(2,2)+}_{mn}\phi^{(1,2)n} +16E^{(2,2)+}_{mr} \overline{\epsilon}^{(1)-r}{}_{s}\phi^{(1,2)s} \nonumber\\ &\quad -4\phi^{(2,2)}E^{(2,2)+}_{mn}\overline{\epsilon}^{(2)n} \nonumber\\ &\quad +4\phi^{(1,1)}\overline{C}^{(1,1)-}_{mn} \overline{\epsilon}^{(2)n} +4\phi^{(2,2)}\overline{C}^{(2,2)-}_{mn} \overline{\epsilon}^{(2)n} \nonumber\\ &\quad +8\overline{K}^{(1,2)-}_{mnr} \overline{\epsilon}^{(2)n}\phi^{(1,2)r} \nonumber\\ &\quad -4g\,\overline{\epsilon}^{(1)} \big[\phi^{(1,1)},\phi^{(1,2)}_m\big] -16g\,\overline{\epsilon}^{(1)-}_{mn} \big[\phi^{(1,1)},\phi^{(1,2)n}\big] \nonumber\\ &\quad -2g\,\big[\phi^{(2,2)},\phi^{(1,1)}\big] \overline{\epsilon}^{(2)}_m +4g\,\big[\phi^{(1,2)}_m,\phi^{(1,2)}_n\big] \overline{\epsilon}^{(2)n} \nonumber\\ &\quad +2g\,\varepsilon_{mnrs} \big[\phi^{(1,2)n},\phi^{(1,2)r}\big]\overline{\epsilon}^{(2)s} \nonumber\\ &\quad -\phi^{(1,1)}\nabla_m\epsilon^{(2)} +4\phi^{(1,1)}\nabla_n\epsilon^{(2)+}_{m}{}^{n} -\phi^{(1,2)}_{n}\nabla_m\epsilon^{(1)n} \nonumber\\ &\quad +\phi^{(1,2)n}\nabla_n\epsilon^{(1)}_m -\phi^{(1,2)}_m\nabla_n\epsilon^{(1)n} +\varepsilon_{mnrs}\phi^{(1,2)s}\nabla^n\epsilon^{(1)r} \, . \end{align}
\paragraph{Scalars.}

\begin{align}
\delta \phi^{(1,1)}
&=
-2\overline{\epsilon}^{(2)m}\overline{\lambda}^{(2)}_m
+2\epsilon^{(1)m}\lambda^{(1)}_m ,\\
\delta \phi^{(2,2)}
&=
-2\epsilon^{(2)}\lambda^{(2)}
+2\overline{\epsilon}^{(1)}\overline{\lambda}^{(1)}
-8\epsilon^{(2)+}_{mn}\lambda^{(2)+mn}
+8\overline{\epsilon}^{(1)-}_{mn}\overline{\lambda}^{(1)-mn} \, , \\
\delta \phi^{(1,2)}_m
&=
-\lambda^{(2)}\epsilon^{(1)}_m
+\overline{\lambda}^{(1)}\overline{\epsilon}^{(2)}_m
+\overline{\epsilon}^{(1)}\overline{\lambda}^{(2)}_m
-\epsilon^{(2)}\lambda^{(1)}_m
\nonumber\\
&\quad
-4\epsilon^{(1)n}\lambda^{(2)+}_{nm}
+4\epsilon^{(2)+}_{mn}\lambda^{(1)n}
-4\overline{\epsilon}^{(1)-}_{mn}\overline{\lambda}^{(2)n}
+4\overline{\epsilon}^{(2)n}\overline{\lambda}^{(1)-}_{nm} .
\end{align}

\section{Matching of the fuzzy sphere solution}\label{sec: fuzzy matching}
In this Appendix we show that the background discussed in section \ref{sec: fuzzy sphere background} in flat space corresponds to the one outlined in \cite{Kato:2011yh}. We will do so by explicitly matching the deformation term in their action. In order to do that, we shall write the equations in form notation using the dictionary in table \ref{tab:fuzzy sphere form dictionary}.

Let us parametrise a single mass deformation by $M$
\begin{align}
    s=3M \, ; \quad t=\frac{M}{48} \, .
\end{align}
The mass deformation to the action of $\mathcal{N}=4$ super Yang-Mills reads \cite{Maxfield:2016lok}
\begin{equation}\label{eq: mass action}
\begin{aligned}
    S_{\text{mass}}=& \int \dd^4 x \sqrt{|g|} \text{Tr} \left(\frac{1}{2}\phi_{ij}(M_\phi)^{ij}{}_{kl} \phi^{kl} +\frac12 \lambda^i (M_{\lambda})_{ij}\lambda^j -\frac12 \overline \lambda_i (\overline M_{\overline \lambda})^{ij} \overline \lambda_j \right) \, , \\
    (M_\phi)^{ij}{}_{kl}=& -\overline T_{ab}{}^{ij} \overline T^{ab}{}_{kl} + \frac12 D^{ij}{}_{kl} -\frac{1}{12} \delta^{[i}_{k} \delta ^{j]}_{l} \overline E^{mn}E_{mn} \, ,\\
    (M_{\lambda})_{ij}=& \frac12 E_{ij} \, ,\\
    (\overline M_{\overline\lambda})^{ij} =& \frac12 \overline E^{ij} - \frac12 \overline T_{ab}{}^{ij} \gamma^{ab} \, ,
\end{aligned}
\end{equation}
where we have simplified the mass terms according to the fields which are vanishing on our background. In flat space, the dilatino equation is solved provided that we turn on the following
\begin{equation}
\begin{aligned}
D^{ij}{}_{kl} &= \frac{8st}{3}\Big( \Sigma^{12\,ij}\Sigma^{12}{}_{kl} -\Sigma^{13\,ij}\Sigma^{13}{}_{kl} -\Sigma^{24\,ij}\Sigma^{24}{}_{kl} +\Sigma^{34\,ij}\Sigma^{34}{}_{kl} -\Sigma^{14\,ij}\Sigma^{23}{}_{kl} -\Sigma^{23\,ij}\Sigma^{14}{}_{kl} \Big) \, ,
\end{aligned}
\end{equation}
where
\begin{align}
\Sigma^{pq}{}_{ij} := \delta_i^p\delta_j^q-\delta_i^q\delta_j^p \, .
\end{align}
The mass term in the action in twisted variables and with the renaming proposed in table \ref{tab:fuzzy sphere form dictionary} becomes\footnote{In deriving this result we used the action of \cite{Maxfield:2016lok}. Bilinears with a negative-chirality spinor on the left, here the $\overline \lambda \overline \lambda$ mass terms, are transcribed with the same compensating sign as in \eqref{eq: matter equations euclidean}, so that this result does not depend on the choice of charge conjugation matrix. However, there may be sign differences, which we were not able to check, due to the different reality condition chosen on $\phi^{ij}$. Possible sign differences would not spoil the final conclusion.}
\begin{align} S_{\mathrm{mass}}={}& \int \dd^4 x \sqrt{|g|} \text{Tr} \left( -4M^2 \left( v^2 +\phi\overline{\phi} \right) +\frac49M^2 \left\langle v^-,v^-\right\rangle  -2M \iota_{C} \psi +2M\,\lambda\chi +\frac23M\,\left\langle\lambda^-,\chi^-\right\rangle \right) \, ,
\end{align}
where
\begin{align}
    \left \langle  X,Y \right \rangle = \frac12 X_{mn} Y^{mn} \, .
\end{align}
This is the mass term in eq. 2.35 of \cite{Kato:2011yh} obtained after integrating out the auxiliary field $h$ and at the point $\widetilde{m}=-\frac{M}{3}$, which makes the action positive semi-definite. 

In comparing to \cite{Kato:2011yh}, we are disregarding the commutator terms and the terms proportional to the field strength: such terms come from interaction terms in the action which can be found in \cite{Maxfield:2016lok}. In the present computation we are focusing on the mass terms only \eqref{eq: mass action}. Opposite Clifford algebra conventions cause our two-forms to be anti-self-dual as opposed to self-dual.

\begin{table}[t]
\centering \begin{tabular}{@{}cc|cc@{}} \toprule Coordinate & Form & Coordinate & Form \\
\midrule
$\phi^{(1,1)}$ & $\phi$ & $\lambda^{(1)}_m$ & $2\,C$\\
$\phi^{(2,2)}$ & $-\frac{12}{5}\, \overline \phi$ & $\lambda^{(2)}_m$ & $\frac13\, \psi$ \\
$\phi^{(1,2)}$ & $\sqrt{\frac{12}{5}} v$ & $\overline \lambda^{(1)}$ & $ \lambda$\\
$\phi^{(1,2)-}_{mn}$ & $\iu \sqrt{\frac{1}{38}}\, v^-$ & $\overline \lambda^{(2)}$ & $\chi$ \\
&& $\overline \lambda^{(1)-}_{mn}$ & $\lambda^-$\\
&& $\overline \lambda^{(2)-}_{mn}$ & $\frac18\,\chi^-$\\
&& $\overline \epsilon^{(1)}$ & $w$\\
&& $\overline{\epsilon}^{(2)}$ & $u$ \\
\bottomrule
\end{tabular} \caption{Dictionary between coordinate and form notation of the twisted matter fields used to obtain the mass deformation term in the action. Bosonic fields are in the left column and fermionic fields and supersymmetry parameters in the right one.} \label{tab:fuzzy sphere form dictionary}
\end{table}

\bibliographystyle{JHEP}
\bibliography{biblio}
\addcontentsline{toc}{section}{References}

\end{document}